\pdfoutput=1

\documentclass[11pt,twoside,a4paper,cmspaper,final,collab]{cms-tdr}
\def\svnVersion{220875}\def\cmsCernNo{2013-168}\def\cmsCernDate{\today}\def\cmsMessage{Published in the Journal of High Energy Physics as \href{http://dx.doi.org/10.1007/JHEP12(2013)030}{\doi{10.1007/JHEP12(2013)030.}}}

\begin{document}\cmsNoteHeader{SMP-13-003}

\hyphenation{had-ron-i-za-tion}
\hyphenation{cal-or-i-me-ter}
\hyphenation{de-vices}

\RCS$Revision: 184558 $
\RCS$HeadURL: svn+ssh://svn.cern.ch/reps/tdr2/papers/SMP-13-003/trunk/SMP-13-003.tex $
\RCS$Id: SMP-13-003.tex 184558 2013-05-03 15:58:41Z hdyoo $
\newcommand{\FEWZ} {\textsc{fewz}\xspace}
\cmsNoteHeader{SMP-13-003} % This is over-written in the CMS environment: useful as preprint no. for export versions
\title{Measurement of the differential and double-differential Drell--Yan cross sections in proton-proton collisions at $\sqrt{s} = 7$\TeV}

\date{\today}

\abstract{
Measurements of the differential and double-differential Drell--Yan cross sections are presented using an
integrated luminosity of 4.5\,(4.8)\fbinv in the dimuon (dielectron) channel
of proton-proton collision data recorded with the CMS detector at the LHC at $\sqrt{s} = 7$\TeV.
The measured inclusive cross section in the \cPZ-peak region (60--120\GeV) is
$\sigma(\ell\ell) = 986.4 \pm 0.6\stat \pm 5.9\,\text{(exp. syst.)} \pm 21.7\,\text{(th. syst.)} \pm  21.7\lum\unit{pb}$ for the combination of the dimuon and dielectron channels.
Differential cross sections $\rd\sigma/\rd{}m$ for the dimuon, dielectron, and combined channels are measured in the mass range 15 to 1500\GeV and corrected to the full phase space.
Results are also presented for the measurement of the double-differential
cross section $\rd^2\sigma/\rd{}m\,\rd\abs{y}$ in the dimuon channel
over the mass range 20 to 1500\GeV and absolute dimuon rapidity
from 0 to 2.4. These measurements are compared to the predictions of perturbative QCD calculations at next-to-leading and next-to-next-to-leading orders
using various sets of parton distribution functions.
}

\hypersetup{%
pdfauthor={CMS Collaboration},%
pdftitle={Measurement of the differential and double-differential Drell-Yan cross sections in proton-proton collisions at 7 TeV},%
pdfsubject={CMS},%
pdfkeywords={LHC, CMS, physics, electroweak, Drell-Yan}}

\maketitle %maketitle comes after all the front information has been supplied
\tableofcontents
\newpage
\section{Introduction}
\label{sec:intro}

The Drell--Yan (DY) lepton pair production in hadron-hadron collisions is described
in the standard model by $s$-channel $\gamma^{*}/\cPZ$ exchange. Theoretical calculations of the
differential cross section $\rd\sigma/\rd{}m$ and the double-differential cross section $\rd^2\sigma/\rd{}m\,\rd\abs{y}$,
where $m$ is the dilepton invariant mass and $\abs{y}$ is the absolute value of the dilepton rapidity,
are well established
up to next-to-next-to-leading order (NNLO) in quantum chromodynamics
(QCD)~\cite{QCDNNLO, DYNNLO, DYNNLO1, DY-Theory}.
The rapidity distributions of the gauge bosons $\gamma^{*}/\cPZ$ are sensitive to the parton content of the proton,
and the very high energy of the Large Hadron Collider (LHC) allows the parton distribution functions (PDFs)
to be probed in a wide region of Bjorken $x$ and $Q^2$:
$0.0003 < x < 0.5$ and $500 < Q^2 < 90000\GeV^2$ in the double-differential cross section measurement.
The differential cross section $\rd\sigma/\rd{}m$ is measured in an even higher $Q^2$ region up to $1.2 \times 10^6\GeV^2$.
The large center of mass energy at the LHC allows a substantial extension of the range of Bjorken $x$ and $Q^2$
covered compared to previous experiments~\cite{bib:ZEUS,bib:SLAC,bib:FNAL1,bib:FNAL3,bib:FNAL_late1,bib:FNAL_late2}.

The rapidity $y$ and the invariant mass $m$ of the dilepton system produced in proton-proton collisions are related at leading order
(LO) to the momentum fraction $x_+$ ($x_-$) carried by the
parton in the forward-going (backward-going) proton as described by
the formula \( x_\pm=(m/\sqrt{s}) e^{\pm y}\),
where the forward direction is defined as the positive $z$ direction of the CMS detector coordinate system.
Therefore, the rapidity and mass distributions are sensitive to the PDFs
of the interacting partons.
Since the $y$ distribution is symmetric around zero in proton-proton collisions, we consider
only the differential cross section in $\abs{y}$ in order to reduce the statistical errors.
The measurements of the double-differential cross section $\rd^2\sigma/\rd{}m\,\rd\abs{y}$  in DY
production are particularly important since they provide quantitative tests of perturbative QCD and help to constrain
the quark and antiquark flavor content of the proton. Precise experimental measurements of these
cross sections also allow comparisons to different PDF sets and the underlying theoretical models
and calculations~\cite{Ball:2010de}. In addition, measuring DY lepton-pair production is  important for other
LHC physics analyses because it is a major source of background for various interesting processes,
such as $\ttbar$ and
diboson production, as well as for searches for new physics beyond the standard model, such as the
production of high-mass dilepton resonances.

The existing PDFs are derived from fixed-target and collider
measurements of deep inelastic scattering (DIS), neutrino-nucleon scattering,
inclusive jet production, and vector boson production from H1 and ZEUS~\cite{bib:ZEUS},
SLAC~\cite{bib:SLAC}, FNAL E665, E772, E866~\cite{bib:FNAL1,bib:FNAL3}, and
the CDF and D0~\cite{bib:FNAL_late1,bib:FNAL_late2} experiments.
These experiments
covered the following ranges of dilepton invariant mass and Bjorken scale parameter $x$: $m \le 20$\GeV and $x > 0.01$.
Previous DY measurements from the fixed-target experiments contributed substantially
to the understanding of the quark and antiquark distributions in the proton. Collider vector boson
production data contribute to constraining the $\cPqd/\cPqu$ ratio at high $x$ and the valence quark distributions.
 These data are also important in reducing the theoretical uncertainties in the determination of the $\PW$-boson mass at hadron colliders~\cite{bib:Wmass}.
The current status of the PDFs and the importance of the LHC measurements are reviewed in
Ref.~\cite{bib:Review,bib:Benchmarking}, and the DY differential cross section
has been measured by CMS, LHCb, and ATLAS~\cite{Paper2010,bib:LHCbDY,bib:ATLAS}.

This paper presents measurements of the DY differential cross section $\rd\sigma/\rd{}m$ in the dimuon and dielectron
channels in the mass range $15 < m < 1500$\GeV
 and the double-differential cross section $\rd^2\sigma/\rd{}m\,\rd\abs{y}$ in the dimuon channel  for the mass range $20 < m < 1500$\GeV.
These measurements are performed with the Compact Muon Solenoid (CMS) detector at the LHC
using proton-proton collision data at $\sqrt{s} = 7$\TeV.
The differential cross section measurements are normalized to the $\Z$-peak region (60--120\GeV).
This normalization cancels out the effect of multiple interactions per bunch crossing (pileup) on the reconstruction,
and the uncertainty in the integrated luminosity, acceptance, and efficiency evaluation.
The measurements in this paper are corrected
for the effects of resolution, which cause event migration between bins in mass and rapidity.
The observed dilepton invariant mass is also corrected for final-state photon radiation (FSR).
This effect is most pronounced below the $\Z$ peak.
The differential cross sections
are measured separately for both lepton flavors within the detector acceptance and are extrapolated to the full phase space.
The consistency of the muon and electron channels enables them to be combined and compared with
the NNLO QCD predictions calculated with \FEWZ~\cite{bib:FEWZ} using the CT10 PDF set.
The $\rd^2\sigma/\rd{}m\,\rd\abs{y}$ measurement is compared to the \FEWZ
next-to-leading-order (NLO)
prediction calculated with CT10 PDFs
and the NNLO theoretical predictions as computed with \FEWZ
 using the CT10, NNPDF2.1, MSTW2008, HERAPDF15, JR09, ABKM09, and CT10W PDFs~\cite{CT10, NNPDF, MSTW, HERAPDF, JR09, ABKM, PDF4LHC}.

\section{CMS detector}
\label{sec:CMS-detector}

A right-handed coordinate system is used in
CMS, with the origin at the nominal collision point, the $x$ axis
pointing to the center of the LHC ring, the $y$ axis pointing up
(perpendicular to the LHC plane), and the $z$ axis along the
counterclockwise-beam direction.
The azimuthal angle $\phi$ is the angle relative to the positive $x$ axis
measured in the $x$-$y$ plane.
The central feature of the CMS detector is a superconducting solenoid
providing an axial magnetic field of 3.8\unit{T} and enclosing an all-silicon inner
tracker, a crystal electromagnetic calorimeter (ECAL), and a
brass/scintillator hadron calorimeter.
The tracker is
composed of a pixel detector and a silicon strip tracker, which are used to measure
charged-particle trajectories covering
the full azimuthal angle 
and pseudorapidity interval $\abs{\eta} < 2.5$.
The pseudorapidity $\eta$ is defined as
$\eta = -\ln [\tan (\theta/2)]$, where $\cos \theta = p_z/p$.
Muons are detected in the pseudorapidity range $\abs{\eta}< 2.4$
with four stations of muon chambers.
These muon stations are installed outside the solenoid and
sandwiched between steel layers, which serve both as hadron absorbers and
as a return yoke for the magnetic field flux. They are made using three technologies: drift tubes (DT), cathode
strip chambers (CSC), and resistive-plate chambers. The muons associated with the
tracks measured in the silicon tracker have a transverse momentum (\pt)
resolution of about 1--6\% in the muon \pt range relevant for the analysis
presented in this paper.
Electrons are detected using the energy deposition in the ECAL, which consists of nearly 76\,000 lead tungstate
crystals that are distributed in the barrel region ($\abs{\eta} < 1.479$) and
two endcap regions ($1.479 < \abs{\eta} < 3$). The ECAL
has an ultimate energy resolution better than 0.5\%
for unconverted photons with transverse energies (\ET) above 100\GeV. The
electron energy resolution is better than 3\% for the range of
energies relevant for the measurement reported in this paper.  A detailed
 description of the CMS detector can be found elsewhere~\cite{CMS}.
The CMS experiment uses a two-level trigger system. The
level-1 (L1) trigger, composed of custom processing hardware, selects events
of interest using information from the calorimeters and muon
detectors~\cite{L1TDR}.  The high-level trigger (HLT) is software-based
and further decreases the event collection rate by using the full event
information, including that from the tracker~\cite{HLTTDR}.

\section{Data and Monte Carlo samples}
\label{sec:data}

The measurements reported in this paper are based on pp collision data recorded in 2011
with the CMS detector at the LHC at $\sqrt{s} = 7$\TeV,
corresponding to integrated luminosities of
4.5\fbinv (dimuon channel) and 4.8\fbinv (dielectron channel).

Monte Carlo (MC) samples are used in the analysis for determining
efficiencies, acceptances, and backgrounds from
processes that result in two leptons, and for the determination of systematic uncertainties.
Methods based on control samples in data are used to
determine efficiency correction factors and backgrounds.
The MC samples are produced with a variety of generators, as discussed below. The samples are processed
with the full CMS detector simulation software based on \GEANTfour~\cite{Geant4}, which includes trigger simulation and the full chain of CMS event reconstruction.

The DY signal samples are generated with the NLO generator \POWHEG~\cite{POWHEG} interfaced
with the \PYTHIA v6.4.24~\cite{PYTHIA} parton shower generator (a combination referred to as \POWHEG).
Both $\ttbar$ and single-top-quark samples are produced with the \POWHEG generator,
and the $\tau$-lepton decays are simulated with the \TAUOLA package~\cite{Tauola}.
The $\ttbar$ sample is rescaled to the NLO cross section of 157\unit{pb}.
Diboson samples ($\PW\PW/\PW\Z/\Z\Z$) and QCD multijet background events are produced with \PYTHIA.
An inclusive single-$\PW$-boson sample ($\PW$+jets) is produced using \POWHEG.
The proton structure is defined using the CT10~\cite{CT10} parton distribution functions. All samples are generated using the
\PYTHIA Z2 tune~\cite{Z2} to model the underlying event.
Pileup effects are taken into account in the MC samples, which are generated with
the inclusion of multiple proton-proton interactions
that have timing and multiplicity distributions similar to those observed in data
(average of 9 interactions per bunch crossing).

The \POWHEG MC sample is based on NLO calculations and a correction is added to take NNLO effects into account.
The NNLO effects alter the cross section as a function of the dilepton kinematic variables
and are important in the low-mass region and in normalizing the cross section.
The dilepton correction is determined from the ratio between the
double-differential cross sections (binned in rapidity $y$ and transverse momentum $\pt$)
calculated at NNLO with \FEWZ~\cite{bib:FEWZ} and at NLO with \POWHEG.
The effect of the correction factors on the acceptance is up to 40\% in the low-mass region
and is almost negligible in the high-mass region.
This correction factor $\omega$ is applied on an event-by-event basis. For a given mass range it is defined in bins of dilepton
rapidity $y$ and dilepton transverse momentum $\pt$:

\begin{equation}
\omega(\pt, y) = \frac{(\rd^2\sigma/\rd\pt\,\rd{}y)_\FEWZ}{(\rd^2\sigma/\rd\pt\,\rd{}y)_\POWHEG}.
\label{eq:powheg-reweight}
\end{equation}

The \POWHEG MC events are then reweighted using $\omega$ as defined in Eq.~(\ref{eq:powheg-reweight}).
The reweighted \POWHEG simulation is referred to as NNLO
and is used for all the simulation-based estimations (acceptance, efficiency, FSR corrections) for both the dimuon and dielectron analyses.
The differences between the NNLO reweighted \POWHEG simulations and the \FEWZ predictions, caused by
unavoidable binning/statistics constraints, are used to extract modeling uncertainties.
These modeling uncertainties are shown in the last column of Tables~\ref{tab:syst-muons} and~\ref{tab:syst-electrons}.

\section{Cross section measurements}
\label{sec:analysis}

This analysis measures the DY dimuon and dielectron invariant
mass spectra, $\rd\sigma/\rd{}m$, in the range 15 to 1500\GeV, and then corrects
them for detector geometrical acceptance and kinematic requirements to
obtain the spectra corresponding to the full phase space.
The double-differential cross section $\rd^2\sigma/\rd{}m\,\rd\abs{y}$ is measured in the dimuon channel
within the detector acceptance in the range of
absolute dimuon rapidity from 0 to 2.4 and dimuon invariant mass from 20 to 1500\GeV.
A $\rd^2\sigma/\rd{}m\,\rd\abs{y}$ analysis of the electron channel has not been performed.

The measured cross sections are calculated using the following formula:

\begin{equation}
\label{xsec}
\sigma = \frac{N_\mathrm{u}}{A\cdot\epsilon\cdot\rho\cdot L_\text{int}},
\end{equation}

where $N_\mathrm{u}$ denotes the background-subtracted yield obtained using a matrix inversion unfolding technique to correct for the effects of
the migration of events in mass due to the detector resolution.
The acceptance $A$ and the efficiency $\epsilon$ are both estimated from MC simulation, while $\rho$, the correction (scale) factor accounting for the differences in the efficiency between data and simulation,
is extracted using a technique described in Section~\ref{sec:eff}.
Complete details of all corrections,
background estimations, and the effects of the detector resolution and FSR are contained in later sections of this paper.
The cross sections for these measurements are normalized to the $\Z$-peak region ($60 < m < 120$\GeV)
and thus the integrated luminosity $L_\text{int}$ is only used for the
$\Z$-boson production cross section discussed in Section~\ref{sec:results}.
The differential $\rd\sigma/\rd{}m$ cross section measurements are performed over a mass range from 15 to 1500\GeV
in 40 variable-width mass bins chosen to provide reasonable statistical precision in each bin.

The double-differential cross section measurement is performed in dimuon rapidity space
by choosing a bin size of 0.1--0.2
to reduce migration among the rapidity bins. The mass bins for the
measurement of the double-differential cross section, $\rd^2\sigma/\rd{}m\,\rd\abs{y}$,
are determined on the basis of optimization of
physics background subtraction, and also the number of events per bin. The low-mass region (20--60\GeV),
where QCD processes contribute the most and the FSR
effects are significant, is divided into three bins.
The $\Z$-peak region (60--120\GeV) is a single bin, because in this region the
DY production is dominated by $\Z$-boson exchange,
and this binning is convenient for both normalization and comparison with other measurements.
The high-mass region (120--1500\GeV) is mapped onto two bins based on the number of events available.
The binning is also chosen to make the systematic uncertainties comparable
to the statistical uncertainties away
from the $\Z$-peak region. 
Six invariant mass bins are used, with bin edges 20, 30, 45, 60, 120, 200, and 1500\GeV. For each mass bin,
24 bins of width 0.1 in $\abs{y}$ are defined, except for the
highest mass bin, where only 12 absolute dimuon rapidity bins of width 0.2 are used.

\subsection{Event selection}
\label{sec:evsel}

The experimental signature of DY production is two isolated and oppositely charged
leptons originating from the same primary vertex. The analysis presented in this paper is
based on the dilepton data samples selected by a variety of inclusive double-lepton triggers.

\subsubsection{Muon selection}
The first step in the muon selection is the trigger.
The muon trigger thresholds depend on the instantaneous luminosity, and, since the instantaneous luminosity increased during 2011 data taking period,
the trigger thresholds also increased. In the L1 trigger and HLT processing the data from the multiple detection
layers of the CSC and DT muon chambers are analyzed to provide an estimate of the muon momentum.
For data taken in the earlier part of the 2011 run, the trigger selects dimuon events where each muon
has a transverse momentum of at least 6\GeV.  For the subsequent running periods,
the trigger selects events where one muon has $\pt > 13$\GeV and the other muon has $\pt > 8$\GeV.
The HLT then matches these candidate muon tracks to hits in the silicon tracker to form HLT muon candidates.
In the offline analysis, data from the CSC and DT muon chambers are matched and fitted to data from the
silicon tracker to form global muon candidates.

The muons are required to pass the standard CMS muon identification and quality control criteria, which are
based on the number of hits found in the tracker, the response of the muon chambers, and a set of matching
criteria between the muon track parameters as measured by the CMS tracker and
those measured in the muon chambers~\cite{bib:MUO10004}. Both muons are required to match the HLT trigger objects.
Cosmic-ray muons that traverse the CMS detector close to the interaction point can appear as back-to-back dimuons;
 these are removed by requiring both muons to have an impact parameter in the
transverse plane of less than 2~mm with respect to the center of the interaction region. Further, the
opening angle between the two muons is required to differ from $\pi$ by more than 5\unit{mrad}.
In order to reject muons from pion and kaon decays, a common vertex
for the two muons is required. An event is rejected if the dimuon vertex probability
is smaller than 2\%. More details on muon reconstruction and identification can be found in Ref.~\cite{bib:MUO10004}.

To suppress the background contributions due to muons
originating from heavy-quark decays and nonprompt muons from hadron decays,
both muons are required to be isolated from other particles.
The muon isolation criterion is based on the sum of the transverse momenta
of the particles reconstructed with the CMS particle-flow
algorithm~\cite{bib:PF}
within a cone of size $\Delta R = 0.3$ centered on the muon direction,
where $\Delta R = \sqrt{(\Delta\eta)^2+(\Delta\phi)^2}$;
photons and the muon candidate itself are excluded from the sum.
The ratio of the summed transverse momenta to the transverse momentum of the muon candidate
is required to be less than 0.2.

Each muon is required to be within the acceptance of the muon subsystem ($\abs{\eta} < 2.4$).
The leading muon in the event is required to have a transverse momentum $\pt > 14$\GeV and the trailing muon
$\pt > 9$\GeV, which allows us to operate on the plateau region of the trigger efficiency.
Events are selected for further analysis if they contain opposite-charge muon pairs  meeting the above requirements.
If more than one dimuon candidate passes these selections, the pair with the highest $\chi^2$ probability
for a kinematic fit to the dimuon vertex is selected.
No attempt has been made in this analysis to use the radiated photons detected in the
ECAL to correct the muon energies for possible FSR.
(Section~\ref{sec:fsr} contains a discussion of FSR effects.)

\subsubsection{Electron selection}
Dielectron events are selected when triggered by two electrons with minimum $\ET$ requirements
of 17\GeV for one of the electrons and 8\GeV for the other.
The triggers are the lowest threshold double-electron triggers in the 2011 data and
allow one to probe the lowest possible dielectron mass.
The selection of events at the trigger level, based on the isolation and the quality of
an electron candidate, made it possible for the thresholds to remain unchanged
throughout the full period of 2011 data taking in spite of the rapidly increasing luminosity.

The dielectron candidates are selected online by requiring two clusters in the ECAL, each with a transverse energy $\ET$
exceeding a threshold value. The offline reconstruction of the electrons starts by building
superclusters~\cite{bib:EGM11001} in the ECAL
in order to collect the energy radiated by bremsstrahlung in the tracker material. A specialized tracking
algorithm is used to accommodate changes of the curvature caused by the bremsstrahlung. The superclusters are then matched to the
electron tracks. The electron candidates are required to have a minimum
$\ET$ of 10\GeV after correction for the ECAL energy scale. In order to avoid the inhomogeneous response
at the interfaces between the ECAL barrel and endcaps, the electrons are further required to be detected
within the pseudorapidity ranges $\abs{\eta} < 1.44$ or $1.57 < \abs{\eta} < 2.5$.

The reconstruction of an electron is based on the CMS particle-flow algorithm~\cite{bib:PF}.
The electrons are identified by  means of shower shape variables while the electron isolation criterion is based on a variable that
combines tracker and calorimeter information.
For isolation, the transverse momenta of the particles within a cone of
$\Delta R < 0.3$ are summed, excluding the electron candidate itself.
The ratio of the summed transverse momenta to the transverse momentum of the electron
candidate is required to be less
than 0.15 for all the electrons, except for those with $\ET < 20$\GeV in the endcaps, where the requirement
is tightened to be less than 0.10.
The isolation criteria are optimized to maximize the rejection of misidentified electrons from QCD multijet
production and the nonisolated electrons from the semileptonic decays of heavy quarks.
The electron candidates are required to be consistent
with particles originating from the primary vertex in the event.
Electrons originating from photon conversions
are suppressed by requiring that there be no missing tracker hits before the first hit on the
reconstructed track matched to the electron, and also by rejecting a candidate if it forms a pair
with a nearby track that is consistent with a conversion. Additional details on electron reconstruction
and identification can be found in Ref.~\cite{bib:EGM11001}.

Both electrons are selected with the impact parameter requirements $\abs{d_{xy}} < 0.02$\unit{cm} and $\abs{d_z} < 0.1$\unit{cm} with respect to the primary vertex.
The leading electron candidate in an event is required to have a transverse momentum of $\pt > 20$\GeV,
while the  trailing electron candidate must have $\pt >10$\GeV. As with muons, electrons are required to match
HLT trigger objects, but no charge requirement is imposed on the electron pairs
to avoid efficiency loss due to nonnegligible charge misidentification.
\subsection{Background estimation}
\label{sec:bkg}

There are several physical and instrumental backgrounds that contribute to the sample of dilepton candidates.
The main backgrounds in the region of high invariant masses (above the $\Z$ peak)
are due to $\ttbar$ and diboson production followed by
leptonic decays, while the DY production of $\Pgt^{+}\Pgt^{-}$ pairs
is the dominant source of background in the region just below the $\Z$ peak. At low values of the dimuon invariant mass (up to 40\GeV), most of the background events
are due to QCD events with multiple jets (QCD multijet). The situation is slightly different for electrons in the final state. At low values of dielectron invariant mass most of the background events are from  $\Pgt^{+}\Pgt^{-}$ and $\ttbar$ processes, whereas the contribution
from the QCD multijet process is small due to the stricter selection for electrons compared to muons.

A combination of techniques is used to determine contributions from various background processes.
Wherever feasible, the background rates are estimated from data,
reducing the uncertainties related to simulation of these sources. The remaining contributions
are evaluated using simulation. The background estimation is performed by
following the same methods for both the $\rd\sigma/\rd{}m$ and $\rd^2\sigma/\rd{}m\,\rd\abs{y}$ measurements.

\subsubsection{Dimuon background estimation}
\label{sec:DimuonBkgrnd}
In the dimuon channel, the QCD multijet background is evaluated using control data samples. This method makes use of the muon isolation and the sign of the charge as two independent discriminating variables to identify a signal region and three background regions in the two-dimensional space defined by the muon charge sign and the isolation. The background estimation is then based on the ratio between the number of signal and background events in the above regions~\cite{bib:CMS_WZ}.

The $\ttbar$ background, which is the dominant process at high masses, is estimated from data using a sample of events with $\Pe\Pgm$ pairs.
The estimated number of $\Pgmp\Pgmm$ events can be expressed as a function of observed $\Pe^\pm\Pgm^\mp$ events based on acceptance and efficiencies determined from simulation.
The electron and muon candidates in the $\Pe\Pgm$ sample are required to satisfy
the $\mathrm{DY}\to \Pep\Pem$ and $\mathrm{DY}\to \Pgmp\Pgmm$ selection criteria, respectively.
The electron candidates are required to have $\ET > 20$\GeV, the muon candidates $\pt>15$\GeV,
and both candidates are required to be within the range $\abs{\eta}<2.4$.
They are further required to pass the lepton quality criteria.
The number of expected $\Pgmp\Pgmm$ events is calculated bin by bin as a function of the dilepton mass.
Deviations from the MC simulation are used for assessing the systematic uncertainties.
All other backgrounds are estimated using MC simulation, although
an estimation of all non-QCD multijet backgrounds has been performed with the $\Pe\Pgm$
method of data analysis as a cross-check.

\begin{figure}[htpb]
\centering
\includegraphics[width=0.49\textwidth]{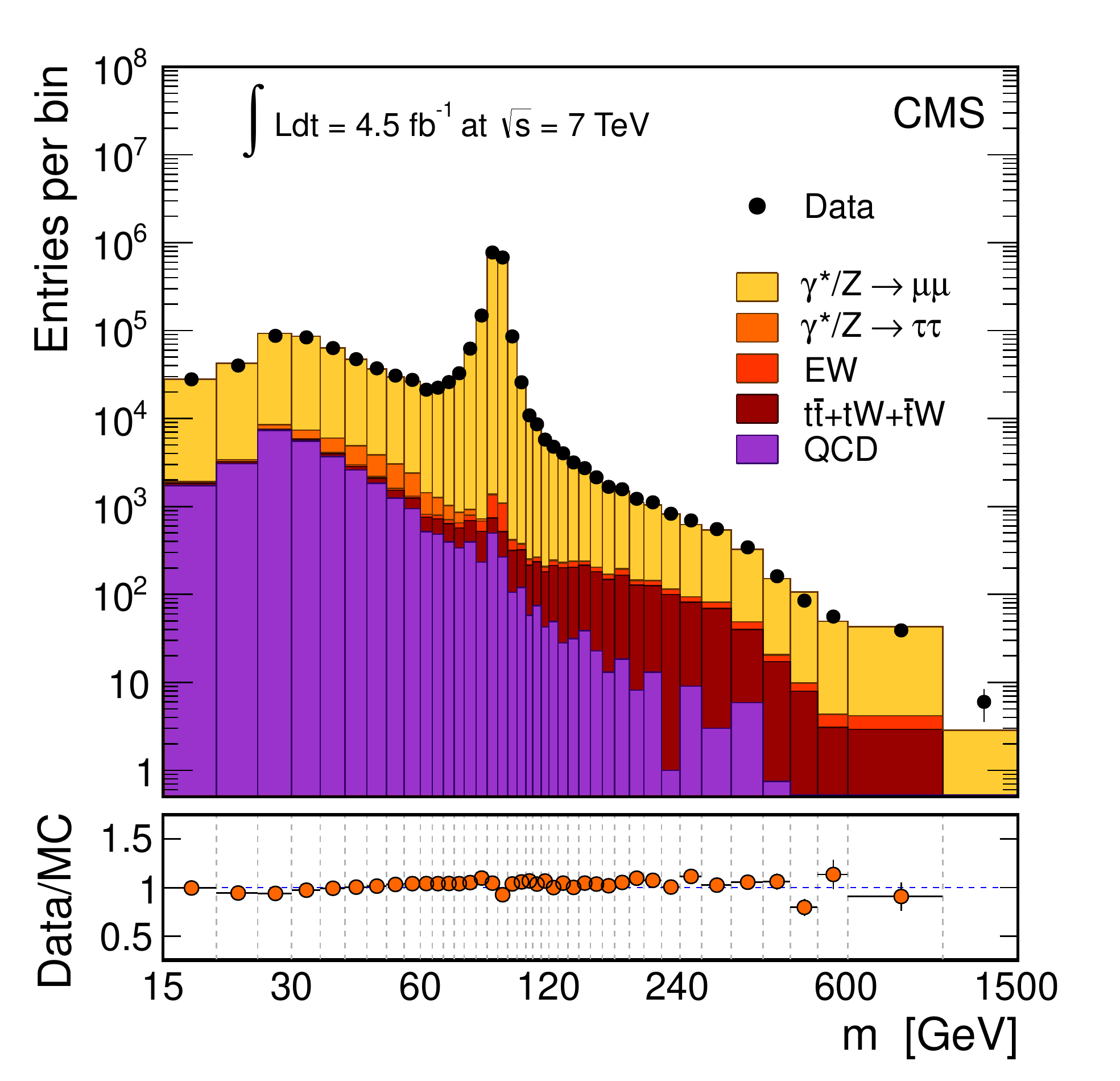}
\includegraphics[width=0.49\textwidth]{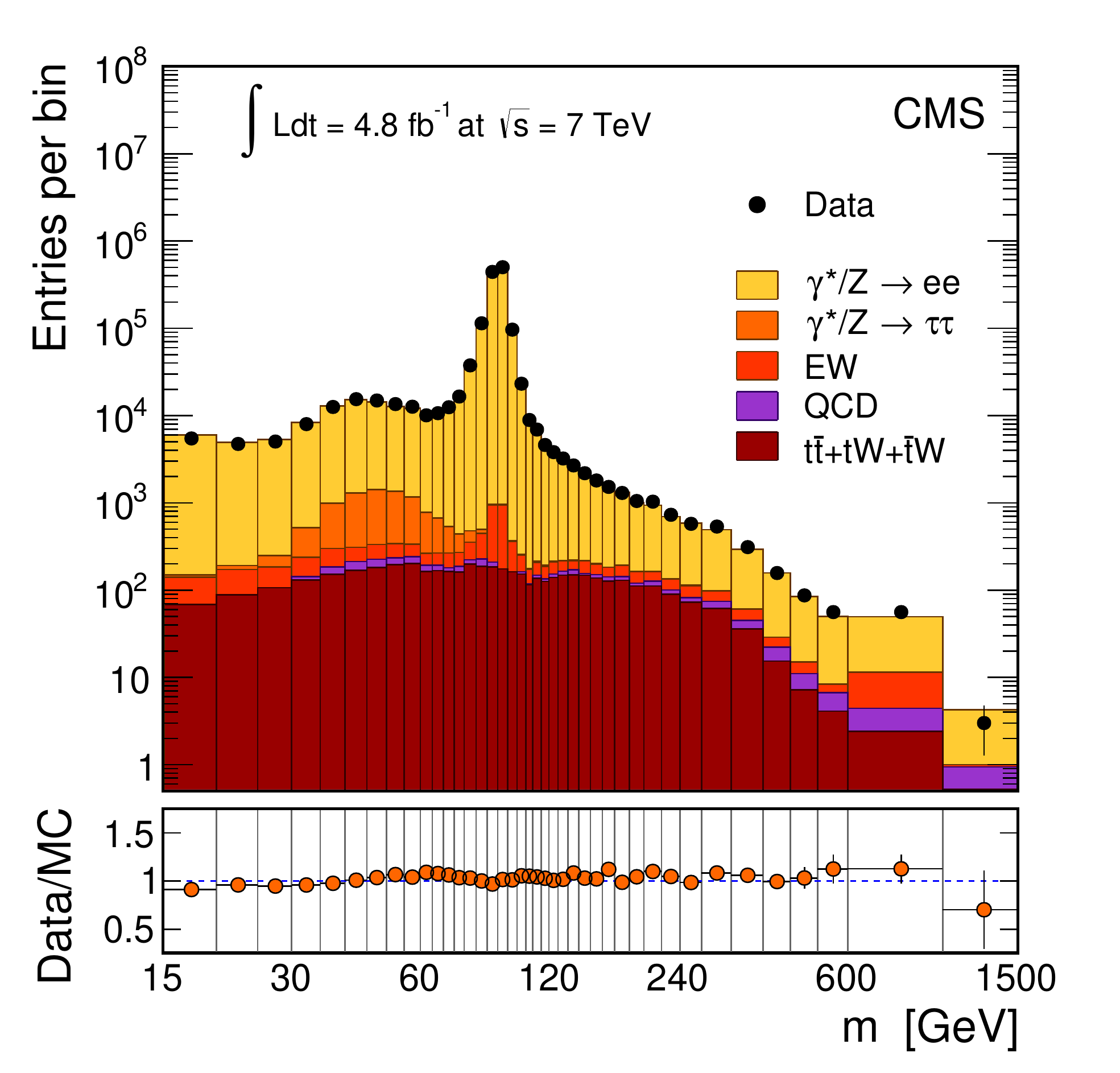}
\caption{
\label{fig:yields1}
The observed dimuon (left) and dielectron (right) invariant mass spectra
for data and MC events and the corresponding ratios of observed to expected yields.
The QCD multijet and $\ttbar$ background yields in the muon channel and the QCD multijet
contribution in the electron channel
are predicted using control samples in data.
The EW histogram indicates the diboson and $\PW$+jets production.
The NNLO reweighted \POWHEG MC signal sample is used.
No other corrections are applied. Error bars are statistical only.
}
\end{figure}

\begin{figure}[htpb]
\centering
\includegraphics[width=0.45\textwidth]{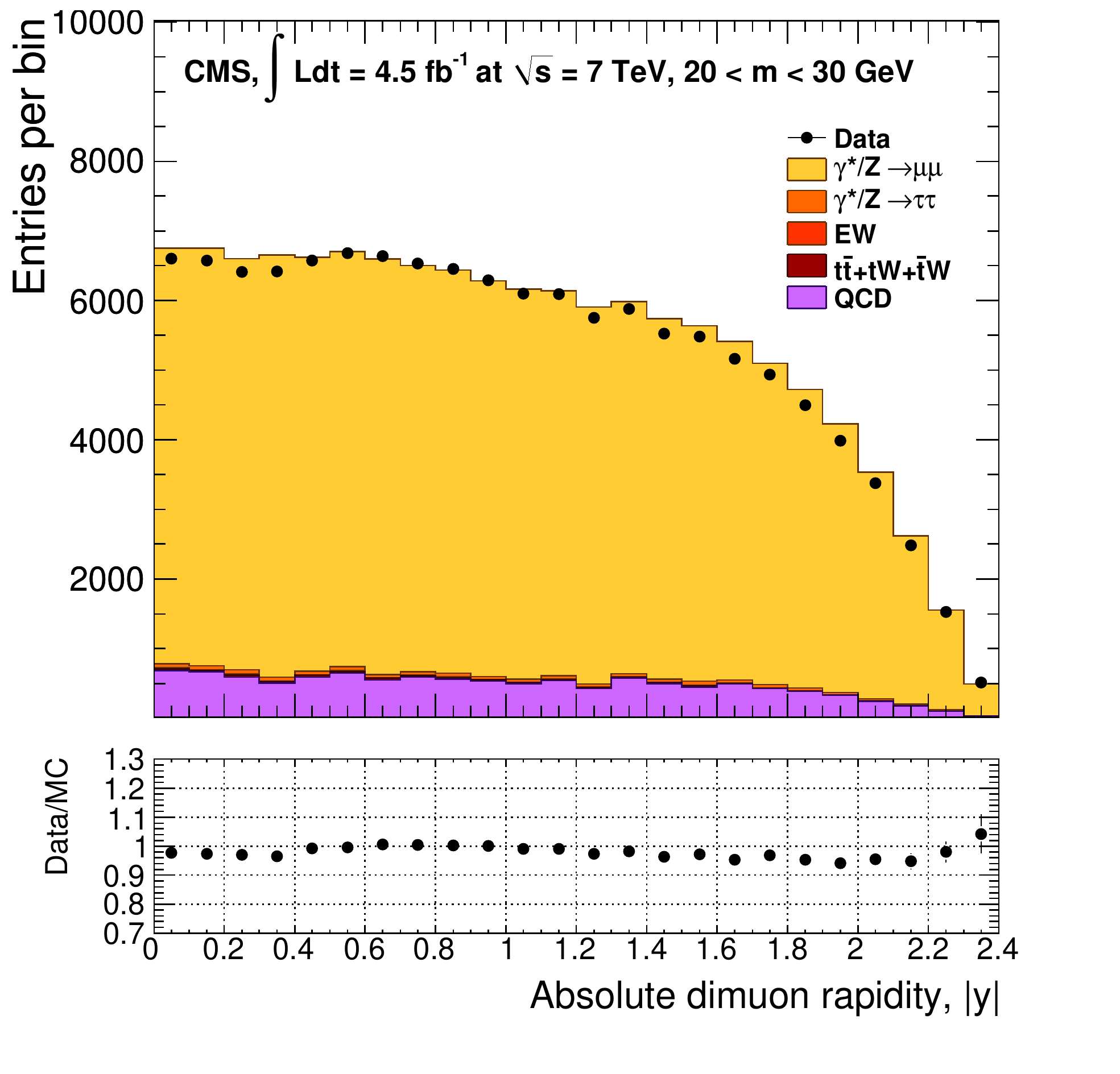}
\includegraphics[width=0.45\textwidth]{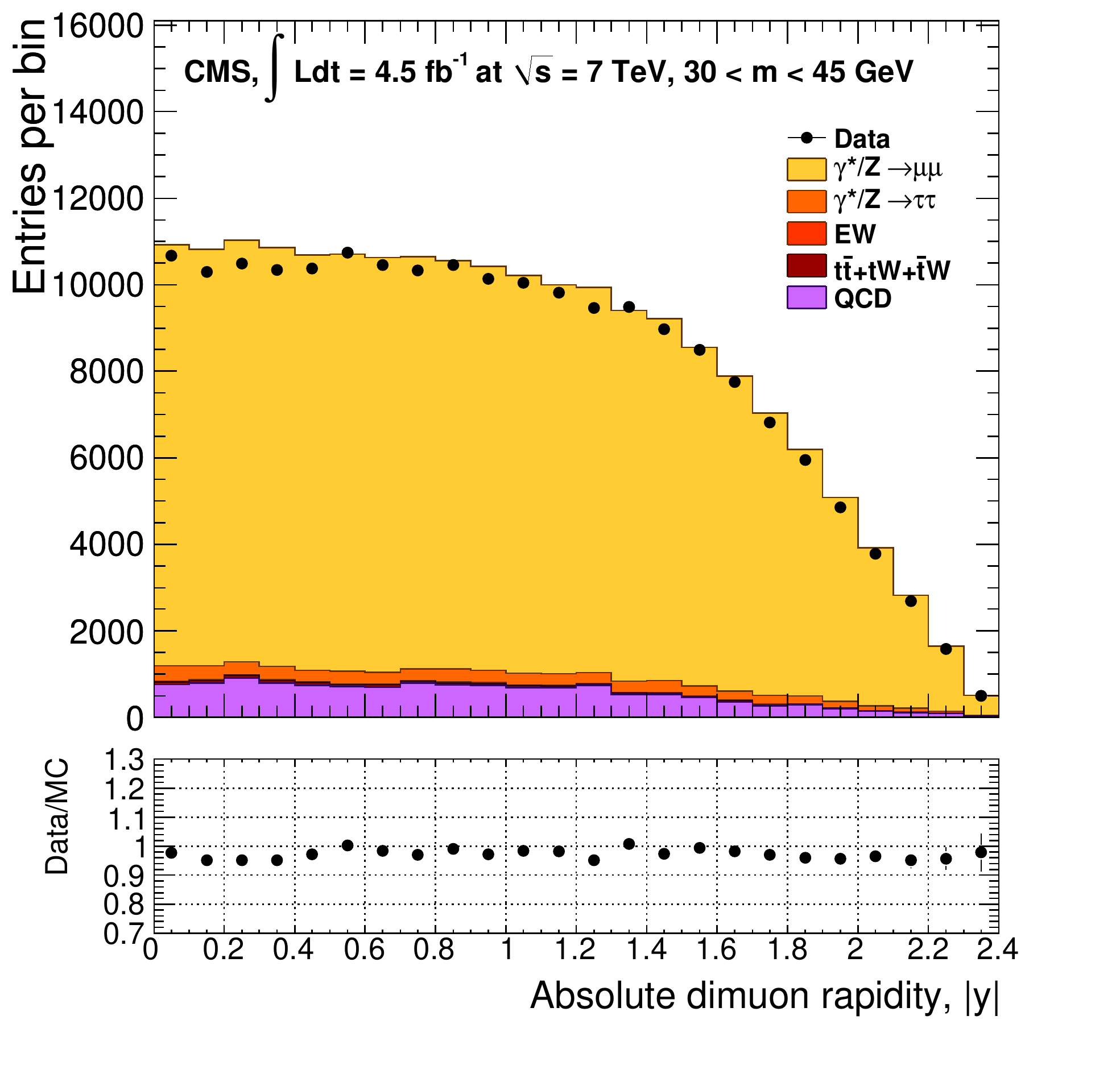}
\includegraphics[width=0.45\textwidth]{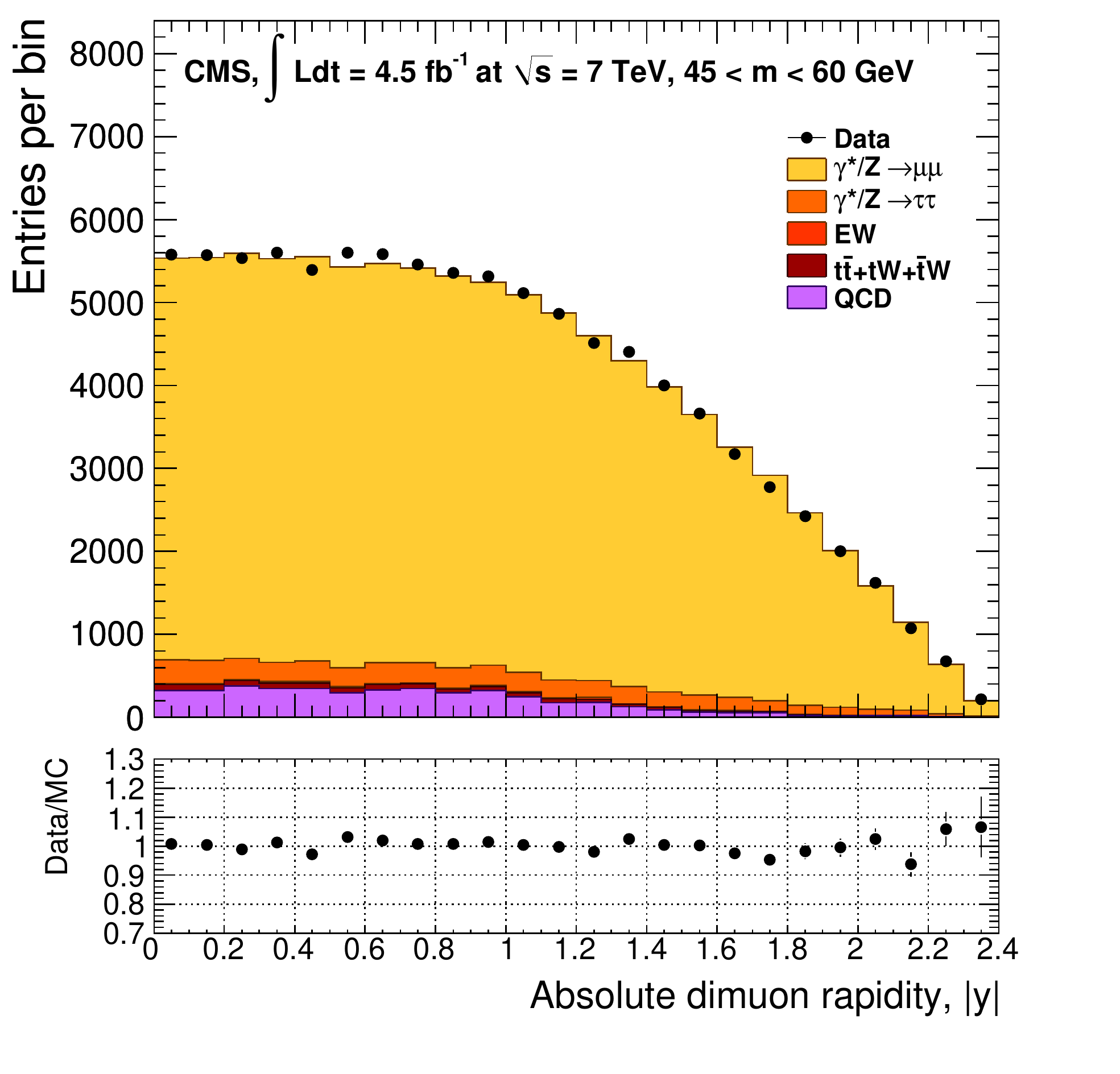}
\includegraphics[width=0.45\textwidth]{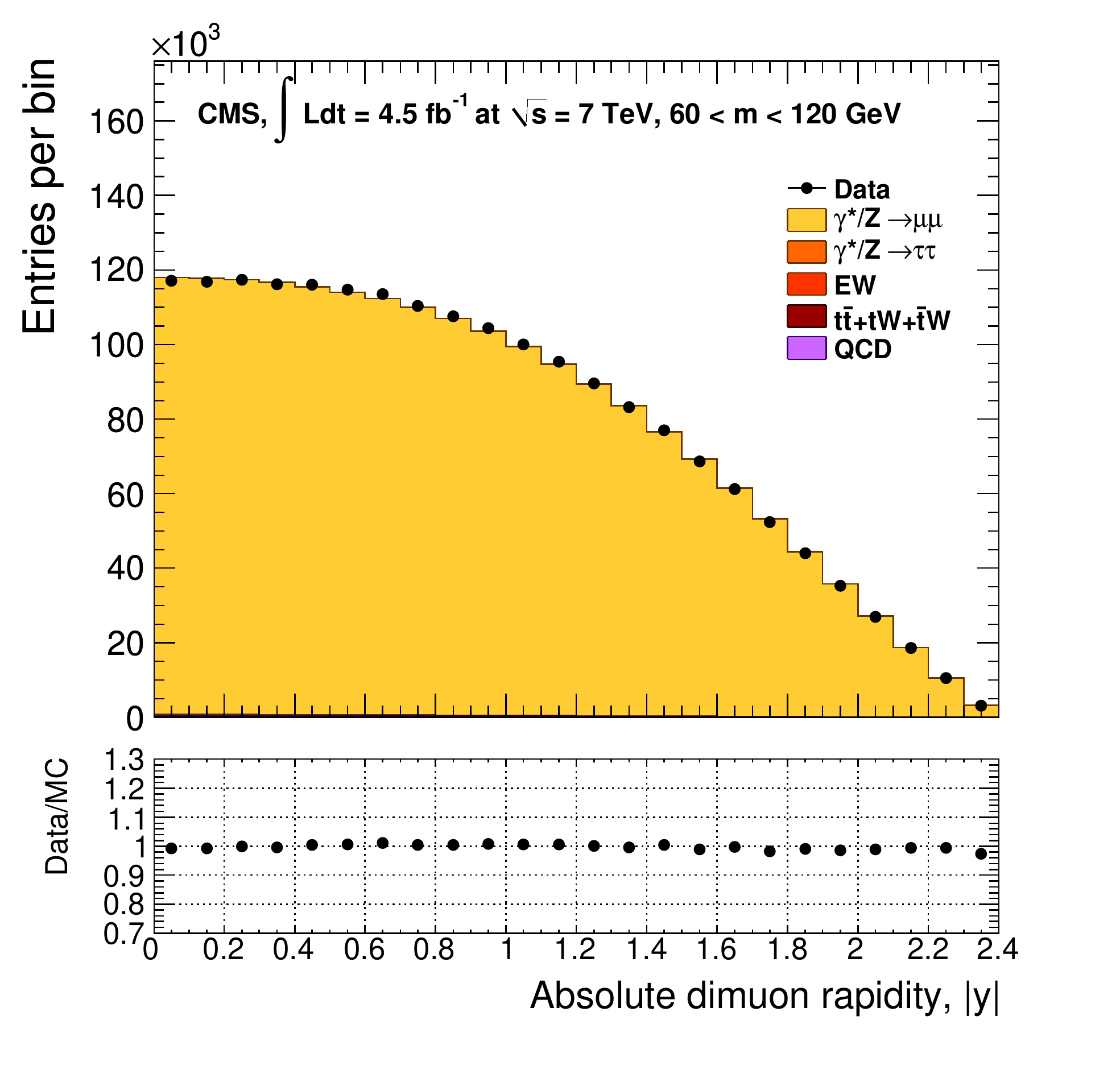}
\includegraphics[width=0.45\textwidth]{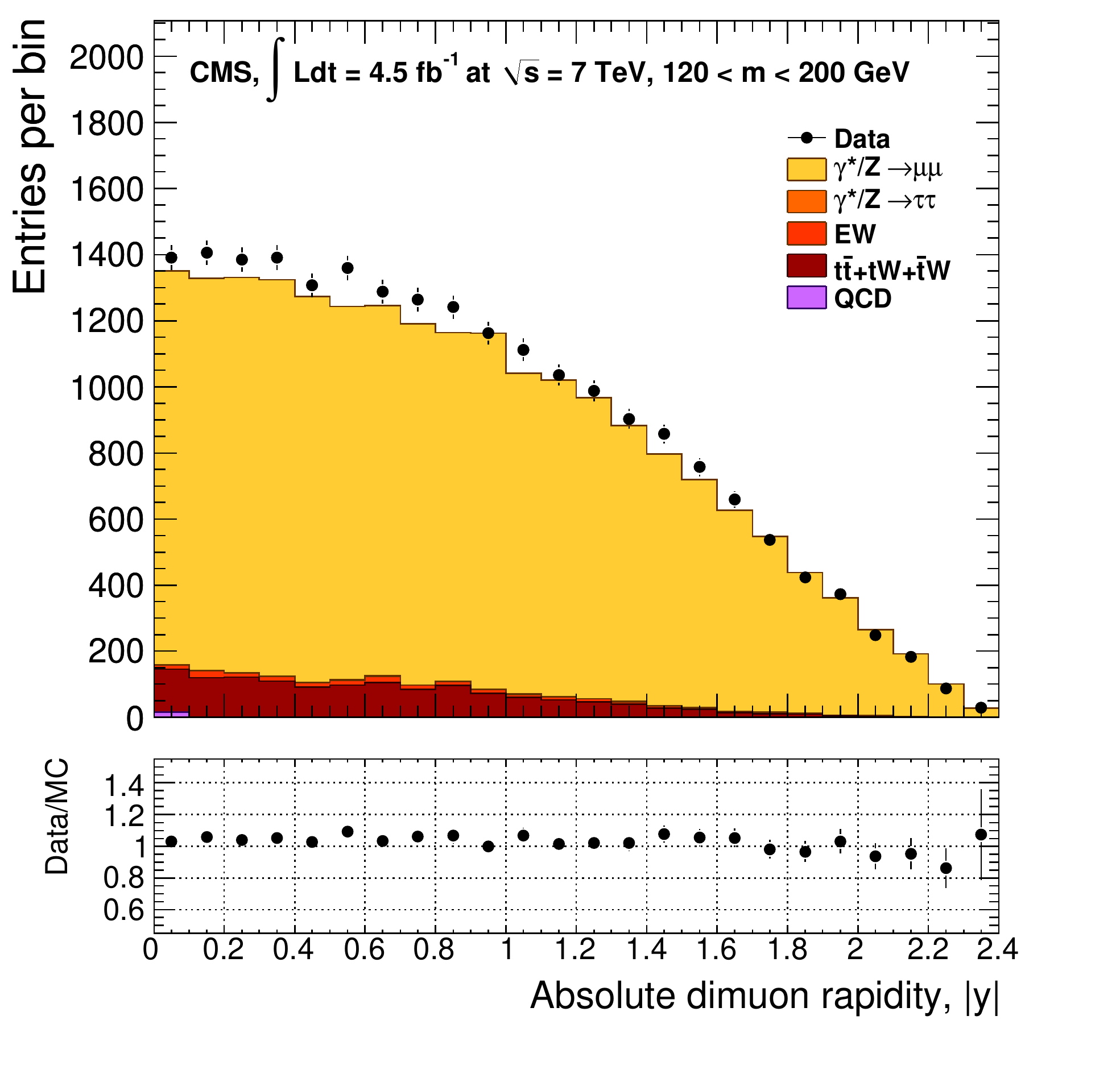}
\includegraphics[width=0.45\textwidth]{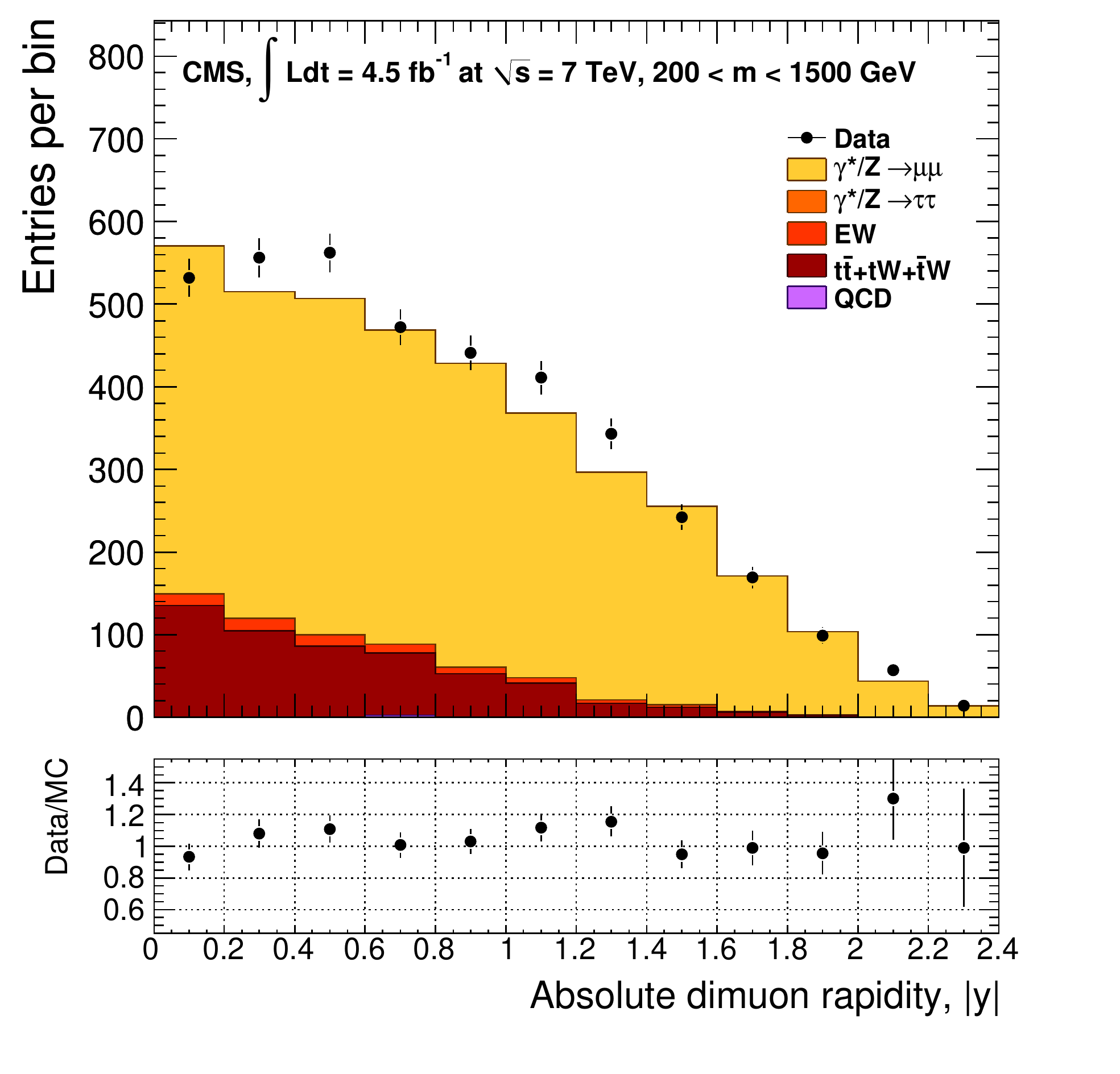}
\caption{
\label{fig:yields3}
The observed dimuon rapidity spectra per invariant mass bin for data and MC events.
There are six mass bins between $20$ and $1500$\GeV, from left to right and from top to bottom.
The NNLO reweighted \POWHEG MC signal sample is used.
The EW histogram indicates the diboson and $\PW$+jets production.
The normalization factors are determined using the number of events in data in the $\Z$-peak region,
and they are applied to all of the mass bins.
Error bars are statistical only.
}
\end{figure}

The expected shapes and relative dimuon yields from data and MC events
in bins of invariant mass can be seen in Fig.~\ref{fig:yields1}.
As shown in the figure, the QCD multijet process is the dominant background in the low-mass region, contributing up to about 10\% in the dimuon rapidity distribution.
In the high-mass regions, $\ttbar$ and single-top-quark ($\cPqt\PW$) production processes are dominant
and collectively contribute up to about 20\%.
The expected shapes and relative dimuon yields from data and MC events
in bins of dimuon rapidity, per invariant mass bin, can be seen in Fig.~\ref{fig:yields3}.

\subsubsection{Dielectron background estimation}
\label{sec:DielectronBkgrnd}
In the dielectron channel, the background processes do contain genuine leptons in most cases.
The background can be divided into two categories:  (1) both electrons are genuine,
and (2) one or both electrons are due to misidentification.

The genuine electron background is estimated from data using the $\Pe\Pgm$ method described above.
The dominant electroweak (EW) background from low invariant mass up to the $\Z$ peak is $\mathrm{DY}\to \Pgt^+\Pgt^-$.
Above the $\Z$ peak the background contributions from $\ttbar$ and diboson production become significant, with relatively smaller contributions from the $\cPqt\PW$ process.
All of these processes produce $\Pe^\pm\Pgm^\mp$ final states at twice the rate of $\Pep\Pem$ or $\Pgmp\Pgmm$. Consequently, the backgrounds from these modes can be measured from a sample of  $\Pe^\pm\Pgm^\mp$ after accounting for the  differences in the acceptance and efficiency. The contributions from $\mathrm{DY} \to \Pgt\Pgt, \ttbar, \cPqt\PW$, and the dibosons to the $\Pep\Pem$ spectrum are estimated from $\Pe\Pgm$ data.
The simulation accurately describes the sample of
$\Pe\Pgm$ events, both in terms of the number of events as well as the shape of the invariant-mass spectrum.

In addition to the genuine $\Pep\Pem$ events from EW processes,
there are events in which jets are falsely identified as
electrons. These are either QCD multijet events where two jets pass the
electron selection criteria or $\PW$+jets events where the $\PW$ boson decays to
an electron and a neutrino, and a jet is misidentified as an
electron. The probability for a jet to pass the requirements of the
electromagnetic trigger and to be falsely reconstructed as an
electron is determined from a sample of events collected with the
trigger requirement for a single electromagnetic cluster
in the event. To ensure that this sample is dominated by jets, the
events are required to have a missing transverse energy $\ETslash<10$\GeV,
and events with more than one particle identified as an electron are
rejected. The jet misidentification probability is measured as a
function of jet \ET and absolute pseudorapidity $\abs{\eta}$.

The number of $\Pep\Pem$ background events is then determined from a
different sample, the sample of events collected with the
double-electron trigger in which at least one electron candidate fails
the full electron selection of the analysis. The events from this
sample are assigned weights based on the expected misidentification
probability for the failing electron candidates, and the sum of the
weights yields the prediction for the background from this source. Since
events in this double-electron trigger sample with at least one
electron failing the full selection contain a small fraction of genuine
DY events, the contribution of the latter is subtracted using
simulation.

The expected shapes and the relative  yields of dielectron events
from data and simulation
in bins of invariant mass are shown in Fig.~\ref{fig:yields1} in
the same format as the dimuon channel.
The genuine electron background is largest in high-mass regions,
where it reaches up to 15--20\% of the observed yields due to $\ttbar$ events.
At the lowest masses, the genuine electron background level, which is dominated by the $\mathrm{DY}\to \Pgt^+\Pgt^-$ contribution,
becomes significant at $\sim$50\GeV, where it ranges up to 10\%.
In other mass ranges
the genuine electron background is typically a few percent and, in particular, it is
very small (less than 0.5\%) in the $\Z$-peak region.
The background associated with falsely identified electrons is relatively small
in the full mass range.

\subsection{Resolution and scale corrections}
\label{sec:unfolding}

Lepton energy and momentum measurements can directly affect
the reconstructed dilepton invariant mass and are, therefore, important in obtaining a correct differential cross section.

The momentum resolution of muons with $\pt<200$\GeV comes primarily from the measurements in the silicon tracker.
A residual misalignment remains in the tracker that is not fully reproduced by the simulation.
This misalignment
leads to a bias in the reconstructed muon momenta
which is removed using a momentum scale correction.

The corrections to muon momenta are extracted separately for
positively and negatively charged muons using the average of the $1/\pt$ spectra
of muons and the dimuon mass from $\Z$ boson decays in bins of muon charge, the polar angle $\theta$, and the azimuthal angle $\phi$.
The same procedure is followed for both data and MC samples.
The correction to $1/\pt$ has two components: an additive component that removes
the bias originating from tracker misalignment, and a multiplicative component
that corrects for residual mismodeling of the magnetic field.
For a 40\GeV muon, the additive correction varies from 0.4\% at small $\abs{\eta}$ to 9\%
at large $\abs{\eta}$. The multiplicative correction is typically much smaller (about 1.0002).

The average reconstructed $\Z$-boson mass is found to be independent of muon $\phi$.
The position of the $\Z$-boson mass peak
in the corrected distribution is different from the expected
$\Z$-boson mass~\cite{PDG} by only ($0.10 \pm 0.01$)\% in data and ($0.00\pm0.01$)\% in
simulation. The small remaining shift in data is corrected by
an additional overall scale correction.
A detailed description of the correction for the muon momentum is given in Ref.~\cite{bib:momcor}.

The electron energy is derived primarily from the measurements of the energy deposited
by the electrons in the ECAL. The energy of these deposits is subject to a set of
corrections following the standard CMS procedures~\cite{bib:EGM11001}.
In addition, energy scale corrections are obtained from the
analysis of the $\Z\rightarrow \Pep\Pem$ peak according to the procedure
described in Ref.~\cite{bib:CMS_WZ}.
These energy scale corrections, which
go beyond the standard CMS electron reconstruction,
range from 0\% to 2\% depending on the pseudorapidity of the electron.

\subsubsection{Unfolding}
\label{sec:unfoldingProcedure}
The effects of detector resolution that cause migration of events among
the analysis bins are corrected through an unfolding procedure~\cite{bib:blob}.
This procedure maps the true lepton distribution onto the measured one,
while taking into account migration of events into and out of the mass and rapidity range of this measurement.
The unfolding procedure used
for the differential and double-differential cross section calculations is described below.

The unfolding of the detector resolution effects is performed prior to corrections for FSR.
The  response matrix $T_{ik}$ for the unfolding,
which gives the fraction of events from bin $k$ of the true distribution
that ends up reconstructed in bin $i$,
is calculated from simulation:

\begin{equation}
N^\text{obs}_i = \sum_{k} T_{ik}N^\text{true}_k.
   \label{eq:unfolding_response}
\end{equation}

In the case of the measurement of $\rd\sigma/\rd{}m$, the matrix is nearly diagonal with a few significant
off-diagonal elements located adjacent to the main diagonal. The effect of regularization on the
unfolding is tested using simulation and found to be negligible.
Therefore, both the dimuon and dielectron response matrices are inverted without regularization.

For the double-differential cross section measurement, a specific procedure has been developed in order to take into
account the effect of migration in bins of dilepton rapidity.
Within the framework of the unfolding method for the double-differential cross section measurement,
a two-dimensional yield distribution (matrix) in bins of dilepton invariant mass and
rapidity is transformed into a one-dimensional distribution by mapping onto a one-dimensional vector.
This procedure amounts to a simple index transformation without any loss of information.
Once the one-dimensional distribution is obtained,
the unfolding procedure follows closely the standard technique for the differential $\rd\sigma/\rd{}m$ measurement
described in~\cite{Paper2010}.
The unfolding response matrix $T_{ik}$
is calculated from simulation corresponding
to the one-dimensional yield vector in Eq.~(\ref{eq:unfolding_response}).
The structure of the response matrix is quite different from the corresponding
matrix derived using the yields binned in invariant mass only.
The matrix consists primarily of three
diagonal-dominated blocks.
There are two types of off-diagonal elements in this response matrix.
The elements adjacent to the diagonals originate from migration
between rapidity bins within the same mass bin.
Two additional sets of diagonal dominated blocks originate as a result of
migration between adjacent mass bins.
The response matrix is inverted and used to unfold the one-dimensional spectrum:

\begin{equation}
N^\mathrm{u}_k = N^\text{true}_k = \sum_{i}(T^{-1})_{ki}N^\text{obs}_i.
   \label{eq:unfolding_invert}
\end{equation}

Finally, the unfolded distribution is mapped back into the two-dimensional invariant
mass-rapidity distribution by performing an index transformation.

A set of tests was performed to validate this unfolding procedure. A closure test,
performed using simulation, confirmed the validity of the procedure. The
robustness of the method with respect to statistical fluctuations in the matrix
elements was checked with a test on an ensemble of MC pseudo-experiments, described in Section~\ref{sec:syst}.

The effects of the unfolding correction in the differential cross section measurement are approximately 30\%
(dimuon) and 60\% (dielectron) due to the detector resolution in the $\Z$-peak region,
where the invariant mass spectrum changes steeply.
In other regions they are less significant, on the order of 5\% (dimuon) and 10\% (dielectron).
The effect in
the double-differential cross section measurement is less pronounced since both the invariant mass
and rapidity bin sizes are wider than the respective detector resolutions,
but it reaches 5\% in the high-rapidity region, $\abs{y} > 2.0$.
\subsection{Efficiency}
\label{sec:eff}

The event efficiency $\varepsilon$ is defined as the probability for an event within the acceptance to pass the reconstruction procedure and the selection process.
The event efficiency is obtained from simulation and is corrected by an efficiency scale factor $\rho$, which is a ratio of efficiencies and takes into account differences between data and simulation.
The determination of the event efficiency is based on the signal MC samples described in Section~\ref{sec:data}.
It is calculated as the ratio of the number of events that pass full reconstruction and selection to the number of events that are found within
 the acceptance at the generator level.

The event efficiency is significantly affected by the pileup in the event.
The average pileup depends on the data taking conditions
and typically increased throughout the data taking in 2011.
The pileup affects primarily the electron isolation efficiency (up to 5\%) whereas the effect on the muon isolation efficiency is less than 1\%.
The procedures outlined below are used to extract the efficiency corrections for
both the $\rd\sigma/\rd{}m$ and the $\rd^2\sigma/\rd{}m\,\rd\abs{y}$ measurements from data.

\subsubsection{Dimuon efficiency}
\label{sec:DimuonEff}
The scale factor $\rho$ accounts for the differences in both the single-muon and the dimuon selections.
The single-muon properties (including the trigger) are determined using $\Z\to \Pgmp\Pgmm$
 events in data and simulation, where one muon, the tag, satisfies the tight selection requirements,
and the selection criteria are applied to the other muon as a probe
(tag-and-probe method~\cite{bib:CMS_WZ}).
An event sample with a single-muon trigger (the tag) is used to evaluate this scale factor.
The dimuon selection scale factor is based on the dimuon vertex efficiency as measured
in data and simulation after the rest of the selection is applied.

The total event selection efficiency in the dimuon channel is factorized in the following way:

\begin{equation}
 \varepsilon = \varepsilon(\Pgm_1)\cdot \varepsilon(\Pgm_2)\cdot\varepsilon(\text{dimuon})\cdot\varepsilon(\text{event, trig}),
 \label{eq:event_eff_fac_muons}
\end{equation}

where
\begin{itemize}
\item $\varepsilon(\Pgm)$ is the single muon efficiency;
\item $\varepsilon(\text{dimuon})$ is the efficiency that the two muon
              tracks of the selected dimuon event come from a common vertex and satisfy the angular requirement between them;
\item $\varepsilon(\text{event, trig})$ is the efficiency of triggering an event in both L1 and HLT.
It includes the efficiency of matching an identified muon to a trigger object.
\end{itemize}

The single-muon efficiency is factorized into the following three factors:

\begin{equation}
  \varepsilon(\Pgm)= \varepsilon(\text{track})\cdot\varepsilon(\text{reco+id})\cdot\varepsilon(\text{iso}),
\end{equation}

where
\begin{itemize}
\item
$\varepsilon(\text{track})$ is the offline track reconstruction efficiency, \ie, the efficiency that a muon track is identified in the tracker;
\item
$\varepsilon(\text{reco+id})$ is the muon reconstruction and  identification efficiency, \ie, the efficiency that the reconstructed track passes all the offline muon quality requirements;
\item
$\varepsilon(\text{iso})$ is the muon isolation efficiency, \ie, the efficiency of an identified
muon to pass the isolation requirement.
\end{itemize}
The double-muon trigger has asymmetric $\pt$ selections for the two legs and, therefore, the efficiency for a muon to trigger the high-$\pt$ leg (leg 1) is different from the efficiency
for a muon to trigger the low-$\pt$ leg (leg 2). We define single-leg efficiencies where
$\varepsilon(\Pgm, \text{trig1})$ is the efficiency of a muon selected offline to be matched to one leg of the double-muon trigger,
and $\varepsilon(\Pgm, \text{trig2})$ is the efficiency of a muon selected offline to be matched to the other leg of the  double-muon trigger.
The efficiency factor $\varepsilon(\Pgm, \text{trig1})$ corresponds to a muon matched to the leg of the double-muon trigger that has the higher $\pt$ threshold.
The double-muon trigger efficiency can then be factorized with single-muon trigger efficiencies in the following way, which takes into account the different efficiencies for the two legs:

\begin{equation}
\begin{split}
\label{trig_fac}
  \varepsilon(\text{event, trig}) & = 1 - P(\text{one leg, failed}) - P(\text{two legs, failed}) \\
  & = \varepsilon(\Pgm_1, \text{trig1})\cdot\varepsilon(\Pgm_2, \text{trig2})
 + \varepsilon(\Pgm_1, \text{trig2})\cdot\varepsilon(\Pgm_2, \text{trig1})  \\
& \hspace{0.45cm} -\varepsilon(\Pgm_1, \text{trig1})\cdot\varepsilon(\Pgm_2, \text{trig1}),
\end{split}
\end{equation}

where
\begin{itemize}
\item
$P(\text{one leg, failed})$ is the probability that exactly one muon fails to trigger a leg, i.e.,
$\varepsilon(\Pgm_1, \text{trig1})\cdot(1-\varepsilon(\Pgm_2, \text{trig2})) + \varepsilon(\Pgm_2, \text{trig1})\cdot(1-\varepsilon(\Pgm_1, \text{trig2}))$;
\item
$P(\text{two legs, failed})$ is the probability that both muons fail to trigger a leg, i.e.,
$(1-\varepsilon(\Pgm_1, \text{trig1}))\cdot(1-\varepsilon(\Pgm_2, \text{trig1}))$.
\end{itemize}

For these measurements
the combinatorial background of tag-probe pairs not coming from the $\Z$-boson signal are subtracted
using a simultaneous maximum-likelihood fit to the invariant mass spectra
for passing and failing probes with identical signal and background shapes.

Finally, the efficiency scale factor $\rho$
is measured to be 1.00--1.02 in most of the phase space, although it rises to $1.10$ at high dimuon rapidity.

\subsubsection{Dielectron efficiency}
\label{sec:Dielectron Eff}

The factorization of the event efficiency for the electron and the dielectron channel analysis is similar to that of the muon analysis.
The total event selection efficiency is given by

\begin{equation}
    \varepsilon = \varepsilon(\Pe_1) \cdot \varepsilon(\Pe_2) \cdot \varepsilon(\text{event, trig}),
\end{equation}

where the two $\varepsilon(\Pe)$ factors are the single-electron
efficiencies for the two electrons in the candidate and
$\varepsilon(\text{event, trig})$ is the efficiency of triggering on the
event. There is no factor $\varepsilon(\text{dielectron})$ analogous to the
one in Eq.~(\ref{eq:event_eff_fac_muons}) because there is no
requirement in the selection for dielectron candidates that depends on
parameters of both electrons at the same time except for the
requirement to originate from the common vertex. This factor, however, is absorbed
into the single-electron efficiency by requiring for each electron a
small impact parameter with respect to the primary vertex of
the event.

The single-electron efficiency is factorized as

\begin{equation}
    \varepsilon(\Pe) = \varepsilon(\text{reco}) \cdot \varepsilon(\text{id+iso}),
\end{equation}

where
\begin{itemize}
  \item the efficiency to detect a supercluster (SC) is known to be very close to 100\%~\cite{bib:ElectronReco2};
  \item $\varepsilon(\text{reco})$ is the offline electron reconstruction efficiency, i.e., the
           probability that, given a SC is found, an electron is reconstructed and passes
           the offline selection;
  \item $\varepsilon(\text{id+iso})$ is the efficiency to pass the selection criteria specific to this measurement,
            including identification, isolation, and conversion rejection, given that the electron candidate
            has already passed the previous stage of the offline selection.
\end{itemize}
The efficiency for an event to pass the trigger is computed in the following way:

\begin{equation}
    \varepsilon(\text{event, trig}) = \varepsilon(\Pe_1, (\text{trig1}.\mathrm{OR}.\text{trig2})) \cdot \varepsilon(\Pe_2, (\text{trig1}.\mathrm{OR}.\text{trig2})),
   \label{eq:ele_trigger_factorization}
\end{equation}

where $\varepsilon(\Pe_i, (\text{trig1}.\mathrm{OR}.\text{trig2}))$ is the efficiency for each electron to match
either one of the two trigger legs. This factorization
is simpler than that of muons given by Eq.~(\ref{trig_fac}) because for the dielectron
trigger, unlike the case for the dimuon trigger, it is measured that
$\varepsilon(\Pe, \text{trig1}) \approx \varepsilon(\Pe, \text{trig2}) \approx \varepsilon(\Pe, (\text{trig1}.\mathrm{OR}.\text{trig2}))$
so Eq.~(\ref{trig_fac}) simplifies to Eq.~(\ref{eq:ele_trigger_factorization}).

For the electron channel, the efficiencies for electron reconstruction and selection and the trigger
efficiencies are obtained from $\Z\to \Pep\Pem$ data and MC samples following the same tag-and-probe
method described above for the muons.

The efficiency scale factor $\rho$ is measured to be in the range of
0.98--1.02, with the values above 1.00 for dielectron masses
$m<40$\GeV and nearly constant at 0.98 above 45\GeV.
\subsection{Acceptance}
\label{sec:acc}

The geometrical and kinematic acceptance $A$ is defined as the fraction of
simulated signal events with both leptons falling within the detector fiducial volume.
The detector fiducial volume is defined by the nominal $\pt$ and $\eta$ requirements for an
analysis using the simulated leptons after the FSR simulation.
It is determined from simulation using the NNLO reweighted \POWHEG MC sample.

The signal event selection efficiency $\epsilon$ for a given mass bin is the fraction of events inside the
acceptance that pass the full selection. This definition uses the same
generator-level quantities after the FSR correction in both the numerator and denominator
(as in the acceptance definition).
The following equation holds:

\begin{equation}\label{eqn:AccEff}
    A \times \epsilon \equiv { \frac{N^A}{N^\text{gen}}} \cdot
{ \frac{N^{\epsilon}}{N^A} } = {\frac{N^{\epsilon}}{N^\text{gen}}},
\end{equation}

where $N^\text{gen}$ is the number of generated signal events in a given
invariant mass bin, $N^A$ is the number of events inside the geometrical and
kinematic acceptance, and $N^{\epsilon}$ is the number of events passing
the analysis selection.
The efficiency is estimated using the NNLO reweighted \POWHEG simulation.

The acceptance calculation depends on higher-order QCD corrections and the choice of PDFs.
The use of an NNLO signal MC is essential, especially in the low-mass region where the difference
between the NLO and NNLO predictions is sizable.

\begin{figure}[htb]
\centering
\includegraphics[width=0.49\textwidth]{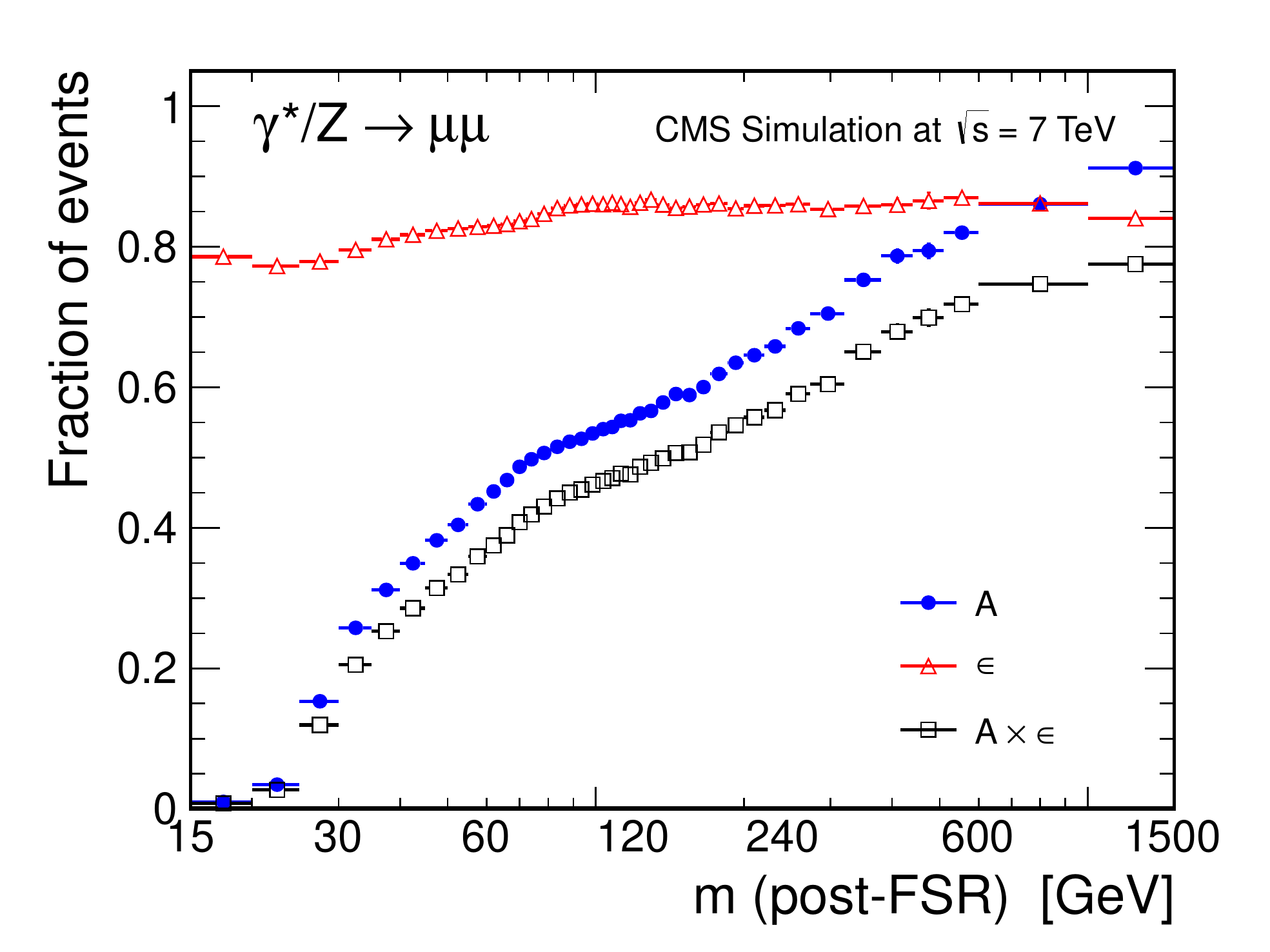}
\includegraphics[width=0.49\textwidth]{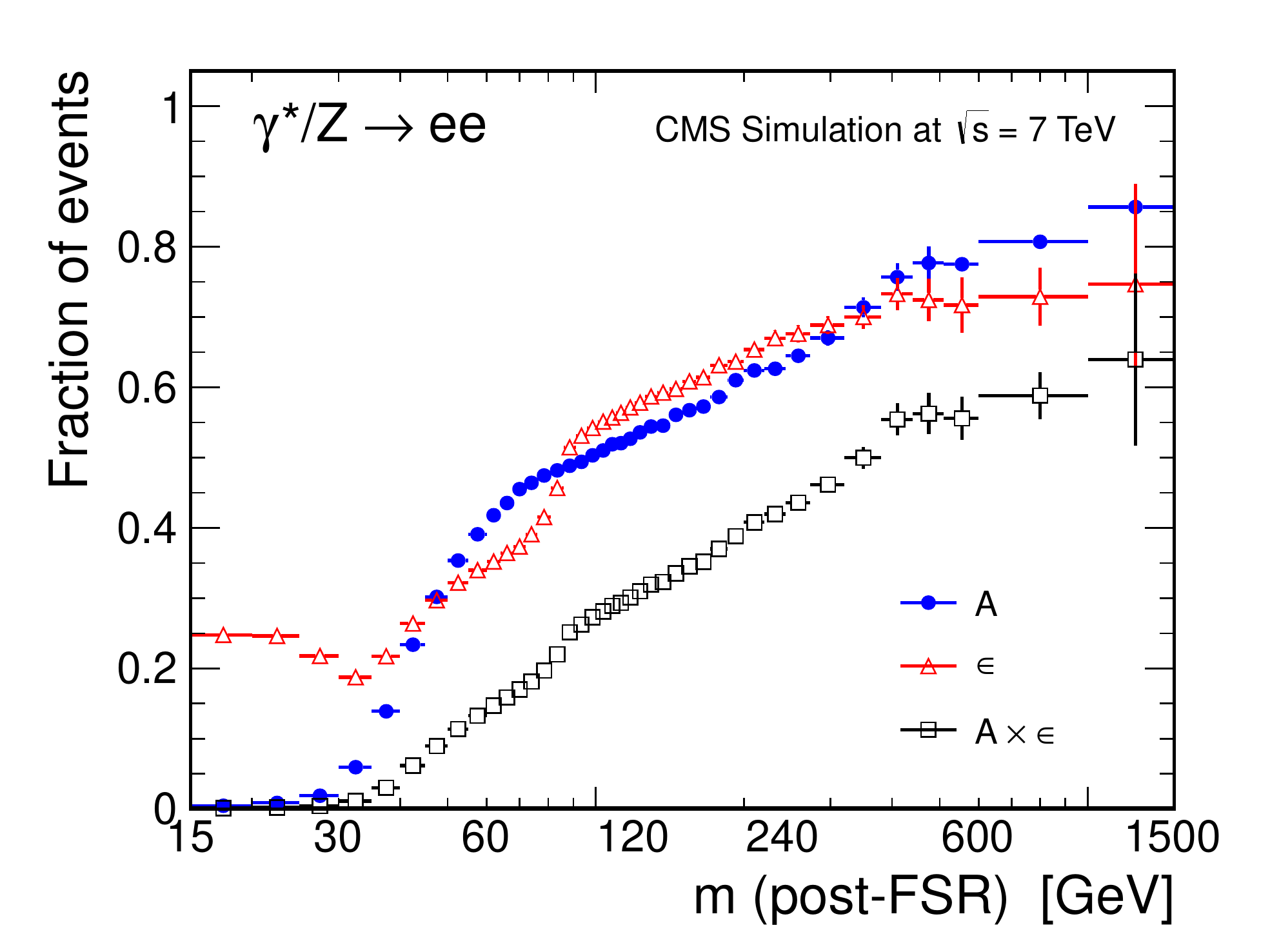}
\caption{
\label{fig:1Dacceff}
The DY acceptance, efficiency, and their product
per invariant mass bin in the dimuon channel (left) and the dielectron channel (right),
where m(post-FSR) means dimuon invariant mass after the FSR.
}
\end{figure}

Figure~\ref{fig:1Dacceff} shows the acceptance, the event efficiency, and $A\times\epsilon$ as functions of the dilepton invariant mass.

\subsection{Final-state QED radiation effects}
\label{sec:fsr}

Leptons can radiate photons in a process referred to as FSR.
This FSR effect changes the observed invariant mass, which is computed from the four-momenta
of the two leptons.
When FSR photons with sizable energy are emitted, the observed mass can
be substantially lower than the original DY mass.
The effect is most pronounced just below the $\Z$ peak, where the `radiative' events in the $\Z$ peak are
shifted lower in mass and become a significant contribution to that mass region.

The correction for FSR is performed separately from the correction for detector resolution.
It aims to transform a post-FSR track (i.e., after radiation and thus
closer to the actual measurement) into a pre-FSR track before any
radiation that is more representative of the original track.
The FSR correction procedure is performed in three steps:
\begin{enumerate}
\item
A bin-by-bin correction for the events in which pre-FSR leptons fail
the acceptance requirements, while post-FSR leptons pass. At
the analysis level we deal with only post-FSR events and this
correction, based on MC simulations, scales back the sample to contain only events
that pass the acceptance requirements in both pre- and post-FSR. The correction is
applied before the FSR unfolding, and is somewhat similar
to a background correction.
\item
An unfolding procedure is used for the events in which both pre- and post-FSR
leptons pass the acceptance requirements,
for which we can construct a response matrix similar
to that of Eq.~(\ref{eq:unfolding_invert}).
\item A bin-by-bin correction is used for the events in which pre-FSR
leptons pass the acceptance requirements, but post-FSR leptons fail those requirements.
These events do not enter the response matrix, but
they need to be taken into account. This correction is applied after the
FSR unfolding, and is similar to an efficiency correction.
\end{enumerate}

The correction for the events from step 1 is quite small, reaching its maximum of
1\% right below the $\Z$ peak.

The unfolding procedure for the events from step 2 follows the unfolding
procedure for the resolution. The response matrix is derived from the
NNLO reweighted \POWHEG MC sample, using pre- and post-FSR yields.

The bin-by-bin correction for the events from step 3 is significant at
low mass, reaching a maximum of 20\% in the lowest mass bin and
decreasing to negligible levels in the $\Z$-peak region.

The same method is applied in the double-differential cross section measurement.
The structure of the response matrix is quite different from the
corresponding matrix derived using the yields binned in invariant mass
only.  The matrix consists of a set of diagonal-dominated blocks, which originate from
migration between mass bins in the pre- and post-FSR distributions.

The effect of the FSR unfolding correction in the differential cross section measurement is significant
in the mass region 50--80\GeV, below the $\Z$ peak. In this region, the magnitude of the effect is of the order of 30--50\%
(40--60\%) for the dimuon (dielectron) channel. In other regions, the effect is of the order of 10--15\% in both channels.
In the double-differential cross section measurement, the effect of FSR unfolding is small,
typically a few percent, due to a larger mass bin size.

\subsection{Systematic uncertainties}
\label{sec:syst}

In this section, we discuss the evaluation of the systematic uncertainties,
which are shown in Tables~\ref{tab:syst-muons}--\ref{tab:sys2D_3}
for both the differential and the double-differential cross section measurements.
The methods used to evaluate the uncertainties are described in Ref.~\cite{Paper2010}.

The estimated uncertainty in the center-of-mass energy is 0.65\% or $46$\GeV
at 7\TeV~\cite{bib:LHCBeam}.
This would result in an additional uncertainty in the absolute differential cross
section of $0.3\%$ in the low-mass region, 0.6\% in the $\Z$-peak region
and $1.0\%$ in the high-mass region on the average.
We do not explicitly include these uncertainties in the systematic uncertainties.

\subsubsection{Dimuon systematic uncertainties}
\label{sec:DimuonSys}

The main uncertainty in the dimuon signal comes from the efficiency scale factor
$\rho$ that reflects systematic deviations that vary up to 2\%
between the data and the simulation.
As discussed in
Section~\ref{sec:eff}, single-muon efficiencies of several types are
measured with the tag-and-probe procedure and are combined into event
efficiency scale factors. The tag-and-probe procedure yields the
efficiency of each type and an associated statistical uncertainty. A variety of
possible systematic biases in the tag-and-probe procedure has been
investigated, such as dependence on binning in single-muon $\pt$ and
$\eta$, dependence on the assumed shape of signal and background in
the fit model, and others. Appropriate systematic uncertainties in the
single-muon efficiency scale factors have been assigned. The effect of
the combined statistical and systematic uncertainties in the event
scale factors $\rho$ on the final result
constitutes the final systematic uncertainty from this source. This uncertainty is
evaluated by recomputing the final result multiple times using an
ensemble of the single-muon efficiency maps where the entries are
modified randomly within ${\pm}1$ standard deviation of the combined statistical and
systematic uncertainties in the map bins. The uncertainties estimated by this method
are available in Tables~\ref{tab:syst-muons}--\ref{tab:sys2D_3}.
The contribution from the dimuon vertex selection is small because its efficiency
scale factor is consistent with being constant; the statistical fluctuations
are treated as systematic.

The uncertainty in the muon momentum scale arises from
the efficiency estimation, the background subtraction,
the detector resolution effect, the modeling of the $\Z$-boson $\pt$ spectrum, and the modeling
of the FSR.
To assign a systematic uncertainty corresponding to the muon momentum scale correction in the measurement,
the correction is shifted by one standard deviation of its total
uncertainty and
the deviation of the differential cross section from the
central value is assigned as the systematic uncertainty.
This uncertainty is used to estimate the systematic uncertainty of  the detector resolution
by the unfolding method.

We assign a systematic uncertainty in the unfolding of detector resolution effects
from two sources:
(1)~up to 1.5\% uncertainty from the momentum scale correction,
which is determined as a difference between the central and shifted
values of the unfolded distribution;
and (2)~up to 0.5\% uncertainty in the momentum scale correction estimation method.
We assign an additional systematic uncertainty to the unfolding procedure, which also
consists of two sources:
(1)~up to 1\% uncertainty due to the systematic difference between data and simulation
(which must be taken into account because the response matrix
is fully determined from simulation),
and (2)~up to 1\% uncertainty in the unfolding method.
To estimate the uncertainty due to the systematic difference between data and simulation,
a bias in unfolding is simulated
by using the migration matrix from simulation in bins of the true
and measured masses, generating ensembles of pseudo-experiments of true and measured
data while holding the response matrix fixed.
Each ensemble is obtained by smearing the initial observed yield vector
with a random Gaussian distribution (taking the width of the Gaussian equal
to 1\% of the yield value in a given bin,
which provides sufficient variation within the detector resolution).
These ensembles of pseudo-experiments
are unfolded and the pull of each ensemble is taken.
The mean of the pulls over the set of ensembles is calculated, and the corresponding systematic uncertainty
is assigned as

\begin{equation}
\frac{\delta N^\text{obs}|_{\text{syst}}}{N_\mathrm{u}} =
\mu_\text{pulls}\cdot\frac{\left.\delta N^\text{obs}\right|_{\text{stat}}}{N_\mathrm{u}}.
\end{equation}
The systematic effect of the unfolding is generally small (less than 1\%), except in the $\Z$-peak region
where it reaches 1--3\%.

The uncertainties in the backgrounds are evaluated using different methods
for the estimates coming from data and simulation.
The QCD multijet, $\ttbar$, and $\cPqt\PW$ background estimates are based on
data, whereas all the other backgrounds
are evaluated from simulation.
For backgrounds derived from data, the uncertainty is based on two sources:
(1)~the Poissonian statistical uncertainty of predicted backgrounds
(which is treated as systematic);
and (2)~the difference between the prediction from the data and simulation.
In the case of an estimate based on simulation, the uncertainty is estimated in a similar way:
(1)~the Poissonian statistical uncertainty from the size of the MC sample (which is treated as systematic);
and (2)~the systematic uncertainty due to the knowledge of the theoretical cross section.
The two components are combined in quadrature in both cases.

The systematic uncertainty due to the model-dependent FSR simulation in
the dimuon channel is estimated using two reweighting techniques.
One is the electroweak radiative correction~\cite{bib:QEDCorr}.
This correction is applied to the electromagnetic coupling constant and
the difference in total event counts between the reweighted and original events
is assigned as a systematic uncertainty.
The second uses photons reconstructed near a muon.
In this case, the additional scale factors are determined by comparing data and simulation using three
distributions: the number of photons, photon energy, and $\Delta R(\Pgm,\Pgg)$.
These factors are applied to the signal MC events.
The effect from the photons is nonnegligible in the low-mass region ($m<45$\GeV)
where a large contribution from falsely identified photons yields an additional systematic uncertainty.

The acceptance times efficiency uncertainty
dominates at low mass. It contains a component related to the statistics
of the MC sample that limits our knowledge of the product $A\times\epsilon$,
which we treat as systematic.
There are two main theoretical uncertainties: the first one arises from our imperfect
knowledge of the nonperturbative PDFs that participate in the hard scattering,
and the second is the modeling of the hard-interaction process, that is,
the effects of higher-order QCD corrections.
These contributions are
largest at low mass ($10\%$) and decrease to less than $1\%$ for masses above the $\Z$-boson peak.
Higher-order EW corrections  are small in comparison to
FSR corrections. They increase for invariant masses in the
TeV region, but are insignificant compared to the experimental
precision for the whole mass range under study.

The PDF uncertainties for the differential and double-differential cross section measurements
are calculated using the \textsc{lhaglue}
interface to the PDF library
{LHAPDF 5.8.7}~\cite{Bourilkov:2003kk,Whalley:2005nh}, by
applying a reweighting technique with asymmetric uncertainties
as described in Ref.~\cite{Bourilkov:2006cj}.
The PDF uncertainty in the acceptance and the modeling
is not considered as a part of the resulting uncertainty in the measurement, but rather is used to facilitate comparison with theoretical models.
The modeling uncertainty is discussed in Section~\ref{sec:data}.

The systematic uncertainties in the dimuon channel are summarized in
Table~\ref{tab:syst-muons} for the $\rd\sigma/\rd{}m$ differential cross
section and in Tables~\ref{tab:sys2D_1}--\ref{tab:sys2D_3} for the
$\rd^2\sigma/\rd{}m\,\rd\abs{y}$ double-differential cross section.

\begin{table} [htpb]
\begin{center}
\topcaption{Summary of the systematic uncertainties for the
dimuon channel $\rd\sigma/\rd{}m$ measurement.
The ``Total'' is a quadratic sum of all sources except for the Acc.+PDF and Modeling.
}
\label{tab:syst-muons}
\begin{tabular}{|c|ccccccc|}
\hline
$m$ & Eff. $\rho$ & Det. resol. & Bkgr. est. & FSR & Total & Acc.+PDF & Modeling \\
(\GeVns) & (\%) & (\%) & (\%) & (\%)& (\%) & (\%) & (\%) \\
\hline
15--20 & $  1.90 $ & $  0.03 $ & $  0.28 $ & $  0.54 $ & $  2.09 $ & $  2.29 $ & $  9.70 $\\
20--25 & $  2.31 $ & $  0.24 $ & $  0.63 $ & $  0.47 $ & $  2.47 $ & $  3.15 $ & $  3.10 $\\
25--30 & $  2.26 $ & $  0.27 $ & $  2.95 $ & $  0.40 $ & $  3.76 $ & $  2.73 $ & $  1.90 $\\
30--35 & $  1.48 $ & $  0.17 $ & $  1.94 $ & $  0.46 $ & $  2.50 $ & $  2.59 $ & $  0.70 $\\
35--40 & $  1.19 $ & $  0.09 $ & $  1.26 $ & $  0.66 $ & $  1.88 $ & $  2.61 $ & $  0.50 $\\
40--45 & $  1.12 $ & $  0.07 $ & $  0.97 $ & $  0.30 $ & $  1.54 $ & $  2.49 $ & $  0.30 $\\
45--50 & $  1.10 $ & $  0.07 $ & $  0.86 $ & $  0.44 $ & $  1.50 $ & $  2.51 $ & $  0.10 $\\
50--55 & $  1.07 $ & $  0.10 $ & $  0.67 $ & $  0.58 $ & $  1.42 $ & $  2.44 $ & $  0.10 $\\
55--60 & $  1.07 $ & $  0.15 $ & $  0.69 $ & $  0.77 $ & $  1.52 $ & $  2.36 $ & $  0.20 $\\
60--64 & $  1.06 $ & $  0.19 $ & $  0.35 $ & $  0.94 $ & $  1.50 $ & $  2.27 $ & $  0.20 $\\
64--68 & $  1.06 $ & $  0.22 $ & $  0.24 $ & $  1.06 $ & $  1.55 $ & $  2.22 $ & $  0.30 $\\
68--72 & $  1.06 $ & $  0.30 $ & $  0.20 $ & $  1.13 $ & $  1.60 $ & $  2.20 $ & $  0.20 $\\
72--76 & $  1.05 $ & $  0.51 $ & $  0.15 $ & $  1.13 $ & $  1.65 $ & $  2.18 $ & $  0.20 $\\
76--81 & $  1.06 $ & $  0.94 $ & $  0.25 $ & $  1.01 $ & $  1.77 $ & $  2.15 $ & $  0.20 $\\
81--86 & $  1.11 $ & $  1.56 $ & $  0.10 $ & $  0.69 $ & $  2.06 $ & $  2.18 $ & $  0.10 $\\
86--91 & $  1.07 $ & $  2.21 $ & $  0.01 $ & $  0.23 $ & $  2.48 $ & $  2.12 $ & $  0.20 $\\
91--96 & $  1.08 $ & $  2.55 $ & $  0.01 $ & $  0.12 $ & $  2.78 $ & $  2.14 $ & $  0.20 $\\
96--101 & $  1.29 $ & $  2.32 $ & $  0.08 $ & $  0.15 $ & $  2.68 $ & $  2.12 $ & $  0.30 $\\
101--106 & $  1.31 $ & $  1.69 $ & $  0.14 $ & $  0.19 $ & $  2.17 $ & $  2.07 $ & $  0.30 $\\
106--110 & $  1.32 $ & $  1.05 $ & $  0.28 $ & $  0.22 $ & $  1.76 $ & $  2.01 $ & $  0.50 $\\
110--115 & $  1.34 $ & $  0.65 $ & $  0.34 $ & $  0.25 $ & $  1.59 $ & $  1.97 $ & $  0.60 $\\
115--120 & $  1.33 $ & $  0.47 $ & $  0.43 $ & $  0.27 $ & $  1.55 $ & $  1.95 $ & $  0.60 $\\
120--126 & $  1.36 $ & $  0.37 $ & $  0.56 $ & $  0.29 $ & $  1.60 $ & $  1.91 $ & $  0.50 $\\
126--133 & $  1.35 $ & $  0.33 $ & $  0.70 $ & $  0.30 $ & $  1.65 $ & $  1.88 $ & $  0.60 $\\
133--141 & $  1.31 $ & $  0.42 $ & $  0.90 $ & $  0.32 $ & $  1.75 $ & $  1.85 $ & $  0.70 $\\
141--150 & $  1.29 $ & $  0.64 $ & $  1.08 $ & $  0.35 $ & $  1.91 $ & $  1.81 $ & $  1.00 $\\
150--160 & $  1.36 $ & $  0.87 $ & $  1.20 $ & $  0.39 $ & $  2.13 $ & $  1.82 $ & $  1.10 $\\
160--171 & $  1.42 $ & $  0.99 $ & $  1.48 $ & $  0.39 $ & $  2.39 $ & $  1.82 $ & $  1.10 $\\
171--185 & $  1.53 $ & $  0.96 $ & $  1.72 $ & $  0.41 $ & $  2.61 $ & $  1.75 $ & $  1.10 $\\
185--200 & $  1.60 $ & $  0.77 $ & $  1.80 $ & $  0.51 $ & $  2.67 $ & $  1.75 $ & $  1.10 $\\
200--220 & $  1.71 $ & $  0.52 $ & $  1.82 $ & $  0.42 $ & $  2.64 $ & $  1.53 $ & $  1.00 $\\
220--243 & $  1.75 $ & $  0.39 $ & $  2.28 $ & $  0.44 $ & $  3.01 $ & $  1.48 $ & $  1.50 $\\
243--273 & $  1.86 $ & $  0.49 $ & $  2.46 $ & $  0.46 $ & $  3.23 $ & $  1.40 $ & $  1.40 $\\
273--320 & $  1.90 $ & $  0.72 $ & $  2.37 $ & $  0.50 $ & $  3.24 $ & $  1.31 $ & $  1.30 $\\
320--380 & $  1.90 $ & $  0.96 $ & $  2.88 $ & $  0.57 $ & $  3.73 $ & $  1.28 $ & $  1.50 $\\
380--440 & $  1.93 $ & $  1.31 $ & $  3.54 $ & $  0.57 $ & $  4.44 $ & $  1.45 $ & $  1.20 $\\
440--510 & $  1.97 $ & $  1.74 $ & $  4.64 $ & $  0.57 $ & $  5.50 $ & $  1.60 $ & $  1.30 $\\
510--600 & $  2.02 $ & $  1.79 $ & $  4.48 $ & $  0.57 $ & $  5.28 $ & $  0.50 $ & $  2.10 $\\
600--1000 & $  2.01 $ & $  1.13 $ & $  5.07 $ & $  0.57 $ & $  5.61 $ & $  0.41 $ & $  2.40 $\\
1000--1500 & $  2.14 $ & $  0.48 $ & $ 15.34 $ & $  0.57 $ & $ 15.51 $ & $  0.24 $ & $  3.10 $\\
\hline
\end{tabular}
\end{center}
\end{table}

\begin{table} [htpb]
\begin{center}
\topcaption{Summary of the systematic uncertainties for the
  dielectron channel $\rd\sigma/\rd{}m$ measurement. E-scale indicates the energy scale uncertainty.
The ``Total'' is a quadratic sum of all sources except for the Acc.+PDF and Modeling.}
\label{tab:syst-electrons}
\begin{tabular}{|c|ccccccc|}
\hline
 $m$ & E-scale & Eff. $\rho$ & Det. resol. & Bkgr. est. & Total & Acc.+PDF & Modeling \\
    (\GeVns) &  (\%)   &  (\%)     &  (\%)      & (\%)     &  (\%) &  (\%)    &  (\%)    \\
\hline
  15--20   &$    1.4  $&$   3.0 $&$   1.9 $&$   0.3 $&$   3.8 $&$   3.0 $&$   9.7 $ \\
  20--25   &$    2.5  $&$   2.3 $&$   3.3 $&$   0.7 $&$   4.8 $&$   2.2 $&$   3.1 $ \\
  25--30   &$    1.5  $&$   2.7 $&$   1.9 $&$   1.1 $&$   3.8 $&$   2.2 $&$   1.9 $ \\
  30--35   &$    1.4  $&$   3.2 $&$   1.4 $&$   4.4 $&$   5.8 $&$   2.2 $&$   0.7 $ \\
  35--40   &$    0.6  $&$   2.3 $&$   1.1 $&$   5.5 $&$   6.1 $&$   2.1 $&$   0.5 $ \\
  40--45   &$    0.7  $&$   1.8 $&$   1.1 $&$   7.1 $&$   7.4 $&$   2.0 $&$   0.3 $ \\
  45--50   &$    0.7  $&$   1.5 $&$   1.3 $&$   8.9 $&$   9.1 $&$   2.0 $&$   0.1 $ \\
  50--55   &$    3.3  $&$   1.2 $&$   1.7 $&$   3.4 $&$   5.2 $&$   2.0 $&$   0.1 $ \\
  55--60   &$    2.8  $&$   1.0 $&$   2.4 $&$   2.5 $&$   4.5 $&$   2.0 $&$   0.2 $ \\
  60--64   &$    6.4  $&$   0.9 $&$   3.8 $&$   2.7 $&$   8.0 $&$   1.9 $&$   0.2 $ \\
  64--68   &$    2.4  $&$   0.9 $&$   4.9 $&$   2.4 $&$   6.0 $&$   1.9 $&$   0.3 $ \\
  68--72   &$    2.1  $&$   0.9 $&$   5.2 $&$   1.8 $&$   5.9 $&$   1.9 $&$   0.2 $ \\
  72--76   &$    1.5  $&$   0.8 $&$   5.3 $&$   1.2 $&$   5.7 $&$   1.8 $&$   0.2 $ \\
  76--81   &$    2.0  $&$   0.8 $&$   3.7 $&$   0.5 $&$   4.4 $&$   1.8 $&$   0.2 $ \\
  81--86   &$    5.9  $&$   0.8 $&$   2.3 $&$   0.2 $&$   6.4 $&$   1.7 $&$   0.1 $ \\
  86--91   &$    8.8  $&$   0.7 $&$   0.7 $&$   0.1 $&$   8.8 $&$   1.7 $&$   0.2 $ \\
  91--96   &$    8.4  $&$   0.7 $&$   0.7 $&$   0.0 $&$   8.4 $&$   1.7 $&$   0.2 $ \\
  96--101  &$   15.6  $&$   0.7 $&$   3.7 $&$   0.2 $&$  16.1 $&$   1.7 $&$   0.3 $ \\
 101--106  &$   17.6  $&$   0.8 $&$   5.8 $&$   0.4 $&$  18.6 $&$   1.7 $&$   0.3 $ \\
 106--110  &$   10.4  $&$   0.9 $&$  13.1 $&$   1.0 $&$  16.7 $&$   1.7 $&$   0.5 $ \\
 110--115  &$    5.5  $&$   0.9 $&$  10.2 $&$   1.2 $&$  11.6 $&$   1.6 $&$   0.6 $ \\
 115--120  &$    2.5  $&$   1.0 $&$  10.2 $&$   1.6 $&$  10.7 $&$   1.6 $&$   0.6 $ \\
 120--126  &$    2.0  $&$   1.1 $&$   8.1 $&$   1.9 $&$   8.6 $&$   1.6 $&$   0.5 $ \\
 126--133  &$    2.9  $&$   1.2 $&$   6.0 $&$   2.1 $&$   7.1 $&$   1.6 $&$   0.6 $ \\
 133--141  &$    4.9  $&$   1.2 $&$   4.7 $&$   2.1 $&$   7.2 $&$   1.6 $&$   0.7 $ \\
 141--150  &$    3.3  $&$   1.3 $&$   4.7 $&$   2.7 $&$   6.5 $&$   1.6 $&$   1.0 $ \\
 150--160  &$    3.5  $&$   1.4 $&$   4.9 $&$   3.1 $&$   6.9 $&$   1.6 $&$   1.1 $ \\
 160--171  &$    6.7  $&$   1.5 $&$   3.9 $&$   2.6 $&$   8.3 $&$   1.7 $&$   1.1 $ \\
 171--185  &$    5.6  $&$   1.6 $&$   4.1 $&$   3.6 $&$   8.0 $&$   1.6 $&$   1.1 $ \\
 185--200  &$    4.1  $&$   1.6 $&$   3.8 $&$   3.4 $&$   6.7 $&$   1.7 $&$   1.1 $ \\
 200--220  &$    2.6  $&$   1.7 $&$   2.9 $&$   3.1 $&$   5.3 $&$   1.6 $&$   1.0 $ \\
 220--243  &$    1.8  $&$   1.9 $&$   3.3 $&$   3.9 $&$   5.7 $&$   1.7 $&$   1.5 $ \\
 243--273  &$    1.6  $&$   2.0 $&$   3.4 $&$   4.0 $&$   5.9 $&$   1.7 $&$   1.4 $ \\
 273--320  &$    1.1  $&$   2.1 $&$   3.0 $&$   4.4 $&$   5.9 $&$   1.7 $&$   1.3 $ \\
 320--380  &$    1.8  $&$   2.5 $&$   3.7 $&$   4.2 $&$   6.4 $&$   1.9 $&$   1.5 $ \\
 380--440  &$    3.3  $&$   3.2 $&$   5.8 $&$   5.8 $&$   9.4 $&$   2.3 $&$   1.2 $ \\
 440--510  &$    3.2  $&$   3.8 $&$   5.3 $&$   5.0 $&$   8.8 $&$   2.8 $&$   1.3 $ \\
 510--600  &$    3.4  $&$   1.3 $&$   1.2 $&$   3.8 $&$   5.4 $&$   0.6 $&$   2.1 $ \\
 600--1000 &$    1.5  $&$   1.3 $&$   2.2 $&$   7.1 $&$   7.7 $&$   0.5 $&$   2.4 $ \\
1000--1500 &$    7.8  $&$   0.9 $&$   1.3 $&$  33.3 $&$  34.2 $&$   0.4 $&$   3.1 $ \\
\hline
\end{tabular}
\end{center}
\end{table}

\begin{table} [htpb]
\begin{center}
\topcaption{Summary of systematic uncertainties in the dimuon channel for $20<m<30$\GeV and $30<m<45$\GeV bins as a function of $\abs{y}$.
The ``Total'' is a quadratic sum of all sources.
}
\label{tab:sys2D_1}
{\small
\begin{tabular}{|c|ccccc|}
\hline
$\abs{y}$ & Eff. $\rho$ (\%) & Det. resol. (\%) & Bkgr. est. (\%) & FSR (\%) & Total (\%) \\
\hline
& \multicolumn{5}{c|}{$20 < m < 30$\GeV} \\
\hline
0.0--0.1 & 6.21 & 0.29 & 0.57 & 0.76 & 6.29 \\
0.1--0.2 & 6.01 & 0.37 & 0.56 & 0.58 & 6.07 \\
0.2--0.3 & 6.01 & 0.33 & 0.55 & 1.15 & 6.15 \\
0.3--0.4 & 5.57 & 0.41 & 0.48 & 0.57 & 5.63 \\
0.4--0.5 & 5.21 & 0.45 & 0.56 & 0.70 & 5.31 \\
0.5--0.6 & 4.87 & 0.32 & 0.57 & 0.54 & 4.94 \\
0.6--0.7 & 4.51 & 0.33 & 0.52 & 0.64 & 4.60 \\
0.7--0.8 & 3.89 & 0.38 & 0.55 & 0.42 & 3.97 \\
0.8--0.9 & 3.42 & 0.31 & 0.54 & 0.57 & 3.52 \\
0.9--1.0 & 3.14 & 0.26 & 0.53 & 0.77 & 3.29 \\
1.0--1.1 & 2.92 & 0.49 & 0.53 & 0.61 & 3.07 \\
1.1--1.2 & 2.87 & 0.50 & 0.58 & 0.47 & 3.01 \\
1.2--1.3 & 3.09 & 0.44 & 0.51 & 0.46 & 3.20 \\
1.3--1.4 & 3.62 & 0.37 & 0.62 & 0.47 & 3.72 \\
1.4--1.5 & 3.87 & 0.50 & 0.60 & 0.92 & 4.05 \\
1.5--1.6 & 4.12 & 0.55 & 0.59 & 0.44 & 4.22 \\
1.6--1.7 & 4.40 & 0.62 & 0.66 & 0.48 & 4.52 \\
1.7--1.8 & 4.76 & 0.51 & 0.65 & 0.45 & 4.85 \\
1.8--1.9 & 4.82 & 0.76 & 0.71 & 0.69 & 4.98 \\
1.9--2.0 & 4.88 & 0.60 & 0.69 & 0.56 & 4.99 \\
2.0--2.1 & 4.84 & 0.46 & 0.72 & 1.26 & 5.07 \\
2.1--2.2 & 5.22 & 0.67 & 0.89 & 1.68 & 5.59 \\
2.2--2.3 & 6.84 & 1.16 & 1.02 & 3.37 & 7.78 \\
2.3--2.4 & 8.40 & 1.14 & 1.56 & 4.96 & 9.94 \\
\hline
& \multicolumn{5}{c|}{$30 < m < 45$\GeV} \\
\hline
0.0--0.1 & 3.03 & 0.08 & 0.36 & 0.88 & 3.18 \\
0.1--0.2 & 2.72 & 0.03 & 0.38 & 0.82 & 2.87 \\
0.2--0.3 & 2.50 & 0.07 & 0.42 & 0.98 & 2.71 \\
0.3--0.4 & 2.30 & 0.03 & 0.38 & 1.13 & 2.59 \\
0.4--0.5 & 2.21 & 0.11 & 0.38 & 1.03 & 2.47 \\
0.5--0.6 & 2.25 & 0.10 & 0.34 & 0.74 & 2.39 \\
0.6--0.7 & 2.39 & 0.05 & 0.37 & 0.69 & 2.51 \\
0.7--0.8 & 2.46 & 0.05 & 0.40 & 0.89 & 2.65 \\
0.8--0.9 & 2.48 & 0.05 & 0.37 & 0.63 & 2.58 \\
0.9--1.0 & 2.39 & 0.05 & 0.38 & 0.74 & 2.53 \\
1.0--1.1 & 2.32 & 0.11 & 0.39 & 0.80 & 2.48 \\
1.1--1.2 & 2.18 & 0.03 & 0.40 & 0.58 & 2.29 \\
1.2--1.3 & 2.12 & 0.06 & 0.44 & 0.71 & 2.28 \\
1.3--1.4 & 2.04 & 0.04 & 0.34 & 0.53 & 2.13 \\
1.4--1.5 & 2.03 & 0.04 & 0.37 & 0.63 & 2.16 \\
1.5--1.6 & 2.02 & 0.07 & 0.39 & 0.66 & 2.16 \\
1.6--1.7 & 2.02 & 0.12 & 0.36 & 0.87 & 2.24 \\
1.7--1.8 & 2.14 & 0.06 & 0.33 & 0.80 & 2.31 \\
1.8--1.9 & 2.47 & 0.10 & 0.45 & 1.13 & 2.75 \\
1.9--2.0 & 2.74 & 0.20 & 0.45 & 1.08 & 2.99 \\
2.0--2.1 & 3.21 & 0.20 & 0.53 & 1.67 & 3.66 \\
2.1--2.2 & 3.86 & 0.19 & 0.71 & 2.52 & 4.67 \\
2.2--2.3 & 5.36 & 0.21 & 2.30 & 2.88 & 6.51 \\
2.3--2.4 & 6.71 & 0.09 & 2.38 & 6.30 & 9.51 \\
\hline
\end{tabular}
}
\end{center}
\end{table}

\begin{table} [htpb]
\begin{center}
\topcaption{Summary of systematic uncertainties in the dimuon channel for $45<m<60$\GeV and $60<m<120$\GeV bins as a function of $\abs{y}$.
The ``Total'' is a quadratic sum of all sources.
}
\label{tab:sys2D_2}
{\small
\begin{tabular}{|c|ccccc|}
\hline
$\abs{y}$ & Eff. $\rho$ (\%) & Det. resol. (\%) & Bkgr. est. (\%) & FSR (\%) & Total (\%) \\
\hline
& \multicolumn{5}{c|}{$45 < m < 60$\GeV} \\
\hline
0.0--0.1 & 1.75 & 0.02 & 0.48 & 0.93 & 2.04 \\
0.1--0.2 & 1.70 & 0.15 & 0.49 & 1.19 & 2.14 \\
0.2--0.3 & 1.64 & 0.05 & 0.54 & 1.74 & 2.45 \\
0.3--0.4 & 1.52 & 0.07 & 0.50 & 1.60 & 2.26 \\
0.4--0.5 & 1.45 & 0.04 & 0.54 & 3.12 & 3.48 \\
0.5--0.6 & 1.37 & 0.08 & 0.47 & 0.71 & 1.61 \\
0.6--0.7 & 1.38 & 0.04 & 0.50 & 1.09 & 1.83 \\
0.7--0.8 & 1.38 & 0.05 & 0.56 & 1.71 & 2.27 \\
0.8--0.9 & 1.39 & 0.02 & 0.49 & 0.62 & 1.60 \\
0.9--1.0 & 1.44 & 0.07 & 0.54 & 0.70 & 1.69 \\
1.0--1.1 & 1.44 & 0.02 & 0.48 & 1.07 & 1.86 \\
1.1--1.2 & 1.53 & 0.08 & 0.42 & 1.92 & 2.50 \\
1.2--1.3 & 1.63 & 0.10 & 0.47 & 1.25 & 2.11 \\
1.3--1.4 & 1.55 & 0.03 & 0.38 & 0.72 & 1.75 \\
1.4--1.5 & 1.40 & 0.23 & 0.38 & 0.77 & 1.65 \\
1.5--1.6 & 1.31 & 0.03 & 0.33 & 2.29 & 2.66 \\
1.6--1.7 & 1.34 & 0.11 & 0.39 & 1.37 & 1.96 \\
1.7--1.8 & 1.41 & 0.04 & 0.70 & 1.17 & 1.96 \\
1.8--1.9 & 1.52 & 0.07 & 0.30 & 3.04 & 3.42 \\
1.9--2.0 & 1.69 & 0.02 & 0.31 & 4.16 & 4.50 \\
2.0--2.1 & 1.78 & 0.06 & 0.55 & 5.31 & 5.63 \\
2.1--2.2 & 2.21 & 0.31 & 1.27 & 4.42 & 5.11 \\
2.2--2.3 & 2.96 & 0.11 & 0.62 & 9.98 & 10.4 \\
2.3--2.4 & 4.76 & 0.11 & 0.26 & 15.1 & 15.8 \\
\hline
& \multicolumn{5}{c|}{$60 < m < 120$\GeV} \\
\hline
0.0--0.1 & 0.83 & 0.004 & 0.04 & 0.29 & 0.88 \\
0.1--0.2 & 0.83 & 0.01 & 0.04 & 0.29 & 0.88 \\
0.2--0.3 & 0.84 & 0.01 & 0.04 & 0.29 & 0.89 \\
0.3--0.4 & 0.87 & 0.01 & 0.04 & 0.29 & 0.92 \\
0.4--0.5 & 0.89 & 0.01 & 0.04 & 0.29 & 0.94 \\
0.5--0.6 & 0.90 & 0.01 & 0.04 & 0.29 & 0.94 \\
0.6--0.7 & 0.89 & 0.01 & 0.04 & 0.29 & 0.94 \\
0.7--0.8 & 0.89 & 0.02 & 0.04 & 0.29 & 0.94 \\
0.8--0.9 & 0.92 & 0.01 & 0.03 & 0.29 & 0.97 \\
0.9--1.0 & 0.97 & 0.02 & 0.03 & 0.34 & 1.03 \\
1.0--1.1 & 1.03 & 0.03 & 0.04 & 0.30 & 1.08 \\
1.1--1.2 & 1.10 & 0.02 & 0.03 & 0.29 & 1.13 \\
1.2--1.3 & 1.16 & 0.02 & 0.03 & 0.31 & 1.20 \\
1.3--1.4 & 1.20 & 0.04 & 0.03 & 0.32 & 1.24 \\
1.4--1.5 & 1.23 & 0.03 & 0.05 & 0.32 & 1.27 \\
1.5--1.6 & 1.29 & 0.01 & 0.05 & 0.33 & 1.33 \\
1.6--1.7 & 1.40 & 0.02 & 0.08 & 0.43 & 1.47 \\
1.7--1.8 & 1.53 & 0.02 & 0.08 & 0.43 & 1.59 \\
1.8--1.9 & 1.67 & 0.03 & 0.05 & 0.46 & 1.73 \\
1.9--2.0 & 2.06 & 0.04 & 0.05 & 0.36 & 2.09 \\
2.0--2.1 & 2.78 & 0.01 & 0.14 & 0.62 & 2.86 \\
2.1--2.2 & 3.87 & 0.04 & 0.07 & 0.70 & 3.94 \\
2.2--2.3 & 5.34 & 0.02 & 0.02 & 0.91 & 5.41 \\
2.3--2.4 & 6.41 & 0.06 & 0.04 & 2.08 & 6.74 \\
\hline
\end{tabular}
}
\end{center}
\end{table}

\begin{table} [htpb]
\begin{center}
\topcaption{Summary of systematic uncertainties in the dimuon channel for $120<m<200$\GeV and $200<m<1500$\GeV bins as a function of $\abs{y}$.
The ``Total'' is a quadratic sum of all sources.
}
\label{tab:sys2D_3}
{\small
\begin{tabular}{|c|ccccc|}
\hline
$\abs{y}$ & Eff. $\rho$ (\%) & Det. resol. (\%) & Bkgr. est. (\%) & FSR (\%) & Total (\%) \\
\hline
& \multicolumn{5}{c|}{$120 < m < 200$\GeV} \\
\hline
0.0--0.1 & 1.68 & 0.28 & 2.17 & 0.56 & 2.81 \\
0.1--0.2 & 1.60 & 0.16 & 2.03 & 0.72 & 2.68 \\
0.2--0.3 & 1.56 & 0.26 & 2.09 & 1.05 & 2.82 \\
0.3--0.4 & 1.57 & 0.53 & 1.89 & 0.78 & 2.63 \\
0.4--0.5 & 1.49 & 0.27 & 1.67 & 0.67 & 2.35 \\
0.5--0.6 & 1.47 & 0.25 & 1.69 & 0.38 & 2.29 \\
0.6--0.7 & 1.57 & 0.33 & 1.97 & 0.54 & 2.60 \\
0.7--0.8 & 1.43 & 0.39 & 1.62 & 0.37 & 2.22 \\
0.8--0.9 & 1.42 & 0.07 & 1.92 & 0.52 & 2.44 \\
0.9--1.0 & 1.35 & 0.48 & 1.53 & 0.37 & 2.13 \\
1.0--1.1 & 1.31 & 0.16 & 1.37 & 0.41 & 1.94 \\
1.1--1.2 & 1.34 & 0.36 & 1.39 & 0.45 & 2.02 \\
1.2--1.3 & 1.51 & 0.45 & 1.35 & 0.57 & 2.15 \\
1.3--1.4 & 1.82 & 0.06 & 1.26 & 0.40 & 2.25 \\
1.4--1.5 & 2.17 & 0.85 & 1.04 & 0.44 & 2.59 \\
1.5--1.6 & 2.76 & 0.14 & 1.08 & 0.43 & 3.00 \\
1.6--1.7 & 3.44 & 0.30 & 0.83 & 0.39 & 3.57 \\
1.7--1.8 & 4.09 & 0.41 & 0.94 & 1.02 & 4.34 \\
1.8--1.9 & 5.37 & 0.17 & 1.03 & 1.09 & 5.57 \\
1.9--2.0 & 6.62 & 0.10 & 0.84 & 1.20 & 6.78 \\
2.0--2.1 & 8.52 & 0.16 & 0.89 & 0.60 & 8.58 \\
2.1--2.2 & 12.3 & 0.85 & 0.70 & 0.51 & 12.3 \\
2.2--2.3 & 16.8 & 0.41 & 0.95 & 1.91 & 16.9 \\
2.3--2.4 & 20.2 & 0.51 & 1.91 & 1.26 & 20.4 \\
\hline
& \multicolumn{5}{c|}{$200 < m < 1500$\GeV} \\
\hline
0.0--0.2 & 2.18 & 0.30 & 7.51 & 0.56 & 7.85 \\
0.2--0.4 & 1.84 & 0.04 & 5.31 & 0.47 & 5.64 \\
0.4--0.6 & 1.68 & 0.32 & 4.33 & 0.53 & 4.69 \\
0.6--0.8 & 1.70 & 0.07 & 4.57 & 0.58 & 4.91 \\
0.8--1.0 & 1.83 & 0.12 & 3.47 & 0.66 & 3.99 \\
1.0--1.2 & 2.28 & 0.44 & 3.10 & 0.66 & 3.93 \\
1.2--1.4 & 3.50 & 0.08 & 1.92 & 0.59 & 4.03 \\
1.4--1.6 & 5.28 & 0.65 & 2.15 & 0.56 & 5.77 \\
1.6--1.8 & 7.14 & 0.19 & 2.11 & 0.98 & 7.51 \\
1.8--2.0 & 10.4 & 0.86 & 2.17 & 0.61 & 10.6 \\
2.0--2.2 & 17.8 & 0.15 & 0.99 & 0.98 & 17.8 \\
2.2--2.4 & 28.8 & 0.42 & 1.99 & 1.36 & 28.9 \\
\hline
\end{tabular}
}
\end{center}
\end{table}

\subsubsection{Dielectron systematic uncertainties}
\label{sec:DielectronSys}
In the dielectron channel, the leading systematic uncertainty is
associated with the energy scale corrections for individual
electrons. The corrections affect both the placement of a given
candidate in a particular invariant mass bin and the likelihood of
surviving the kinematic selection. The energy scale correction itself
is calibrated to 1--2\% precision.
Several sources of systematic uncertainties due to the energy scale
correction are considered: (1)~the uncertainty in the energy scale corrections;
(2)~the residual differences in simulated and measured distributions;
(3)~the choice of line shape modeling; and (4)~the choice of $\eta$
binning.
The associated
uncertainty in the signal yield is calculated by varying the energy scale
correction value within its uncertainty and remeasuring the yield. The electron
energy scale uncertainty takes its largest values for the bins near
the central $\Z$-peak bin because of sizable event migration. This
uncertainty for the electron channel is roughly 20 times
larger than the momentum scale uncertainty for muons, for which the
associated systematic uncertainties in the cross section are rather
small.

Another significant uncertainty for electrons results from the
uncertainty in the efficiency scale factors.  The systematic
uncertainty in the scale factors as well as the resulting uncertainty in the
normalized cross section are found with the same procedure as for the
muon channel.

The uncertainty
associated with the unfolding procedure in the electron channel comes
primarily from the uncertainty in the unfolding matrix elements due to
imperfect simulation of detector resolution. This simulation
uncertainty for electrons is significantly larger than for muons,
leading to a larger systematic uncertainty in the normalized cross
section.
The dielectron background uncertainties are evaluated by comparing the
background yields calculated as described in Section~\ref{sec:bkg}
with predictions from simulation. These uncertainties become dominant
at the higher invariant masses above the $\Z$-boson peak.

The systematic effects due to the FSR simulation uncertainty for the
electron channel primarily affect the detector resolution unfolding
procedure. The impact of these effects is higher for the electron channel
than for the muon channel because of the partial recovery of FSR photons
in the supercluster energy reconstruction
as well as the overall stronger FSR effects for the electron channel.
To evaluate the FSR uncertainty for electrons, we adopt a more conservative approach
than the one used for the dimuons. The final results of the measurement
are recomputed using a detector resolution unfolding matrix prepared
by varying the fraction of events with significant FSR ($>$1\GeV) in
simulation by $\pm$5\%, and taking the spread as the systematic uncertainty.
This component is absorbed into the total detector resolution unfolding
systematic uncertainties. The effect of the FSR simulation on other analysis steps for the
electron  channel is negligible in comparison to other systematic effects
associated with those steps.

The PDF uncertainties affecting the acceptance
are computed in the same way as described for the muon channel.

The systematic uncertainties for the electron channel are summarized
in Table~\ref{tab:syst-electrons}.

\subsubsection{Covariance matrix}
A covariance matrix gives the uncertainties of the measurements together with the correlations
between the analysis bins and different systematic sources.
There are several distinctive steps in the covariance analysis.

For the muon data sample the measured spectrum is unfolded, which redistributes
the signal and background events according to the unfolding matrix $T^{-1}$,
described in Section~\ref{sec:unfoldingProcedure}.

The total uncertainty before the unfolding is given by a diagonal matrix $V_I$ describing
all the analysis bins. The mathematical description of the procedure to obtain the covariance matrix $V_\mathrm{UNF}$
associated with the unfolding is given in Ref.~\cite{bib:blob}:

\begin{equation}
V_\mathrm{UNF} = T^{-1} V_I  {T^{-1}}^T.
\label{eq:covEq}
\end{equation}

The common normalization to the $\Z$-boson peak does not change the results of the unfolding (matrix) procedure.

After the unfolding, the resulting yield is corrected for detector and reconstruction efficiencies. The
largest effect in the uncertainty comes from the efficiency corrections for the single
leptons, which are estimated with the tag-and-probe method.  A large part of this uncertainty comes
from systematic effects related to data/MC variations, together with
statistical limitations. The single-lepton efficiency corrections and their uncertainties are
propagated to the final results using MC pseudo-experiments, where the
correction values are varied according to their measured
uncertainties.   Similar pseudo-experiments also give the correlations (or
directly---the covariance) resulting from the particular choices of
the tag-and-probe binning.  The normalization to the $\Z$-boson peak is applied by measuring the
efficiency correlation
effects on the normalized yields. The efficiency covariance and correlations are trivially related
by the efficiency correction uncertainties (i.e., by the square roots of the diagonal elements of
the efficiency covariance matrix). The efficiency covariance matrix is denoted by $V_\mathrm{EFF}$.

The last step in the procedure is to apply FSR corrections to the measurement. As described earlier, it is based on the
FSR unfolding matrix and additional bin-by-bin corrections. There are associated uncertainties in the FSR description. As in the first step, the correlations induced by this procedure are described by the FSR unfolding matrix alone
and the covariance matrix $V_\mathrm{FSR}$ is given by Eq.~(\ref{eq:covEq}), but with the FSR related inputs.

The total covariance matrix $V_\text{tot}$ is simply by the sum of the three uncorrelated sources:

\begin{equation}
V_\mathrm{tot} = V_\mathrm{UNF} + V_\mathrm{EFF} + V_\mathrm{FSR}.
\end{equation}

In the electron data sample, the covariance of the post-FSR cross section is calculated as a weighted
covariance of three independent sources. First, the total uncertainty in the
signal yield is propagated through the detector resolution unfolding
matrix, as given by Eq.~(\ref{eq:covEq}). Then the uncertainty is increased by
contributions due to the statistical inaccuracy of the unfolding
matrix elements as well as additional sources of systematic uncertainty associated with
the resolution unfolding (\eg, the electron energy scale uncertainty and
FSR). The latter contribution is taken as diagonal. Second, the
covariance of the efficiency correction factors is evaluated using
pseudo-experiments as described for the muon channel analysis.
In this case, efficiency correction factors contribute significantly to
correlations in the low-mass region.
Third, the
diagonal covariance of each MC efficiency factor is obtained from the
statistical uncertainty. The covariance of the pre-FSR cross section is
obtained from the covariance of the post-FSR cross section via error
propagation.
After the FSR unfolding some covariances with the $\Z$-boson peak region become negative.
The contribution from the statistical
uncertainty of the FSR unfolding matrix is negligible. The covariance
of the normalized cross section is derived from the covariance of the
unnormalized cross section, taking the uncertainty of the $\Z$-peak bin
(Table~\ref{tab:normFactor1D_data}) and assuming no
correlation between the cross section value in a particular mass bin
and the normalization factor.

The covariance matrices are included in the HEPDATA record for this paper.
\section{Results and discussion}
\label{sec:results}

This section provides a summary of the results for the $\rd\sigma/\rd{}m$ cross section measurements in the dielectron and dimuon channels
and the $\rd^2\sigma/\rd{}m\,\rd\abs{y}$ cross section
measurement in the dimuon channel.

\subsection{Differential cross section \texorpdfstring{$\rd\sigma/\rd{}m$}{dsigma/dm} measurement}

The result of the measurement is calculated as the ratio

\begin{equation}
R^{i}_\text{pre-FSR} = \left.\frac{N_\mathrm{u}^{i}}{A^{i}\epsilon^{i}\rho^{i}}\middle/\frac{N_\mathrm{u}^\text{norm}}{A^\text{norm}\epsilon^\text{norm}\rho^\text{norm}}\right.,
\end{equation}

where $N_\mathrm{u}^{i}$ is the number of events after background subtraction and
the unfolding procedure for the detector resolution and FSR correction, $A^i$ is the acceptance,
$\epsilon^i$ is the efficiency, and $\rho^{i}$ is the correction estimated from data
in a given invariant mass bin $i$ as defined earlier.
$N_{u}^{\text{norm}}$, $A^{\text{norm}}$, $\epsilon^{\text{norm}}$, and $\rho^{\text{norm}}$ refer to the $\Z$-peak region.
The DY $\rd\sigma/\rd{}m$ differential cross section
is normalized to the cross section in the $\Z$-peak region ($60<m<120$\GeV).
The results are also normalized to the invariant mass bin widths, $\Delta m^i$, defining the shape
${r}^{i} = R^{i}/\Delta m^{i}$.

\begin{figure}[htb]
\begin{center}
\includegraphics[width=0.49\textwidth]{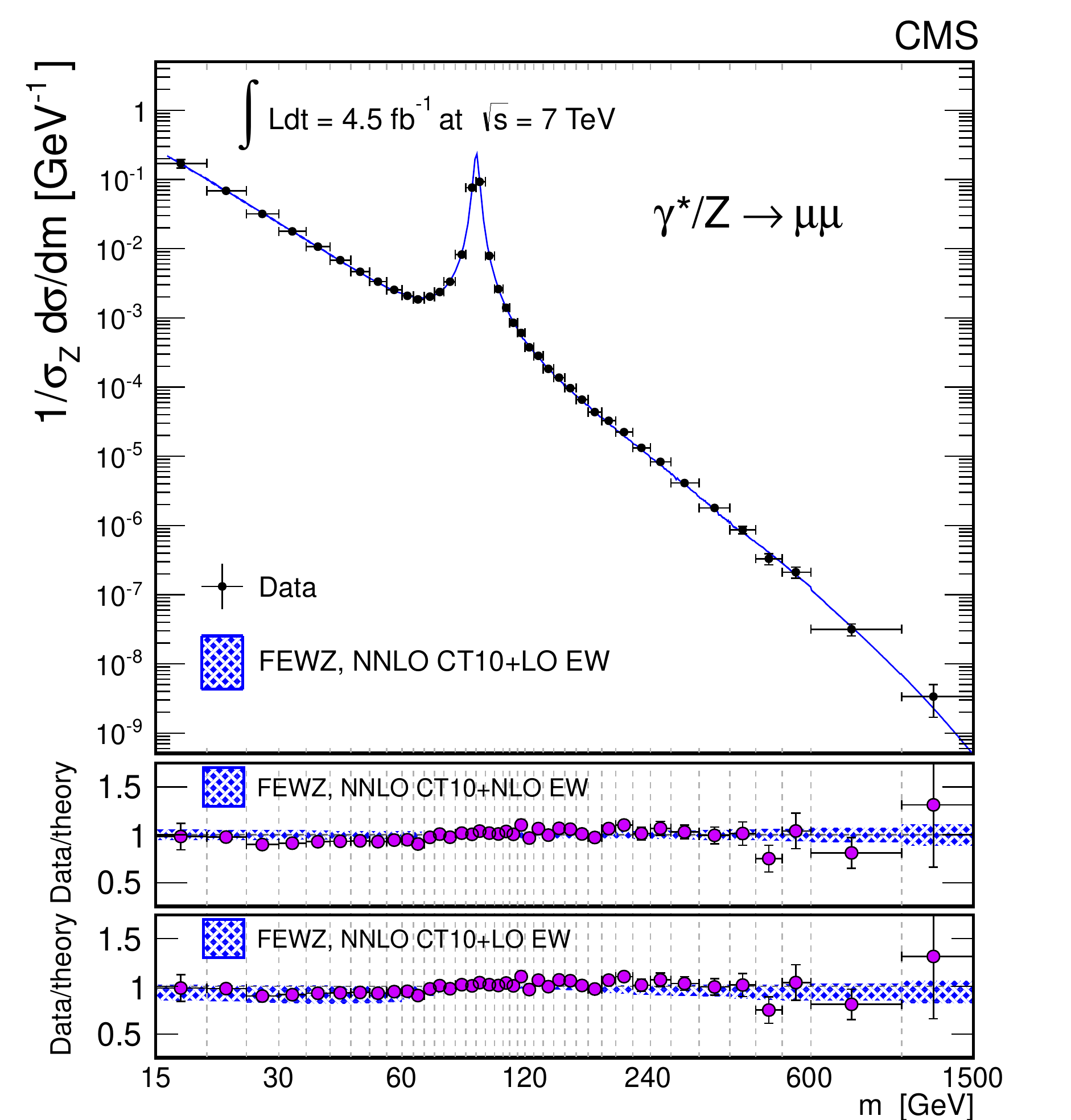}
\includegraphics[width=0.49\textwidth]{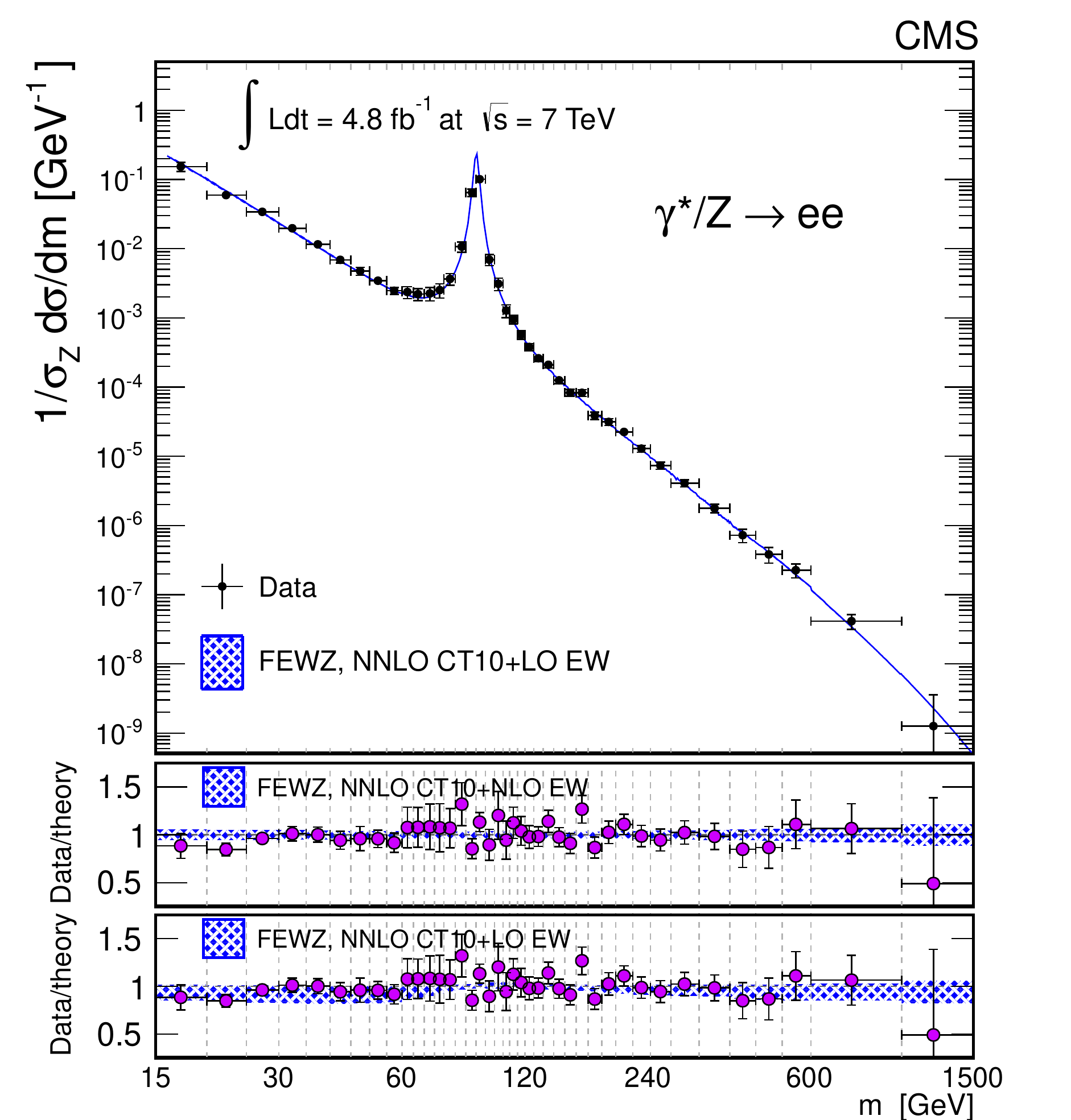}
\caption{
The DY dimuon and dielectron
invariant-mass spectra normalized to the $\Z$-boson production cross section $(1/\sigma_{\Z}\,\rd\sigma/\rd{}m)$,
as measured and as predicted by {\FEWZ}+CT10 NNLO calculations, for the full phase
space.
The vertical error bars for the measurement indicate the experimental (statistical and systematic) uncertainties
summed in quadrature with the theoretical uncertainty resulting from the model-dependent
kinematic distributions inside each bin.
The shaded uncertainty band for the theoretical calculation includes the
statistical uncertainty from the \FEWZ calculation and the 68\% confidence level uncertainty from PDFs combined in quadrature.
The effect of NLO EW correction including
$\Pgg\Pgg$-initiated processes (LO EW correction only)
is shown in the middle (bottom) plot for each channel.
The data point abscissas are computed according to Eq.~(6) in Ref.~\cite{bib:pts_plac}.
}
\label{fig:1Drshape}
\end{center}
\end{figure}

The results of the DY cross section measurement are presented in
Fig.~\ref{fig:1Drshape} for both the muon and
the electron channels.
The $\Z$-boson production cross sections used as normalization factors in
the dimuon and dielectron channels are measured from data.
Their values
are shown in Table~\ref{tab:normFactor1D_data}.
The muon and electron cross sections in the $\Z$-peak region are in good agreement with NNLO predictions
for the full phase space (\eg, a typical NNLO prediction is $970 \pm 30$\unit{pb})
and also with the previous CMS measurements~\cite{WZpaper,bib:CMS_WZ}.

\begin{table} [htb]
\centering
\topcaption{
    Normalization factors for the cross section measurements from the $\Z$-peak region ($60 < m < 120$\GeV)
      with associated uncertainties.
}
\label{tab:normFactor1D_data}
\begin{tabular}{|l|c|c|}
\hline
Muon channel & Cross section in the $\Z$-peak region\\
\hline
pre-FSR full acc. & $989.5 \pm 0.8\,\text{(stat.)} \pm 9.8\,\text{(exp. syst.)} \pm 21.9\,\text{(th. syst.)} \pm  21.8\lum$\unit{pb} \\
post-FSR full acc. & $ 974.8 \pm 0.7\stat \pm 9.2\,\text{(exp. syst.)} \pm 21.6\,\text{(th. syst.)} \pm 21.4\lum$\unit{pb} \\
pre-FSR detector acc. & $524.7 \pm 0.4\stat \pm 5.1\,\text{(exp. syst.)} \pm 1.2\,\text{(th. syst.)} \pm  11.5\lum$\unit{pb} \\
post-FSR detector acc. & $516.5 \pm 0.4\stat \pm 4.9\,\text{(exp. syst.)} \pm 1.1\,\text{(th. syst.)} \pm 11.4\lum$\unit{pb} \\
\hline
Electron channel & Cross section in the $\Z$-peak region\\
\hline
pre-FSR full acc.      & $984.6 \pm 0.9\stat \pm 7.3\,\text{(exp. syst.)} \pm 21.4\,\text{(th. syst.)} \pm 21.7\lum$\unit{pb}\\
post-FSR full acc.     & $950.0 \pm 0.9\stat \pm 7.0\,\text{(exp. syst.)} \pm 20.6\,\text{(th. syst.)} \pm 20.9\lum$\unit{pb} \\
pre-FSR detector acc.  & $480.5 \pm 0.4\stat \pm 3.5\,\text{(exp. syst.)} \pm 1.0\,\text{(th. syst.)} \pm 10.6\lum$\unit{pb}\\
post-FSR detector acc. & $462.3 \pm 0.4\stat \pm 3.4\,\text{(exp. syst.)} \pm 0.9\,\text{(th. syst.)} \pm 10.2\lum$\unit{pb} \\
\hline
Combined channel & Cross section in the $\Z$-peak region\\
\hline
pre-FSR full acc.      & $986.4 \pm 0.6\stat \pm 5.9\,\text{(exp. syst.)} \pm 21.7\,\text{(th. syst.)} \pm 21.7\lum$\unit{pb}\\
\hline
\end{tabular}
\end{table}

The theoretical predictions include leptonic decays of $\Z$ bosons with full spin correlations as well as
the $\Pgg^*/\Z$ interference effects.
The effects of lepton pair production in $\Pgg\Pgg$-initiated processes, where
both initial-state protons radiate a photon, are calculated with
\FEWZ~3.1.b2~\cite{Li:2012wna}.
They are particularly important for the high-mass region
and are included as additional mass-dependent factors to the main calculation,
which takes into account the difference between NLO and LO in the EW correction.
The effect rises to approximately 10\% in the highest-mass bins.
The uncertainties in the theoretical predictions due to the imprecise
knowledge of the PDFs are calculated with the \textsc{lhaglue} interface to the PDF library
{LHAPDF}, using a reweighting technique with asymmetric uncertainties.
The normalization of the spectrum is defined by the number of events in the
$\Z$-boson mass peak, so the uncertainty is calculated for the ratio of events in each bin to the
number in the $\Z$-boson mass peak.

The result of the measurement is in good agreement
with the NNLO theoretical predictions as computed with \FEWZ~2.1.1 using CT10.
The uncertainty band in Fig.~\ref{fig:1Drshape} for the theoretical calculation includes the
statistical uncertainty from the \FEWZ calculation and the 68\% confidence level (CL) uncertainty from PDFs combined in
quadrature. The effect of the higher-order EW correction computed with
\FEWZ~3.1.b2 (described above) is included
as an additional correction factor and the ratio between data and theoretical prediction
is shown in the middle plot.
Differences between NLO and NNLO values in theoretical expectations are significant in the low-mass region,
as reported in~\cite{Paper2010}. Although this measurement is sensitive to NNLO effects,
it does not provide sufficient sensitivity to distinguish between different PDFs.

In addition to the fully corrected $\rd\sigma/\rd{}m$ measurement, we report the cross sections within the detector acceptance ($r^i_\text{pre-FSR, det}$)
and the post-FSR cross sections (${r}^{i}_\text{post-FSR}, {r}^{i}_\text{post-FSR, det}$).
The corresponding definitions are

\begin{equation}
{r}^{i}_\text{post-FSR} = \frac{1}{\Delta m^i} \cdot \left(\left. \frac{N_\mathrm{u}^{'i}}{A^{i}\epsilon^{i}\rho^{'i}} \middle/ \frac{N_\mathrm{u}^{'\text{norm}}}{A^\text{norm}\epsilon^\text{norm}\rho^{'\text{norm}}}\right.\right),
\end{equation}

\begin{equation}
{r}^{i}_\text{pre-FSR, det} = \frac{1}{\Delta m^i} \cdot \left(\left. \frac{N_\mathrm{u}^{i}}{\epsilon^{i}\rho^{i}} \middle/ \frac{N_\mathrm{u}^\text{norm}}{\epsilon^\text{norm}\rho^\text{norm}}\right.\right),
\end{equation}

\begin{equation}
{r}^{i}_\text{post-FSR, det} = \frac{1}{\Delta m^i} \cdot \left(\left. \frac{N_\mathrm{u}^{'i}}{\epsilon^{i}\rho^{'i}} \middle/ \frac{N_\mathrm{u}^{'\text{norm}}}{\epsilon^\text{norm}\rho^{'\text{norm}}} \right.\right),
\end{equation}

where the quantities labeled with primes do not contain the FSR correction.
All the ${r}$ shape measurements for dimuons are summarized in
Table~\ref{tab:rshapes-muons}. The corresponding results
for the dielectron channel can be found in Table~\ref{tab:rshapes-electrons}.

\begin{table} [htpb]
\begin{center}
\topcaption{
  The DY cross section measurements for the muon channel normalized to the $\Z$-peak region,
  pre- and post-FSR, as measured in the full acceptance and for the CMS detector acceptance.
  The uncertainty indicates the experimental (statistical and systematic) uncertainties
  summed in quadrature with the theoretical uncertainty resulting from the model-dependent
  kinematic distributions inside each bin.
  The results presented are in $\GeVns{}^{-1}$ units.
}
\label{tab:rshapes-muons}
\begin{tabular}{|c|rrrr|}
\hline
$m$ (\GeVns) & \multicolumn{1}{c}{${r}^{i}_\text{pre-FSR}$} & \multicolumn{1}{c}{${r}^{i}_\text{post-FSR}$} &  \multicolumn{1}{c}{${r}^{i}_\text{pre-FSR, det}$} & \multicolumn{1}{c|}{${r}^{i}_\text{post-FSR, det}$} \\
\hline
15--20 & $\! ( 17.1\! \pm\!   1.7)\! \times\! 10^{-2} $\! & $\! ( 15.9\! \pm\!   1.6)\! \times\! 10^{-2} $\! & $\! (325.2\! \pm\!   7.7)\! \times\! 10^{-5} $\! & $\! (303.2\! \pm\!   7.0)\! \times\! 10^{-5} $\\
20--25 & $\! ( 68.5\! \pm\!   3.5)\! \times\! 10^{-3} $\! & $\! ( 66.6\! \pm\!   3.4)\! \times\! 10^{-3} $\! & $\! ( 44.8\! \pm\!   1.1)\! \times\! 10^{-4} $\! & $\! ( 44.0\! \pm\!   1.1)\! \times\! 10^{-4} $\\
25--30 & $\! ( 31.8\! \pm\!   1.6)\! \times\! 10^{-3} $\! & $\! ( 31.5\! \pm\!   1.6)\! \times\! 10^{-3} $\! & $\! ( 92.2\! \pm\!   3.5)\! \times\! 10^{-4} $\! & $\! ( 91.3\! \pm\!   3.4)\! \times\! 10^{-4} $\\
30--35 & $\! (177.7\! \pm\!   6.5)\! \times\! 10^{-4} $\! & $\! (177.0\! \pm\!   6.5)\! \times\! 10^{-4} $\! & $\! ( 87.0\! \pm\!   2.2)\! \times\! 10^{-4} $\! & $\! ( 86.2\! \pm\!   2.1)\! \times\! 10^{-4} $\\
35--40 & $\! (106.8\! \pm\!   3.5)\! \times\! 10^{-4} $\! & $\! (108.1\! \pm\!   3.5)\! \times\! 10^{-4} $\! & $\! ( 63.6\! \pm\!   1.2)\! \times\! 10^{-4} $\! & $\! ( 63.6\! \pm\!   1.2)\! \times\! 10^{-4} $\\
40--45 & $\! ( 68.2\! \pm\!   2.0)\! \times\! 10^{-4} $\! & $\! ( 70.9\! \pm\!   2.1)\! \times\! 10^{-4} $\! & $\! (452.4\! \pm\!   7.4)\! \times\! 10^{-5} $\! & $\! (467.6\! \pm\!   7.5)\! \times\! 10^{-5} $\\
45--50 & $\! ( 46.3\! \pm\!   1.4)\! \times\! 10^{-4} $\! & $\! ( 50.5\! \pm\!   1.5)\! \times\! 10^{-4} $\! & $\! (330.3\! \pm\!   5.4)\! \times\! 10^{-5} $\! & $\! (364.3\! \pm\!   5.8)\! \times\! 10^{-5} $\\
50--55 & $\! (333.5\! \pm\!   9.8)\! \times\! 10^{-5} $\! & $\! ( 39.6\! \pm\!   1.1)\! \times\! 10^{-4} $\! & $\! (246.0\! \pm\!   4.1)\! \times\! 10^{-5} $\! & $\! (301.8\! \pm\!   4.7)\! \times\! 10^{-5} $\\
55--60 & $\! (254.1\! \pm\!   7.5)\! \times\! 10^{-5} $\! & $\! (328.0\! \pm\!   9.4)\! \times\! 10^{-5} $\! & $\! (205.5\! \pm\!   3.7)\! \times\! 10^{-5} $\! & $\! (275.5\! \pm\!   4.5)\! \times\! 10^{-5} $\\
60--64 & $\! (208.4\! \pm\!   6.3)\! \times\! 10^{-5} $\! & $\! (309.0\! \pm\!   8.8)\! \times\! 10^{-5} $\! & $\! (173.9\! \pm\!   3.5)\! \times\! 10^{-5} $\! & $\! (270.6\! \pm\!   4.8)\! \times\! 10^{-5} $\\
64--68 & $\! (184.9\! \pm\!   5.7)\! \times\! 10^{-5} $\! & $\! (316.0\! \pm\!   9.2)\! \times\! 10^{-5} $\! & $\! (159.8\! \pm\!   3.5)\! \times\! 10^{-5} $\! & $\! (287.7\! \pm\!   5.5)\! \times\! 10^{-5} $\\
68--72 & $\! (202.6\! \pm\!   6.2)\! \times\! 10^{-5} $\! & $\! ( 36.0\! \pm\!   1.0)\! \times\! 10^{-4} $\! & $\! (180.5\! \pm\!   3.9)\! \times\! 10^{-5} $\! & $\! (335.7\! \pm\!   6.3)\! \times\! 10^{-5} $\\
72--76 & $\! (236.5\! \pm\!   7.2)\! \times\! 10^{-5} $\! & $\! ( 44.7\! \pm\!   1.3)\! \times\! 10^{-4} $\! & $\! (217.3\! \pm\!   4.7)\! \times\! 10^{-5} $\! & $\! (426.2\! \pm\!   7.8)\! \times\! 10^{-5} $\\
76--81 & $\! (333.1\! \pm\!   9.8)\! \times\! 10^{-5} $\! & $\! ( 64.1\! \pm\!   1.8)\! \times\! 10^{-4} $\! & $\! (315.3\! \pm\!   6.4)\! \times\! 10^{-5} $\! & $\! ( 62.2\! \pm\!   1.1)\! \times\! 10^{-4} $\\
81--86 & $\! ( 82.2\! \pm\!   2.5)\! \times\! 10^{-4} $\! & $\! (134.4\! \pm\!   4.0)\! \times\! 10^{-4} $\! & $\! ( 80.4\! \pm\!   1.7)\! \times\! 10^{-4} $\! & $\! (132.7\! \pm\!   2.7)\! \times\! 10^{-4} $\\
86--91 & $\! ( 76.2\! \pm\!   2.5)\! \times\! 10^{-3} $\! & $\! ( 78.4\! \pm\!   2.6)\! \times\! 10^{-3} $\! & $\! ( 76.2\! \pm\!   1.9)\! \times\! 10^{-3} $\! & $\! ( 78.5\! \pm\!   1.9)\! \times\! 10^{-3} $\\
91--96 & $\! ( 92.5\! \pm\!   3.2)\! \times\! 10^{-3} $\! & $\! ( 78.6\! \pm\!   2.8)\! \times\! 10^{-3} $\! & $\! ( 93.3\! \pm\!   2.6)\! \times\! 10^{-3} $\! & $\! ( 79.3\! \pm\!   2.2)\! \times\! 10^{-3} $\\
96--101 & $\! ( 78.8\! \pm\!   2.7)\! \times\! 10^{-4} $\! & $\! ( 70.0\! \pm\!   2.4)\! \times\! 10^{-4} $\! & $\! ( 80.5\! \pm\!   2.2)\! \times\! 10^{-4} $\! & $\! ( 71.6\! \pm\!   1.9)\! \times\! 10^{-4} $\\
101--106 & $\! (260.8\! \pm\!   8.2)\! \times\! 10^{-5} $\! & $\! (237.5\! \pm\!   7.4)\! \times\! 10^{-5} $\! & $\! (269.4\! \pm\!   6.4)\! \times\! 10^{-5} $\! & $\! (245.7\! \pm\!   5.8)\! \times\! 10^{-5} $\\
106--110 & $\! (139.9\! \pm\!   4.4)\! \times\! 10^{-5} $\! & $\! (129.2\! \pm\!   4.0)\! \times\! 10^{-5} $\! & $\! (145.2\! \pm\!   3.5)\! \times\! 10^{-5} $\! & $\! (134.3\! \pm\!   3.2)\! \times\! 10^{-5} $\\
110--115 & $\! ( 84.9\! \pm\!   2.6)\! \times\! 10^{-5} $\! & $\! ( 79.8\! \pm\!   2.5)\! \times\! 10^{-5} $\! & $\! ( 89.5\! \pm\!   2.2)\! \times\! 10^{-5} $\! & $\! ( 84.4\! \pm\!   2.0)\! \times\! 10^{-5} $\\
115--120 & $\! ( 60.5\! \pm\!   2.0)\! \times\! 10^{-5} $\! & $\! ( 56.5\! \pm\!   1.8)\! \times\! 10^{-5} $\! & $\! ( 64.0\! \pm\!   1.7)\! \times\! 10^{-5} $\! & $\! ( 59.8\! \pm\!   1.6)\! \times\! 10^{-5} $\\
120--126 & $\! ( 37.5\! \pm\!   1.3)\! \times\! 10^{-5} $\! & $\! ( 35.7\! \pm\!   1.2)\! \times\! 10^{-5} $\! & $\! ( 40.3\! \pm\!   1.2)\! \times\! 10^{-5} $\! & $\! ( 38.4\! \pm\!   1.1)\! \times\! 10^{-5} $\\
126--133 & $\! (282.8\! \pm\!   9.6)\! \times\! 10^{-6} $\! & $\! (265.4\! \pm\!   9.0)\! \times\! 10^{-6} $\! & $\! (305.9\! \pm\!   9.1)\! \times\! 10^{-6} $\! & $\! (287.4\! \pm\!   8.5)\! \times\! 10^{-6} $\\
133--141 & $\! (183.4\! \pm\!   6.8)\! \times\! 10^{-6} $\! & $\! (174.2\! \pm\!   6.4)\! \times\! 10^{-6} $\! & $\! (200.6\! \pm\!   6.7)\! \times\! 10^{-6} $\! & $\! (191.7\! \pm\!   6.4)\! \times\! 10^{-6} $\\
141--150 & $\! (136.8\! \pm\!   5.4)\! \times\! 10^{-6} $\! & $\! (130.5\! \pm\!   5.1)\! \times\! 10^{-6} $\! & $\! (153.0\! \pm\!   5.5)\! \times\! 10^{-6} $\! & $\! (146.2\! \pm\!   5.3)\! \times\! 10^{-6} $\\
150--160 & $\! ( 96.5\! \pm\!   4.2)\! \times\! 10^{-6} $\! & $\! ( 91.7\! \pm\!   4.0)\! \times\! 10^{-6} $\! & $\! (107.4\! \pm\!   4.4)\! \times\! 10^{-6} $\! & $\! (102.8\! \pm\!   4.2)\! \times\! 10^{-6} $\\
160--171 & $\! ( 65.8\! \pm\!   3.2)\! \times\! 10^{-6} $\! & $\! ( 63.2\! \pm\!   3.1)\! \times\! 10^{-6} $\! & $\! ( 75.6\! \pm\!   3.5)\! \times\! 10^{-6} $\! & $\! ( 72.7\! \pm\!   3.4)\! \times\! 10^{-6} $\\
171--185 & $\! ( 43.5\! \pm\!   2.2)\! \times\! 10^{-6} $\! & $\! ( 41.3\! \pm\!   2.1)\! \times\! 10^{-6} $\! & $\! ( 50.9\! \pm\!   2.6)\! \times\! 10^{-6} $\! & $\! ( 48.6\! \pm\!   2.4)\! \times\! 10^{-6} $\\
185--200 & $\! ( 32.6\! \pm\!   1.8)\! \times\! 10^{-6} $\! & $\! ( 31.2\! \pm\!   1.7)\! \times\! 10^{-6} $\! & $\! ( 39.2\! \pm\!   2.1)\! \times\! 10^{-6} $\! & $\! ( 37.6\! \pm\!   2.0)\! \times\! 10^{-6} $\\
200--220 & $\! ( 22.3\! \pm\!   1.2)\! \times\! 10^{-6} $\! & $\! ( 20.6\! \pm\!   1.1)\! \times\! 10^{-6} $\! & $\! ( 27.0\! \pm\!   1.4)\! \times\! 10^{-6} $\! & $\! ( 25.0\! \pm\!   1.2)\! \times\! 10^{-6} $\\
220--243 & $\! (132.4\! \pm\!   8.3)\! \times\! 10^{-7} $\! & $\! (129.5\! \pm\!   8.1)\! \times\! 10^{-7} $\! & $\! (164.2\! \pm\!   9.7)\! \times\! 10^{-7} $\! & $\! (161.0\! \pm\!   9.5)\! \times\! 10^{-7} $\\
243--273 & $\! ( 83.1\! \pm\!   5.4)\! \times\! 10^{-7} $\! & $\! ( 78.8\! \pm\!   5.1)\! \times\! 10^{-7} $\! & $\! (108.0\! \pm\!   6.7)\! \times\! 10^{-7} $\! & $\! (102.5\! \pm\!   6.4)\! \times\! 10^{-7} $\\
273--320 & $\! ( 41.1\! \pm\!   2.8)\! \times\! 10^{-7} $\! & $\! ( 38.5\! \pm\!   2.6)\! \times\! 10^{-7} $\! & $\! ( 55.3\! \pm\!   3.6)\! \times\! 10^{-7} $\! & $\! ( 52.0\! \pm\!   3.4)\! \times\! 10^{-7} $\\
320--380 & $\! ( 17.9\! \pm\!   1.5)\! \times\! 10^{-7} $\! & $\! ( 17.0\! \pm\!   1.4)\! \times\! 10^{-7} $\! & $\! ( 25.5\! \pm\!   2.1)\! \times\! 10^{-7} $\! & $\! ( 24.2\! \pm\!   2.0)\! \times\! 10^{-7} $\\
380--440 & $\! (  8.6\! \pm\!   1.0)\! \times\! 10^{-7} $\! & $\! ( 77.5\! \pm\!   9.2)\! \times\! 10^{-8} $\! & $\! ( 12.8\! \pm\!   1.5)\! \times\! 10^{-7} $\! & $\! ( 11.5\! \pm\!   1.3)\! \times\! 10^{-7} $\\
440--510 & $\! ( 33.1\! \pm\!   6.1)\! \times\! 10^{-8} $\! & $\! ( 33.3\! \pm\!   6.2)\! \times\! 10^{-8} $\! & $\! ( 49.3\! \pm\!   9.1)\! \times\! 10^{-8} $\! & $\! ( 49.9\! \pm\!   9.2)\! \times\! 10^{-8} $\\
510--600 & $\! ( 21.2\! \pm\!   3.8)\! \times\! 10^{-8} $\! & $\! ( 20.0\! \pm\!   3.5)\! \times\! 10^{-8} $\! & $\! ( 33.0\! \pm\!   5.8)\! \times\! 10^{-8} $\! & $\! ( 31.2\! \pm\!   5.5)\! \times\! 10^{-8} $\\
600--1000 & $\! ( 31.6\! \pm\!   6.1)\! \times\! 10^{-9} $\! & $\! ( 32.8\! \pm\!   6.4)\! \times\! 10^{-9} $\! & $\! ( 51.4\! \pm\!   9.9)\! \times\! 10^{-9} $\! & $\! (  5.3\! \pm\!   1.0)\! \times\! 10^{-8} $\\
1000--1500 & $\! (  3.4\! \pm\!   1.7)\! \times\! 10^{-9} $\! & $\! (  2.9\! \pm\!   1.4)\! \times\! 10^{-9} $\! & $\! (  5.8\! \pm\!   2.9)\! \times\! 10^{-9} $\! & $\! (  5.0\! \pm\!   2.4)\! \times\! 10^{-9} $\\
\hline
\end{tabular}
\end{center}
\end{table}

\begin{table} [htpb]
\begin{center}
\topcaption{
  The DY cross section measurements for the electron channel normalized to the $\Z$-peak region,
  pre- and post-FSR, as measured in the full acceptance and for the CMS detector acceptance.
  The uncertainty indicates the experimental (statistical and systematic) uncertainties
  summed in quadrature with the theoretical uncertainty resulting from the model-dependent
  kinematic distributions inside each bin.
  The results presented are in $\GeVns^{-1}$ units.
}
\label{tab:rshapes-electrons}
\begin{tabular}{|c|rrrr|}
\hline
$m$ (\GeVns) & \multicolumn{1}{c}{${r}^{i}_\text{pre-FSR}$} & \multicolumn{1}{c}{${r}^{i}_\text{post-FSR}$} &  \multicolumn{1}{c}{${r}^{i}_\text{pre-FSR, det}$} & \multicolumn{1}{c|}{${r}^{i}_\text{post-FSR, det}$} \\
\hline
   15--20   &$\! (15.4\! \pm\! 1.7)\! \times\! 10^{-2} $\! &$\! (14.9\! \pm\! 1.6)\! \times\! 10^{-2} $\! &$\! (145.4\! \pm\! 6.5)\! \times\! 10^{-5} $\! &$\! (140.3\! \pm\! 5.9)\! \times\! 10^{-5} $\\
   20--25   &$\! (59.5\! \pm\! 4.2)\! \times\! 10^{-3} $\! &$\! (58.5\! \pm\! 3.9)\! \times\! 10^{-3} $\! &$\! (105.2\! \pm\! 6.2)\! \times\! 10^{-5} $\! &$\! (103.3\! \pm\! 5.6)\! \times\! 10^{-5} $\\
   25--30   &$\! (34.0\! \pm\! 2.0)\! \times\! 10^{-3} $\! &$\! (33.4\! \pm\! 1.8)\! \times\! 10^{-3} $\! &$\! (131.5\! \pm\! 6.7)\! \times\! 10^{-5} $\! &$\! (128.5\! \pm\! 5.8)\! \times\! 10^{-5} $\\
   30--35   &$\! (19.7\! \pm\! 1.5)\! \times\! 10^{-3} $\! &$\! (19.4\! \pm\! 1.3)\! \times\! 10^{-3} $\! &$\! (23.9\! \pm\! 1.7)\! \times\! 10^{-4} $\! &$\! (23.7\! \pm\! 1.5)\! \times\! 10^{-4} $\\
   35--40   &$\! (115.3\! \pm\! 9.1)\! \times\! 10^{-4} $\! &$\! (116.6\! \pm\! 7.8)\! \times\! 10^{-4} $\! &$\! (33.2\! \pm\! 2.5)\! \times\! 10^{-4} $\! &$\! (33.3\! \pm\! 2.1)\! \times\! 10^{-4} $\\
   40--45   &$\! (68.9\! \pm\! 6.9)\! \times\! 10^{-4} $\! &$\! (74.1\! \pm\! 5.8)\! \times\! 10^{-4} $\! &$\! (33.6\! \pm\! 3.2)\! \times\! 10^{-4} $\! &$\! (35.6\! \pm\! 2.7)\! \times\! 10^{-4} $\\
   45--50   &$\! (47.5\! \pm\! 6.3)\! \times\! 10^{-4} $\! &$\! (55.1\! \pm\! 5.3)\! \times\! 10^{-4} $\! &$\! (28.6\! \pm\! 3.7)\! \times\! 10^{-4} $\! &$\! (34.1\! \pm\! 3.2)\! \times\! 10^{-4} $\\
   50--55   &$\! (34.4\! \pm\! 3.3)\! \times\! 10^{-4} $\! &$\! (44.8\! \pm\! 2.7)\! \times\! 10^{-4} $\! &$\! (23.3\! \pm\! 2.2)\! \times\! 10^{-4} $\! &$\! (32.5\! \pm\! 1.8)\! \times\! 10^{-4} $\\
   55--60   &$\! (24.7\! \pm\! 2.7)\! \times\! 10^{-4} $\! &$\! (39.4\! \pm\! 2.2)\! \times\! 10^{-4} $\! &$\! (18.2\! \pm\! 2.0)\! \times\! 10^{-4} $\! &$\! (31.7\! \pm\! 1.6)\! \times\! 10^{-4} $\\
   60--64   &$\! (23.6\! \pm\! 4.7)\! \times\! 10^{-4} $\! &$\! (41.9\! \pm\! 3.7)\! \times\! 10^{-4} $\! &$\! (18.7\! \pm\! 3.9)\! \times\! 10^{-4} $\! &$\! (36.1\! \pm\! 3.1)\! \times\! 10^{-4} $\\
   64--68   &$\! (22.1\! \pm\! 4.2)\! \times\! 10^{-4} $\! &$\! (45.5\! \pm\! 3.3)\! \times\! 10^{-4} $\! &$\! (18.0\! \pm\! 3.6)\! \times\! 10^{-4} $\! &$\! (40.7\! \pm\! 2.8)\! \times\! 10^{-4} $\\
   68--72   &$\! (22.5\! \pm\! 4.9)\! \times\! 10^{-4} $\! &$\! (52.8\! \pm\! 3.7)\! \times\! 10^{-4} $\! &$\! (19.7\! \pm\! 4.4)\! \times\! 10^{-4} $\! &$\! (49.3\! \pm\! 3.3)\! \times\! 10^{-4} $\\
   72--76   &$\! (25.2\! \pm\! 5.9)\! \times\! 10^{-4} $\! &$\! (65.3\! \pm\! 4.3)\! \times\! 10^{-4} $\! &$\! (22.7\! \pm\! 5.3)\! \times\! 10^{-4} $\! &$\! (62.3\! \pm\! 4.0)\! \times\! 10^{-4} $\\
   76--81   &$\! (36.5\! \pm\! 7.0)\! \times\! 10^{-4} $\! &$\! (95.3\! \pm\! 4.9)\! \times\! 10^{-4} $\! &$\! (34.4\! \pm\! 6.5)\! \times\! 10^{-4} $\! &$\! (92.8\! \pm\! 4.4)\! \times\! 10^{-4} $\\
   81--86   &$\! (10.7\! \pm\! 1.8)\! \times\! 10^{-3} $\! &$\! (19.2\! \pm\! 1.3)\! \times\! 10^{-3} $\! &$\! (10.4\! \pm\! 1.7)\! \times\! 10^{-3} $\! &$\! (19.0\! \pm\! 1.2)\! \times\! 10^{-3} $\\
   86--91   &$\! (64.8\! \pm\! 7.8)\! \times\! 10^{-3} $\! &$\! (70.3\! \pm\! 6.4)\! \times\! 10^{-3} $\! &$\! (64.8\! \pm\! 7.7)\! \times\! 10^{-3} $\! &$\! (70.6\! \pm\! 6.3)\! \times\! 10^{-3} $\\
   91--96   &$\! (100.8\! \pm\! 8.8)\! \times\! 10^{-3} $\! &$\! (74.0\! \pm\! 6.4)\! \times\! 10^{-3} $\! &$\! (101.9\! \pm\! 8.8)\! \times\! 10^{-3} $\! &$\! (75.1\! \pm\! 6.4)\! \times\! 10^{-3} $\\
   96--101   &$\! ( 6.9\! \pm\! 1.2)\! \times\! 10^{-3} $\! &$\! (56.8\! \pm\! 9.3)\! \times\! 10^{-4} $\! &$\! ( 7.1\! \pm\! 1.3)\! \times\! 10^{-3} $\! &$\! (58.8\! \pm\! 9.5)\! \times\! 10^{-4} $\\
  101--106   &$\! (31.1\! \pm\! 6.4)\! \times\! 10^{-4} $\! &$\! (25.7\! \pm\! 4.8)\! \times\! 10^{-4} $\! &$\! (32.5\! \pm\! 6.7)\! \times\! 10^{-4} $\! &$\! (26.9\! \pm\! 5.0)\! \times\! 10^{-4} $\\
  106--110   &$\! (12.8\! \pm\! 2.7)\! \times\! 10^{-4} $\! &$\! (11.3\! \pm\! 2.0)\! \times\! 10^{-4} $\! &$\! (13.5\! \pm\! 2.8)\! \times\! 10^{-4} $\! &$\! (12.1\! \pm\! 2.1)\! \times\! 10^{-4} $\\
  110--115   &$\! ( 9.5\! \pm\! 1.4)\! \times\! 10^{-4} $\! &$\! ( 8.2\! \pm\! 1.0)\! \times\! 10^{-4} $\! &$\! (10.1\! \pm\! 1.5)\! \times\! 10^{-4} $\! &$\! ( 8.8\! \pm\! 1.1)\! \times\! 10^{-4} $\\
  115--120   &$\! (57.1\! \pm\! 8.2)\! \times\! 10^{-5} $\! &$\! (50.9\! \pm\! 6.1)\! \times\! 10^{-5} $\! &$\! (61.7\! \pm\! 8.7)\! \times\! 10^{-5} $\! &$\! (55.2\! \pm\! 6.5)\! \times\! 10^{-5} $\\
  120--126   &$\! (38.0\! \pm\! 4.6)\! \times\! 10^{-5} $\! &$\! (34.1\! \pm\! 3.5)\! \times\! 10^{-5} $\! &$\! (41.2\! \pm\! 4.9)\! \times\! 10^{-5} $\! &$\! (37.3\! \pm\! 3.7)\! \times\! 10^{-5} $\\
  126--133   &$\! (26.1\! \pm\! 2.7)\! \times\! 10^{-5} $\! &$\! (23.9\! \pm\! 2.1)\! \times\! 10^{-5} $\! &$\! (29.1\! \pm\! 3.0)\! \times\! 10^{-5} $\! &$\! (26.7\! \pm\! 2.3)\! \times\! 10^{-5} $\\
  133--141   &$\! (21.0\! \pm\! 2.1)\! \times\! 10^{-5} $\! &$\! (18.7\! \pm\! 1.6)\! \times\! 10^{-5} $\! &$\! (23.4\! \pm\! 2.3)\! \times\! 10^{-5} $\! &$\! (20.8\! \pm\! 1.8)\! \times\! 10^{-5} $\\
  141--150   &$\! (12.5\! \pm\! 1.3)\! \times\! 10^{-5} $\! &$\! (115.3\! \pm\! 9.9)\! \times\! 10^{-6} $\! &$\! (14.3\! \pm\! 1.4)\! \times\! 10^{-5} $\! &$\! (13.3\! \pm\! 1.1)\! \times\! 10^{-5} $\\
  150--160   &$\! (83.1\! \pm\! 9.6)\! \times\! 10^{-6} $\! &$\! (80.2\! \pm\! 7.4)\! \times\! 10^{-6} $\! &$\! ( 9.6\! \pm\! 1.1)\! \times\! 10^{-5} $\! &$\! (93.2\! \pm\! 8.4)\! \times\! 10^{-6} $\\
  160--171   &$\! (82.7\! \pm\! 9.4)\! \times\! 10^{-6} $\! &$\! (71.9\! \pm\! 7.2)\! \times\! 10^{-6} $\! &$\! ( 9.6\! \pm\! 1.1)\! \times\! 10^{-5} $\! &$\! (84.0\! \pm\! 8.3)\! \times\! 10^{-6} $\\
  171--185   &$\! (38.8\! \pm\! 4.9)\! \times\! 10^{-6} $\! &$\! (37.1\! \pm\! 3.8)\! \times\! 10^{-6} $\! &$\! (46.6\! \pm\! 5.8)\! \times\! 10^{-6} $\! &$\! (45.1\! \pm\! 4.6)\! \times\! 10^{-6} $\\
  185--200   &$\! (31.4\! \pm\! 3.5)\! \times\! 10^{-6} $\! &$\! (29.4\! \pm\! 2.8)\! \times\! 10^{-6} $\! &$\! (38.4\! \pm\! 4.3)\! \times\! 10^{-6} $\! &$\! (36.8\! \pm\! 3.4)\! \times\! 10^{-6} $\\
  200--220   &$\! (22.5\! \pm\! 2.1)\! \times\! 10^{-6} $\! &$\! (20.3\! \pm\! 1.7)\! \times\! 10^{-6} $\! &$\! (28.8\! \pm\! 2.6)\! \times\! 10^{-6} $\! &$\! (25.7\! \pm\! 2.1)\! \times\! 10^{-6} $\\
  220--243   &$\! (12.9\! \pm\! 1.4)\! \times\! 10^{-6} $\! &$\! (12.2\! \pm\! 1.2)\! \times\! 10^{-6} $\! &$\! (16.5\! \pm\! 1.8)\! \times\! 10^{-6} $\! &$\! (15.8\! \pm\! 1.5)\! \times\! 10^{-6} $\\
  243--273   &$\! (73.6\! \pm\! 8.8)\! \times\! 10^{-7} $\! &$\! (67.2\! \pm\! 7.1)\! \times\! 10^{-7} $\! &$\! ( 9.6\! \pm\! 1.2)\! \times\! 10^{-6} $\! &$\! (90.1\! \pm\! 9.3)\! \times\! 10^{-7} $\\
  273--320   &$\! (40.8\! \pm\! 4.8)\! \times\! 10^{-7} $\! &$\! (37.3\! \pm\! 4.0)\! \times\! 10^{-7} $\! &$\! (56.2\! \pm\! 6.6)\! \times\! 10^{-7} $\! &$\! (51.0\! \pm\! 5.3)\! \times\! 10^{-7} $\\
  320--380   &$\! (17.7\! \pm\! 2.4)\! \times\! 10^{-7} $\! &$\! (15.7\! \pm\! 2.0)\! \times\! 10^{-7} $\! &$\! (26.4\! \pm\! 3.6)\! \times\! 10^{-7} $\! &$\! (23.3\! \pm\! 2.9)\! \times\! 10^{-7} $\\
  380--440   &$\! ( 7.2\! \pm\! 1.6)\! \times\! 10^{-7} $\! &$\! ( 6.7\! \pm\! 1.3)\! \times\! 10^{-7} $\! &$\! (11.3\! \pm\! 2.5)\! \times\! 10^{-7} $\! &$\! (10.5\! \pm\! 2.0)\! \times\! 10^{-7} $\\
  440--510   &$\! (38.2\! \pm\! 9.6)\! \times\! 10^{-8} $\! &$\! (36.0\! \pm\! 8.0)\! \times\! 10^{-8} $\! &$\! ( 6.0\! \pm\! 1.5)\! \times\! 10^{-7} $\! &$\! ( 5.7\! \pm\! 1.2)\! \times\! 10^{-7} $\\
  510--600   &$\! (22.6\! \pm\! 5.1)\! \times\! 10^{-8} $\! &$\! (20.2\! \pm\! 4.2)\! \times\! 10^{-8} $\! &$\! (35.9\! \pm\! 8.1)\! \times\! 10^{-8} $\! &$\! (32.2\! \pm\! 6.7)\! \times\! 10^{-8} $\\
  600--1000   &$\! ( 4.1\! \pm\! 1.0)\! \times\! 10^{-8} $\! &$\! (36.8\! \pm\! 8.9)\! \times\! 10^{-9} $\! &$\! ( 6.8\! \pm\! 1.7)\! \times\! 10^{-8} $\! &$\! ( 6.1\! \pm\! 1.5)\! \times\! 10^{-8} $\\
 1000--1500   &$\! ( 1.3\! \pm\! 2.3)\! \times\! 10^{-9} $\! &$\! ( 1.1\! \pm\! 2.0)\! \times\! 10^{-9} $\! &$\! ( 2.2\! \pm\! 4.0)\! \times\! 10^{-9} $\! &$\! ( 1.9\! \pm\! 3.5)\! \times\! 10^{-9} $\\
\end{tabular}
\end{center}
\end{table}

The measurements in the two channels are combined using the procedure defined in Ref.~\cite{BLUE},
which provides a full covariance matrix for the uncertainties.
Given the cross section measurements in the dimuon and dielectron channels, and their
symmetric and positive-definite covariance matrices, the estimates of the true cross section values are found as
unbiased linear combinations of the input measurements having a minimum variance.

The uncertainties are considered to be uncorrelated between the two analyses. Exceptions are
the modeling uncertainty, which is $100\%$ correlated between channels, and the uncertainty in the acceptance, which originates
mainly from the PDFs. The acceptance is almost identical between the two channels and
the differences in uncertainties between them are negligible. Thus, when combining
the measurements we add the uncertainty in the acceptance (in quadrature) to the
total uncertainty after the combination is done.
The acceptance uncertainty does not include correlations between analysis bins.

\begin{figure}[htb]
\centering
\includegraphics[width=0.90\textwidth]{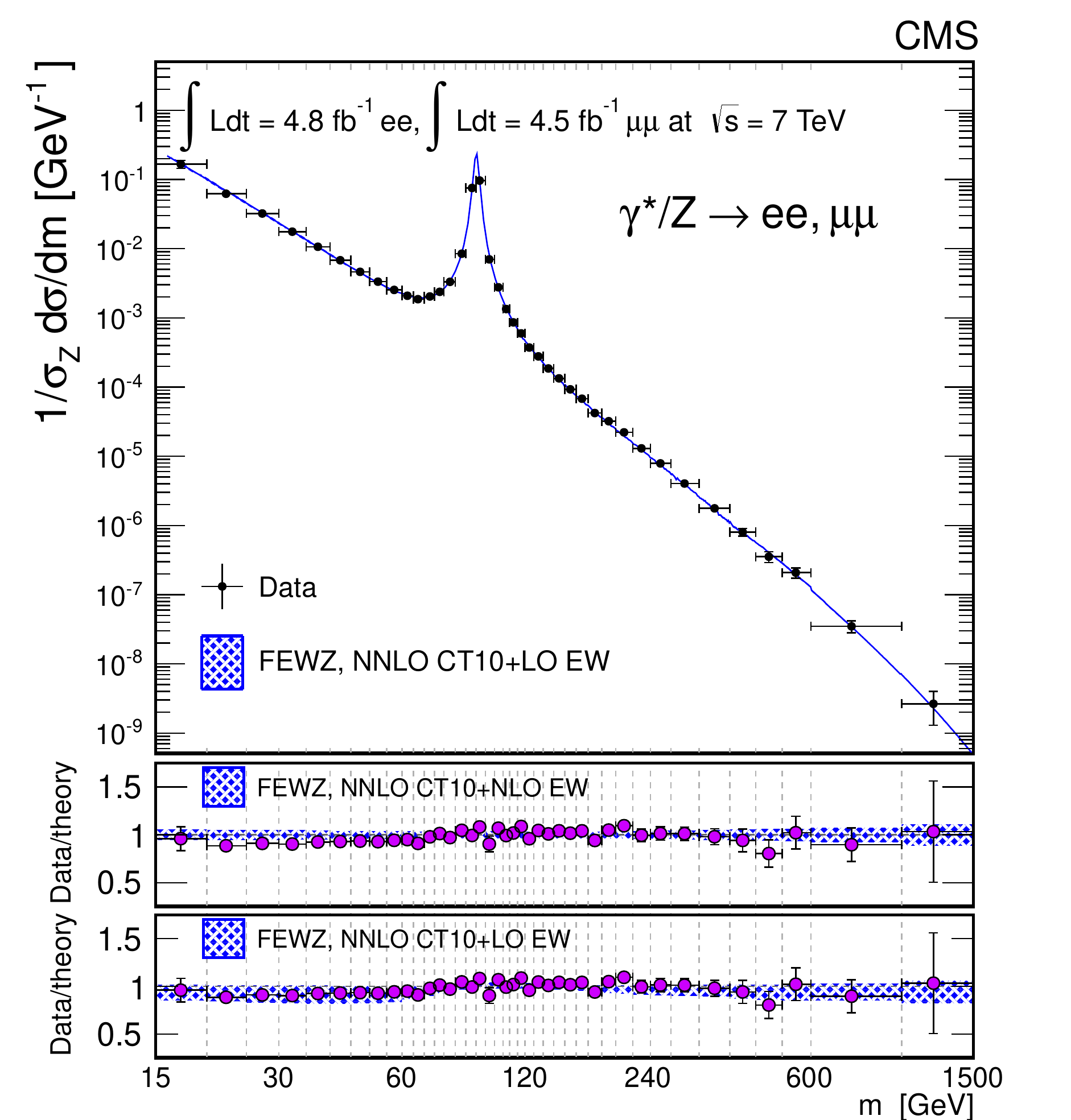}
\caption{
\label{fig:combination}
Combined DY differential cross section measurement in the dimuon and dielectron channels
normalized to the $\Z$-peak region with the FSR effect taken into account.
The data point abscissas are computed according to Eq.~(6) in Ref.~\cite{bib:pts_plac}.
Including the correlations between the two channels,
the normalized $\chi^2$ calculated with total uncertainties on the combined results
is 1.1 between data and the theoretical expectation, with 40 degrees of freedom.
The corresponding $\chi^2$ probability is 36.8\%.
}
\end{figure}

Figure~\ref{fig:combination} shows the DY cross section measurement in the combined dimuon and
dielectron channels normalized to the $\Z$-boson mass peak region with the FSR effect taken into account.
The corresponding results are summarized in Table~\ref{tab:rshapes-comb}.

\begin{table} [htb]
\begin{center}
\topcaption{
  The DY pre-FSR cross section measurements for the combined dimuon and dielectron channels
  normalized to the $\Z$-peak region in the full acceptance.
  The results presented are in $\GeVns^{-1}$ units.
}
\label{tab:rshapes-comb}
\begin{tabular}{|cr|cr|cr|}
\hline
$m$ (\GeVns) & \multicolumn{1}{c|}{${r}^{i}_\text{pre-FSR}$} & $m$ (\GeVns) & \multicolumn{1}{c|}{${r}^{i}_\text{pre-FSR}$} & $m$ (\GeVns) & \multicolumn{1}{c|}{${r}^{i}_\text{pre-FSR}$} \\
\hline
15--20 & $\! ( 16.7\! \pm\!   1.4)\! \times\! 10^{-2} $\! & 81--86 & $\! ( 84.5\! \pm\!   4.5)\! \times\! 10^{-4} $\!   & 171--185 & $\! ( 42.1\! \pm\!   2.4)\! \times\! 10^{-6} $ \\
20--25 & $\! ( 62.2\! \pm\!   2.9)\! \times\! 10^{-3} $\! & 86--91 & $\! ( 75.3\! \pm\!   2.7)\! \times\! 10^{-3} $\!   & 185--200 & $\! ( 32.1\! \pm\!   1.9)\! \times\! 10^{-6} $ \\
25--30 & $\! ( 32.2\! \pm\!   1.2)\! \times\! 10^{-3} $\! & 91--96 & $\! ( 96.4\! \pm\!   3.3)\! \times\! 10^{-3} $\!   & 200--220 & $\! ( 22.2\! \pm\!   1.2)\! \times\! 10^{-6} $ \\
30--35 & $\! (175.9\! \pm\!   4.7)\! \times\! 10^{-4} $\! & 96--101 & $\! ( 70.0\! \pm\!   6.3)\! \times\! 10^{-4} $\!  & 220--243 & $\! (130.1\! \pm\!   8.4)\! \times\! 10^{-7} $  \\
35--40 & $\! (106.5\! \pm\!   2.8)\! \times\! 10^{-4} $\! & 101--106 & $\! ( 27.6\! \pm\!   1.5)\! \times\! 10^{-4} $\! & 243--273 & $\! ( 78.9\! \pm\!   5.3)\! \times\! 10^{-7} $  \\
40--45 & $\! ( 68.0\! \pm\!   1.9)\! \times\! 10^{-4} $\! & 106--110 & $\! (134.1\! \pm\!   5.9)\! \times\! 10^{-5} $\! & 273--320 & $\! ( 40.4\! \pm\!   2.7)\! \times\! 10^{-7} $  \\
45--50 & $\! ( 46.2\! \pm\!   1.2)\! \times\! 10^{-4} $\! & 110--115 & $\! ( 86.0\! \pm\!   3.2)\! \times\! 10^{-5} $\! & 320--380 & $\! ( 17.7\! \pm\!   1.5)\! \times\! 10^{-7} $  \\
50--55 & $\! ( 33.3\! \pm\!   1.1)\! \times\! 10^{-4} $\! & 115--120 & $\! ( 59.7\! \pm\!   2.1)\! \times\! 10^{-5} $\! & 380--440 & $\! (  8.0\! \pm\!   1.0)\! \times\! 10^{-7} $  \\
55--60 & $\! (253.5\! \pm\!   7.7)\! \times\! 10^{-5} $\! & 120--126 & $\! ( 37.3\! \pm\!   1.4)\! \times\! 10^{-5} $\! & 440--510 & $\! ( 35.4\! \pm\!   6.2)\! \times\! 10^{-8} $  \\
60--64 & $\! (208.7\! \pm\!   6.2)\! \times\! 10^{-5} $\! & 126--133 & $\! ( 27.8\! \pm\!   1.0)\! \times\! 10^{-5} $\! & 510--600 & $\! ( 20.8\! \pm\!   3.3)\! \times\! 10^{-8} $  \\
64--68 & $\! (186.2\! \pm\!   6.2)\! \times\! 10^{-5} $\! & 133--141 & $\! (185.7\! \pm\!   7.7)\! \times\! 10^{-6} $\! & 600--1000 & $\! ( 34.9\! \pm\!   6.7)\! \times\! 10^{-9} $  \\
68--72 & $\! (203.7\! \pm\!   6.7)\! \times\! 10^{-5} $\! & 141--150 & $\! (133.6\! \pm\!   5.8)\! \times\! 10^{-6} $\! & 1000--1500 & $\! (  2.7\! \pm\!   1.4)\! \times\! 10^{-9} $  \\
72--76 & $\! (237.9\! \pm\!   8.2)\! \times\! 10^{-5} $\! & 150--160 & $\! ( 92.7\! \pm\!   4.6)\! \times\! 10^{-6} $\! & & \\
76--81 & $\! ( 33.2\! \pm\!   1.1)\! \times\! 10^{-4} $\! & 160--171 & $\! ( 67.9\! \pm\!   3.7)\! \times\! 10^{-6} $\! & & \\
\hline
\end{tabular}
\end{center}
\end{table}

\subsection{Double-differential cross section \texorpdfstring{$\rd^2\sigma/\rd{}m\,\rd\abs{y}$}{dsigma/dmdy} measurement}

The result of the double-differential cross section measurement for the dimuon channel is presented as the following ratio:

\begin{equation}
R^{ij}_\text{pre-FSR, det} = \left. \frac{N_\mathrm{u}^{ij}}{\epsilon^{ij}\rho^{ij}} \middle/ \frac{N_\mathrm{u}^\text{norm}}{\epsilon^\text{norm}\rho^\text{norm}} \right..
\end{equation}

The quantities $N_\mathrm{u}^{ij}$, $\epsilon^{ij}$, $\rho^{ij}$
are defined in a given bin $(i,~j)$, with $i$ corresponding to the binning
in invariant mass, and $j$ corresponding to the binning
in absolute rapidity; $N^\text{norm}_\mathrm{u}$,
$\epsilon^\text{norm}$, and $\rho^\text{norm}$ refer to the $\Z$-peak region within $\abs{y} < 2.4$ in the muon acceptance.
The normalization factors from our measurement and each theoretical prediction from various PDF sets
are available in Table~\ref{tab:normFactor2D}. As shown in this table,
the normalization factors
between data and theoretical predictions are in good agreement within the uncertainty, except for JR09.
These results are normalized to the dimuon absolute rapidity bin widths, $\Delta y^{j}$, defining the shape
${r}^{ij} = R^{ij}/(\Delta y^{j}).$
An acceptance correction to the full phase space would not increase the sensitivity to PDFs.
Therefore, this measurement is performed within the detector acceptance in order to reduce
model dependence.
We use the NNLO reweighted \POWHEG sample in this measurement, which is discussed in Section~\ref{sec:data}.
This sample is used to derive the selection efficiency and to produce response matrices for
detector resolution and FSR corrections.

\begin{table} [htb]
\begin{center}
\topcaption{
  Normalization factors for the measurement in the $\Z$-peak region ($60 < m < 120$\GeV and $\abs{y} < 2.4$)
  and the detector acceptance for the dimuon channel.
  The row for the data corresponds to the pre-FSR, detector acceptance result in
  Table~\ref{tab:normFactor1D_data} for the muon channel.
  The uncertainty in the theoretical cross sections indicates the statistical calculation uncertainty
  and PDF uncertainty in \FEWZ.
}
\label{tab:normFactor2D}
\begin{tabular}{|l|c|c|c|c|c|c|}
\hline
& Cross section in the $\Z$-peak region in the detector acceptance\\
& ($60 < m < 120$\GeV, $\abs{y} < 2.4$)\\
\hline
Data & $524.7 \pm 0.4\stat \pm 5.1\,\text{(exp. syst.)} \pm 1.2\,\text{(th. syst.)} \pm  11.5\lum$\unit{pb} \\
CT10 NNLO & 534.29 $\pm$ 0.36 (stat) $\pm$ 16.60\,(PDF)\unit{pb}\\
NNPDF2.1 NNLO & 524.76 $\pm$ 0.68 (stat) $\pm$ 6.38\,(PDF)\unit{pb} \\
MSTW2008 NNLO & 524.02 $\pm$ 0.38\stat $\pm$ 17.46\,(PDF)\unit{pb} \\
JR09 NNLO & 485.97 $\pm$ 0.36\stat $\pm$ 11.78\,(PDF)\unit{pb} \\
ABKM09 NNLO & 534.69 $\pm$ 0.43\stat $\pm$ 9.30\,(PDF)\unit{pb} \\
HERAPDF15 NNLO & 531.92 $\pm$ 0.23\stat $\pm$ 6.25\,(PDF)\unit{pb} \\
\hline
\end{tabular}
\end{center}
\end{table}

\begin{figure}[htpb]
{\centering
\includegraphics[width=0.45\textwidth]{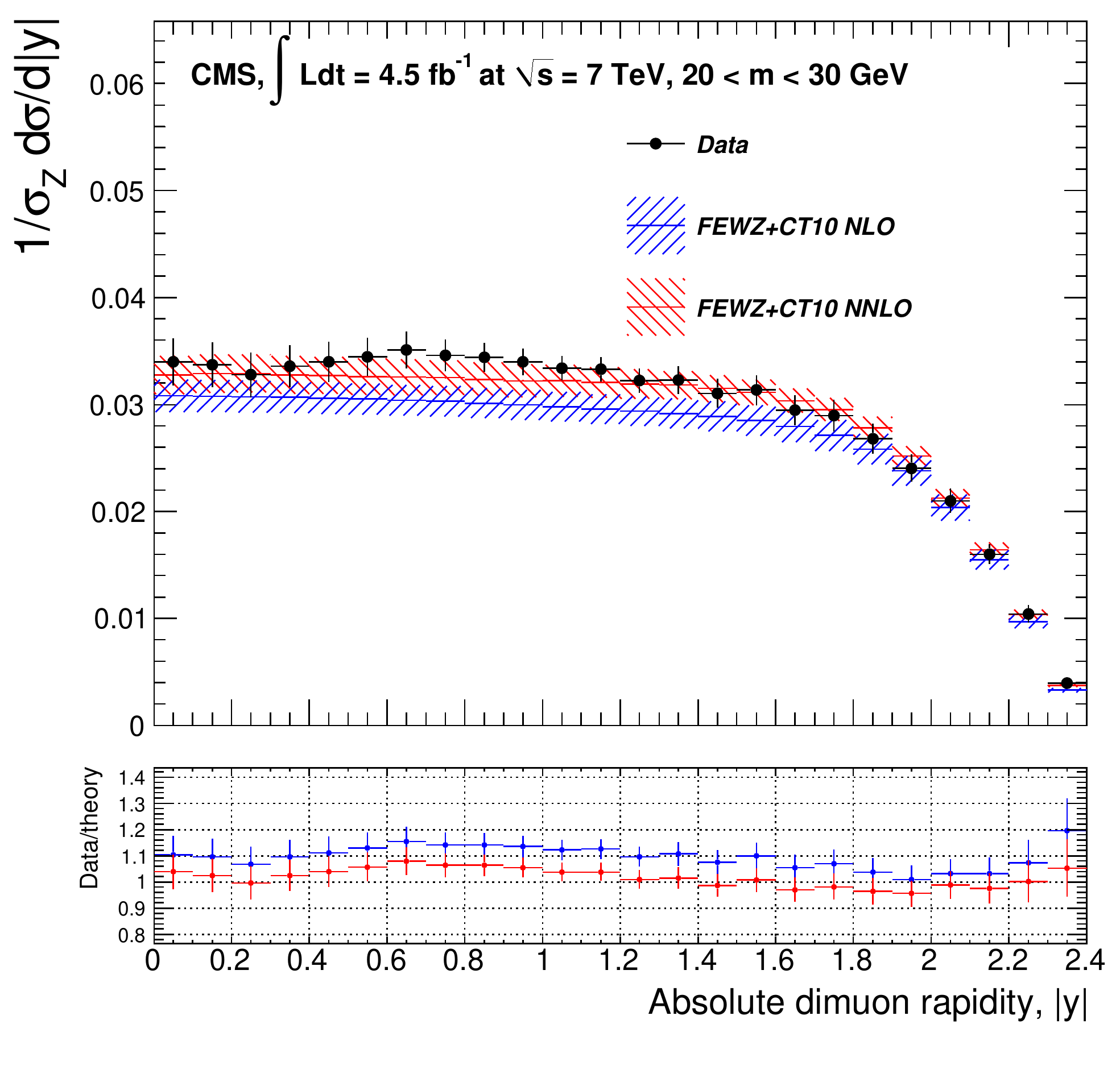}
\includegraphics[width=0.45\textwidth]{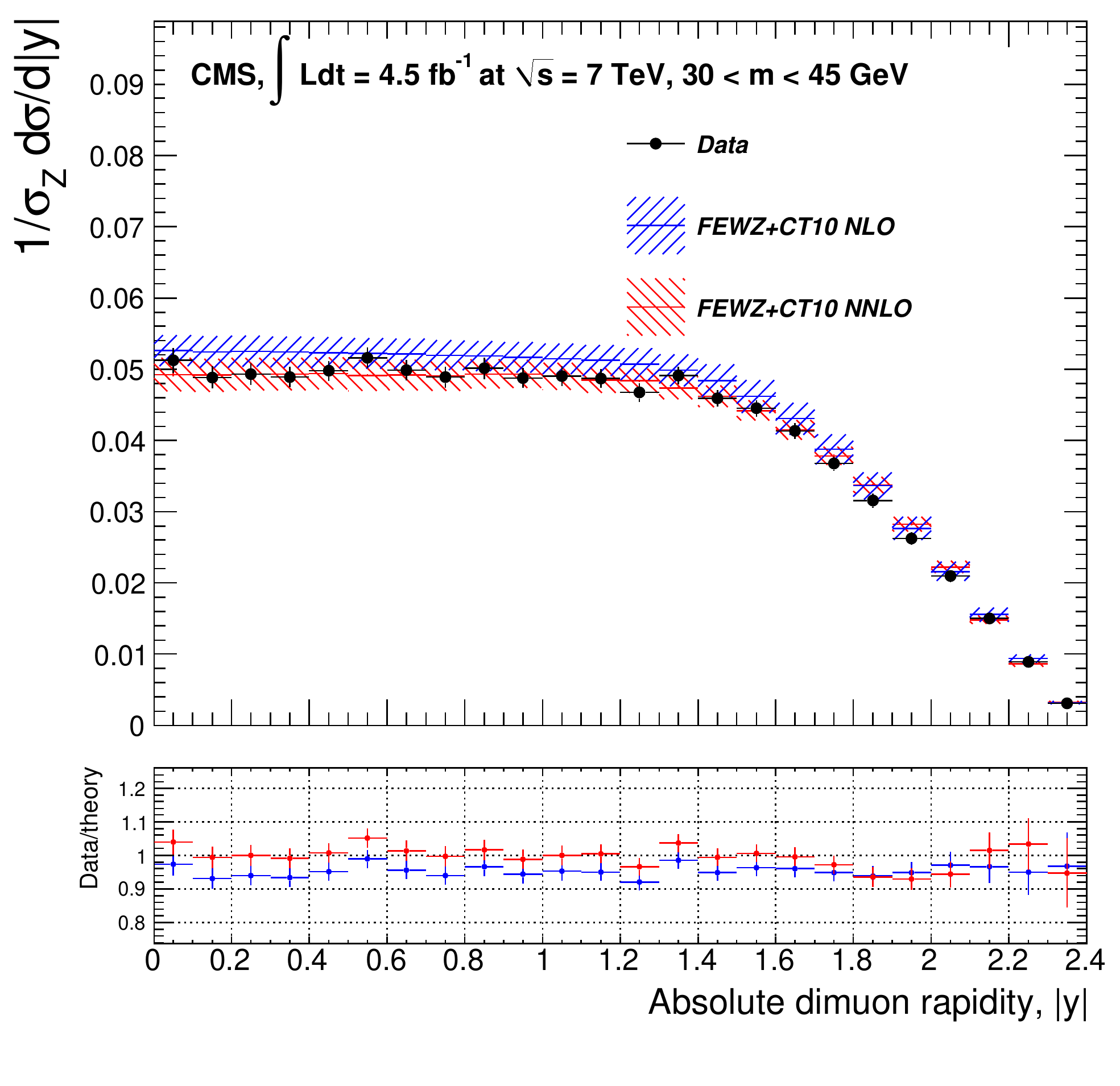}
\includegraphics[width=0.45\textwidth]{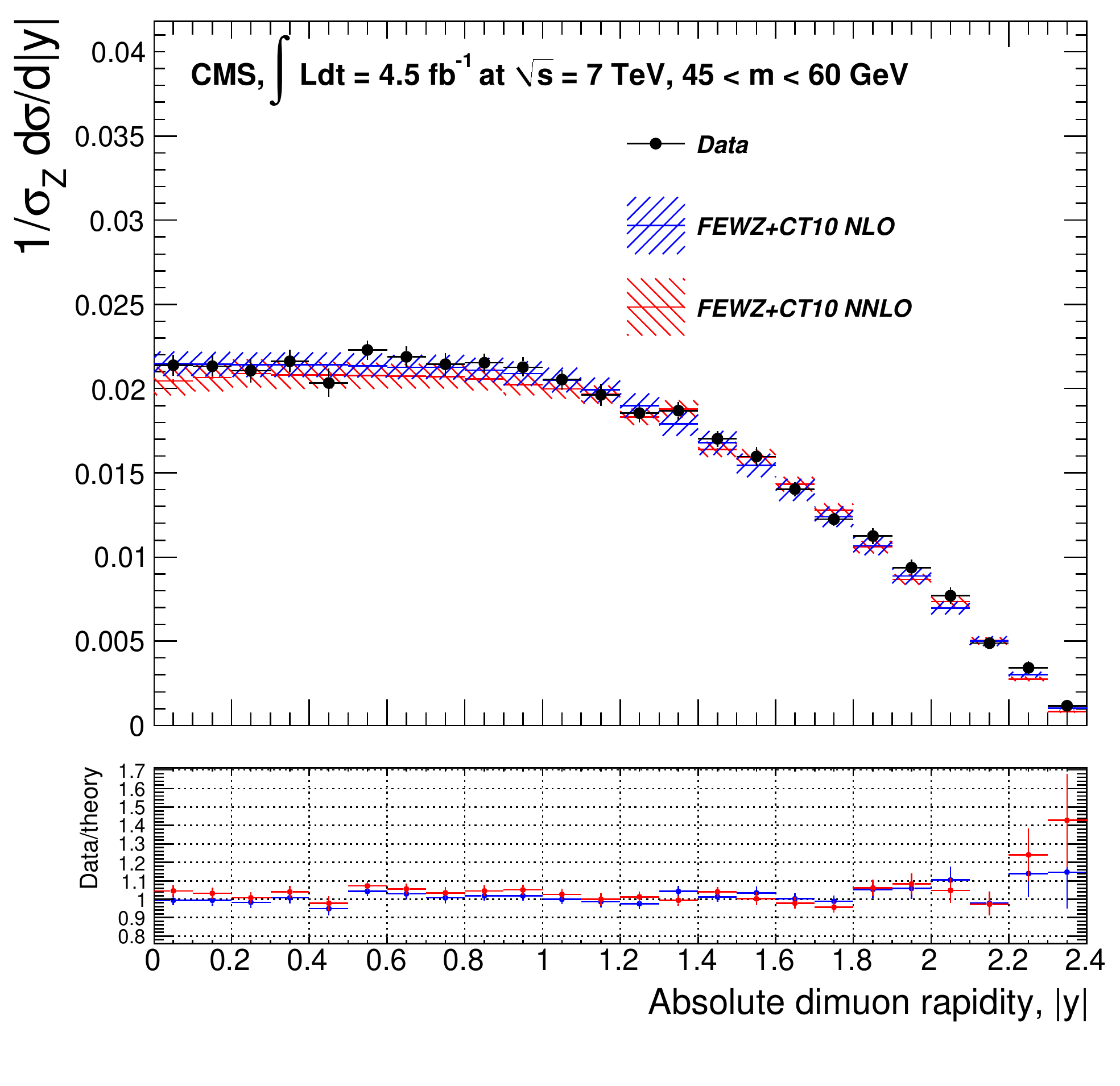}
\includegraphics[width=0.45\textwidth]{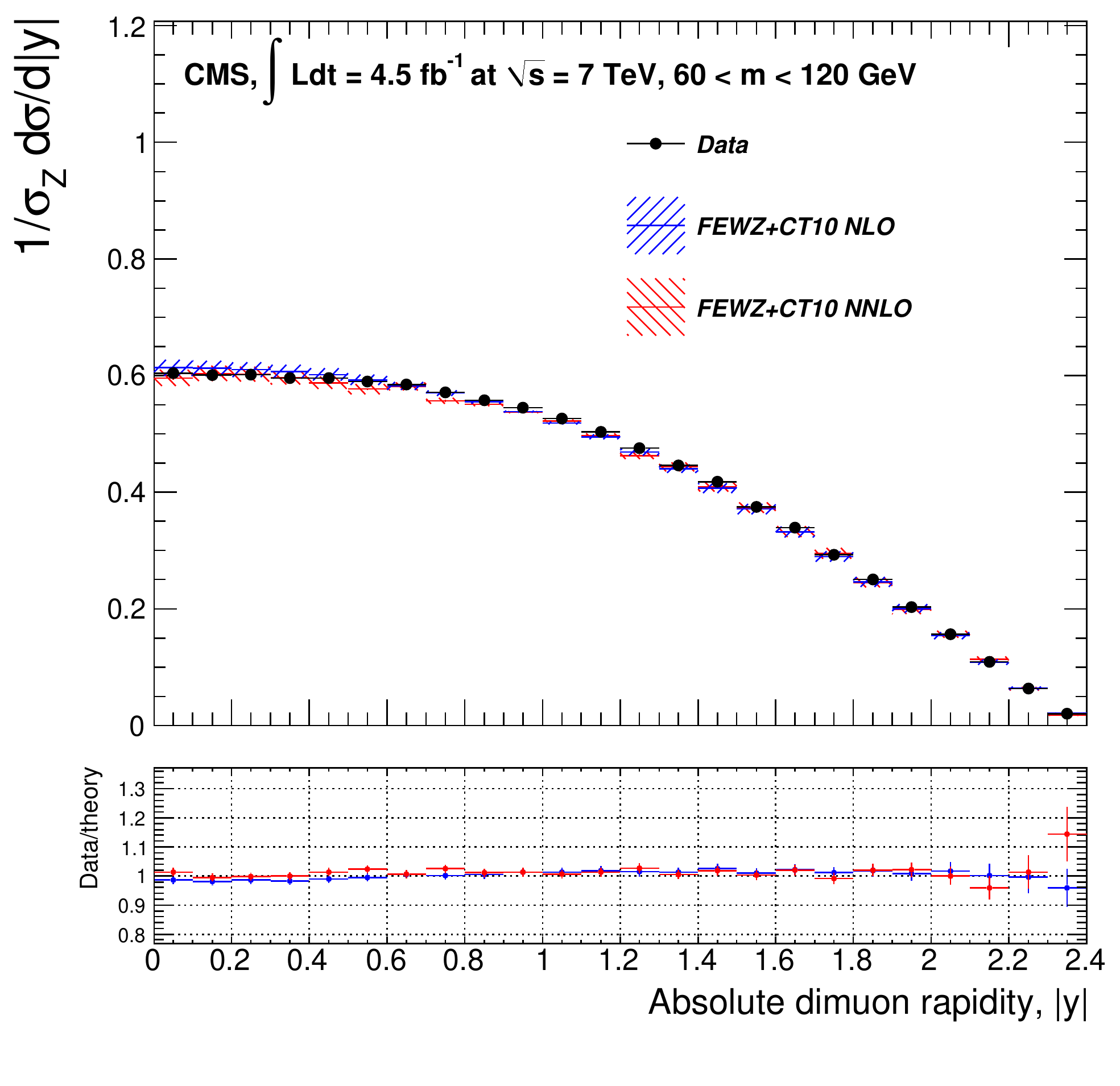}
\includegraphics[width=0.45\textwidth]{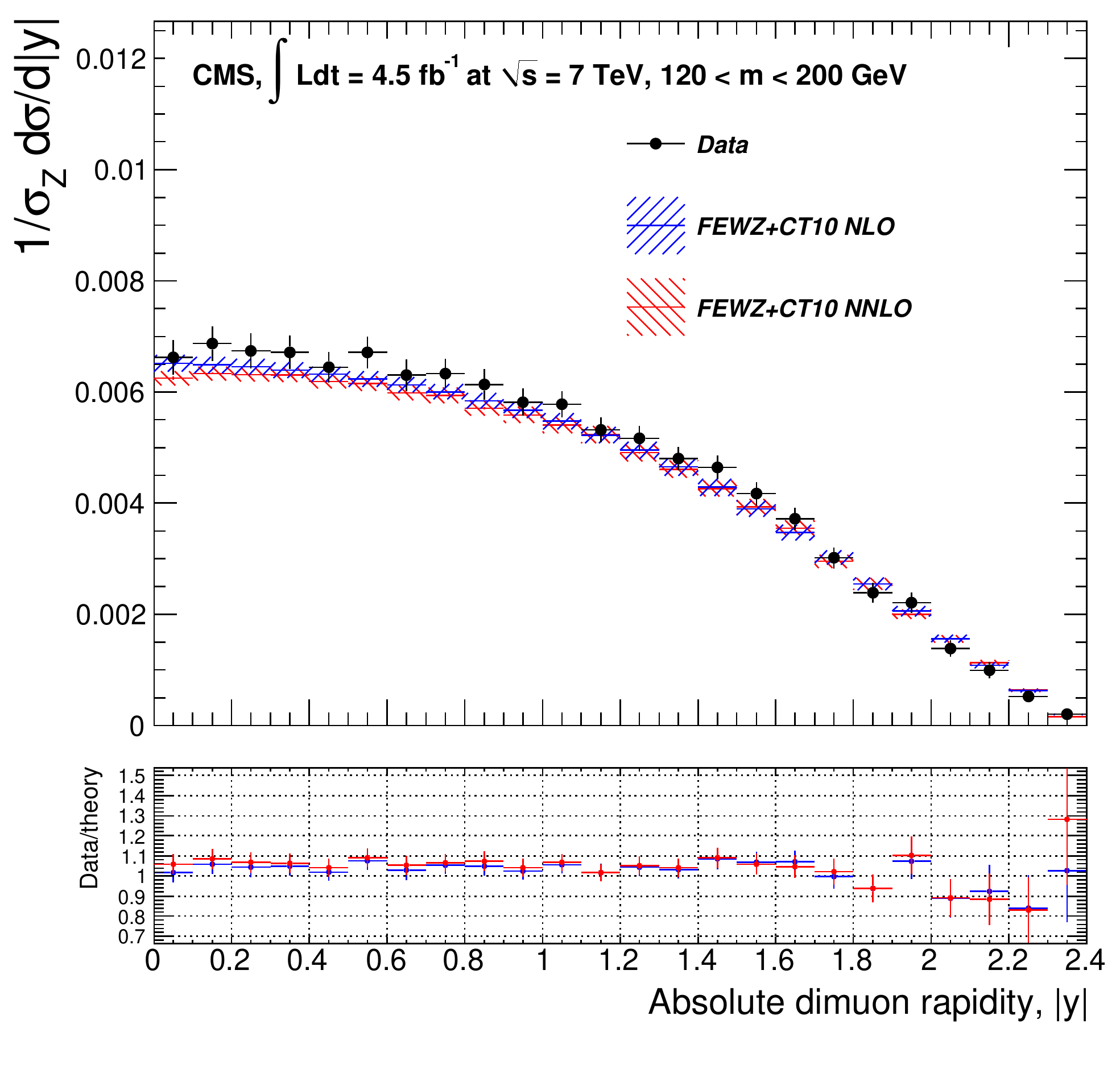}
\includegraphics[width=0.45\textwidth]{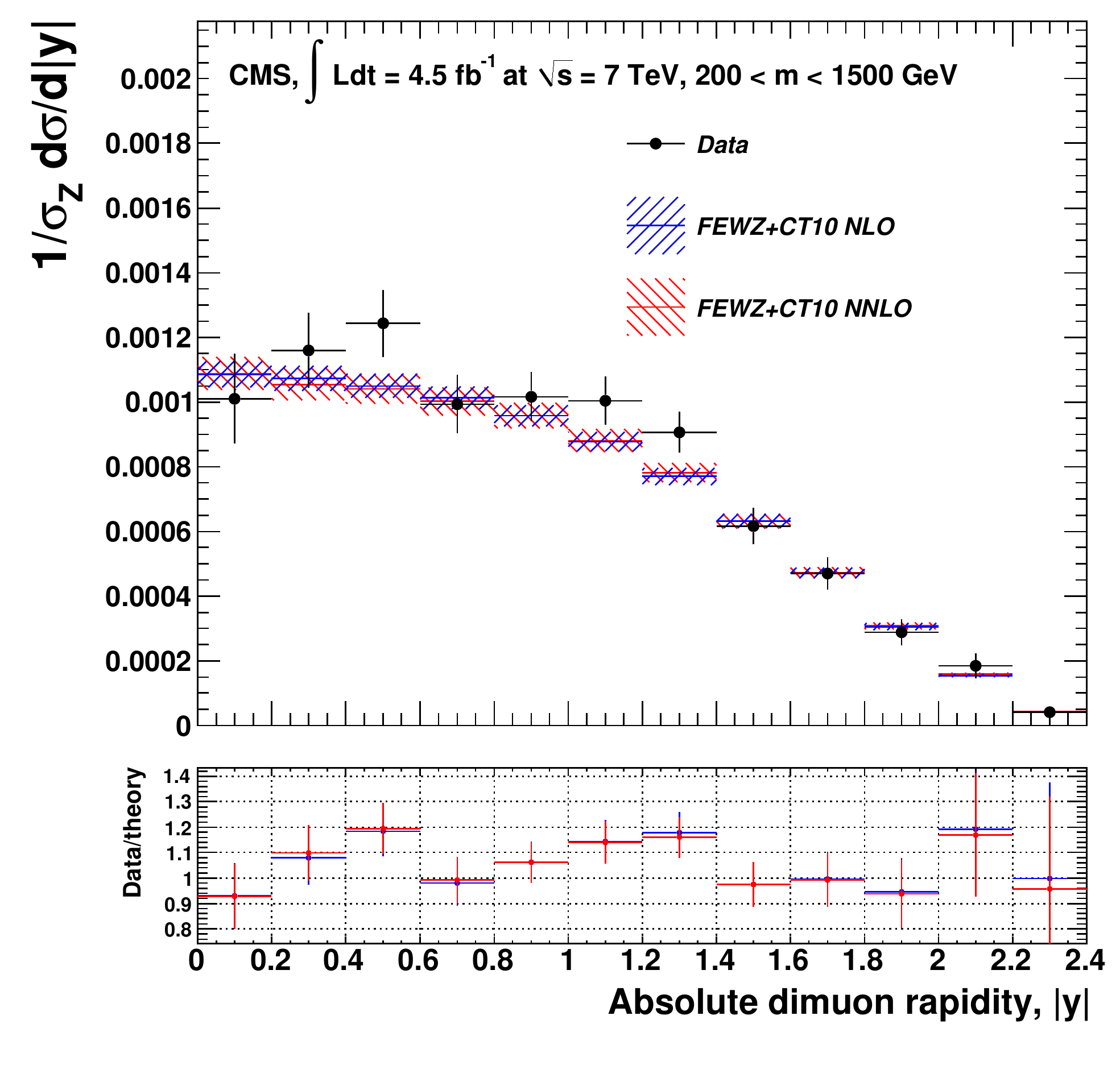}
\caption{
\label{fig:2Drshape}
The DY rapidity spectrum normalized to the $\Z$-peak region $(1/\sigma_{\Z}\hspace{0.05cm}\rd^2\sigma/d\abs{y})$,
plotted for different mass regions within the detector acceptance,
as measured and as predicted by NLO {\FEWZ}+CT10 PDF and NNLO {\FEWZ}+CT10 PDF calculations.
There are six mass bins between $20$ and $1500$\GeV, from left to right and from top to bottom.
The uncertainty bands in the theoretical predictions
combine the statistical and the PDF uncertainties (shaded bands).
The statistical component is negligible.
The smaller plots show the ratio of data to theoretical expectation.
}
}
\end{figure}

Figure~\ref{fig:2Drshape} shows the results for the double-differential cross section.
The results are compared to the {\FEWZ}+CT10 NLO PDF and {\FEWZ}+CT10 NNLO
PDF theoretical calculations. The results of the measurement are in a better agreement
with CT10 NNLO predictions than with CT10 NLO ones. The CT10 PDF set is a general-purpose
NLO PDF set with 52 eigenvectors that
uses a variable strong coupling $\alpha_{s} (M_{\Z})$ in the range 0.116--0.120 and
0.112--0.127.
The CT10 (NNLO) is also a general purpose PDF set.
It includes a part of the data sample for the D0 $\PW$-charge asymmetry measurement~\cite{bib:FNAL_late2} that is not included in the CT10 NLO fit.
The $\PW$-charge asymmetry data primarily modifies the slope of the ratio $d(x,Q^{2})/u(x,Q^{2})$ at large $x$.
The CT10 (NNLO) 
PDF set uses a variable strong coupling $\alpha_{s} (M_{\Z})$ in the range of
0.116--0.120 and 0.110--0.130.
We have chosen CT10 (NLO) and CT10 (NNLO) to compare with our measurement in Fig.~\ref{fig:2Drshape} because we have used the CT10 (NLO) for the \POWHEG MC signal sample.
The uncertainty bands in the theoretical expectations include the statistical and the PDF uncertainties
from the \FEWZ calculations summed in quadrature (shaded band).
The statistical uncertainty (solid band) is smaller than the PDF uncertainty and the latter
is the dominant uncertainty in the \FEWZ calculations.
In general, the PDF uncertainty assignment is different
for each PDF set. For instance, CT10 (NLO) and CT10 (NNLO) PDF uncertainties correspond to a 90\% CL, so, to get a consistent
comparison to other PDF sets the uncertainties are scaled to the 68\% CL.

In the low-mass region and the $\Z$-peak region, we observe good agreement between data
and theory. The NNLO effects are more significant in the low-mass region.
The corrections for the $\Pgg\Pgg$-initiated processes calculated with \FEWZ~3.1.b2
are negligible in the double-differential
cross section measurement, because the effects are approximately constant over the
investigated rapidity range and statistical fluctuations or other systematic uncertainties
are much larger across the invariant-mass range of the measurement.

\begin{figure}[htpb]
{\centering
\includegraphics[width=0.45\textwidth]{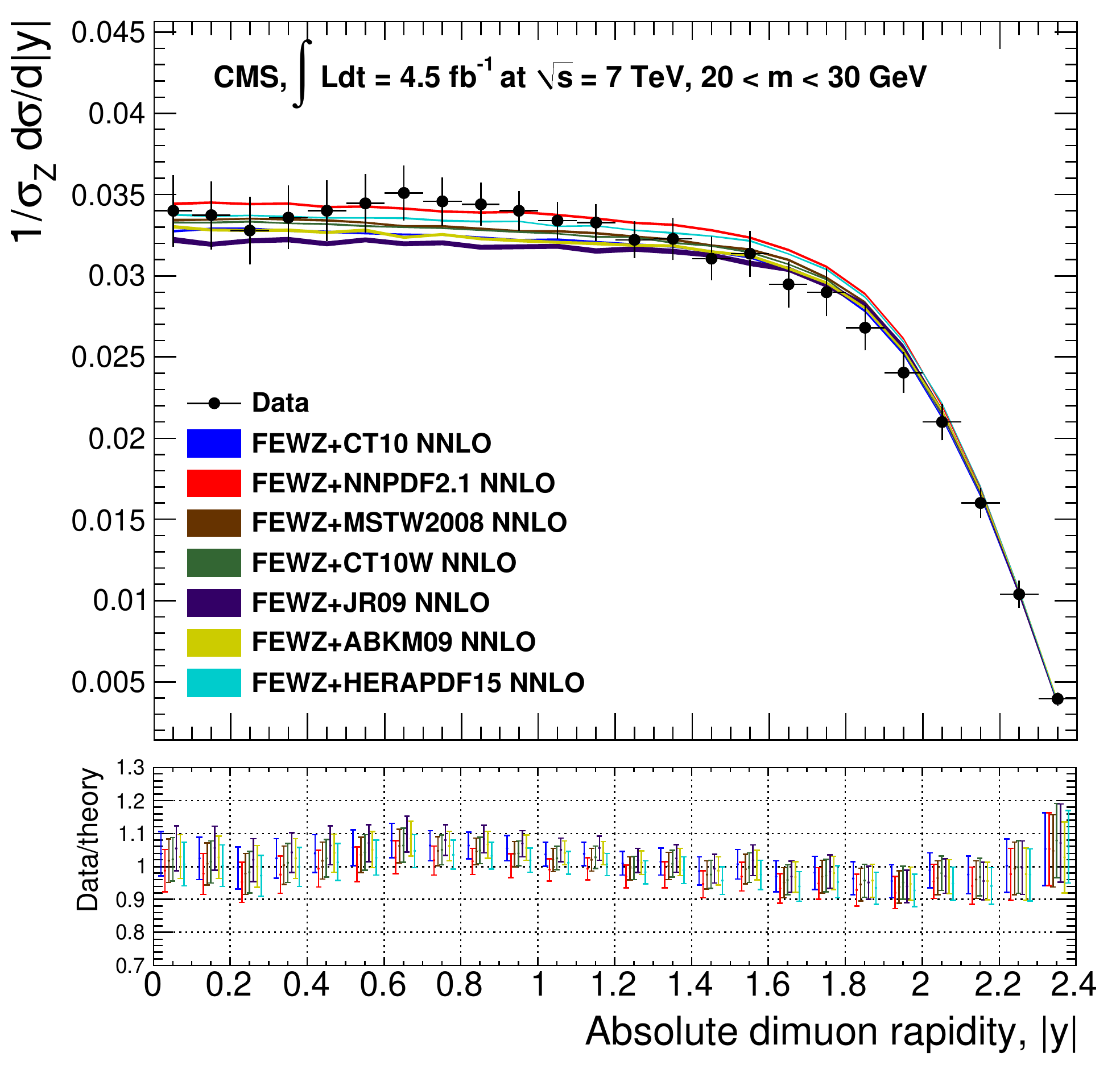}
\includegraphics[width=0.45\textwidth]{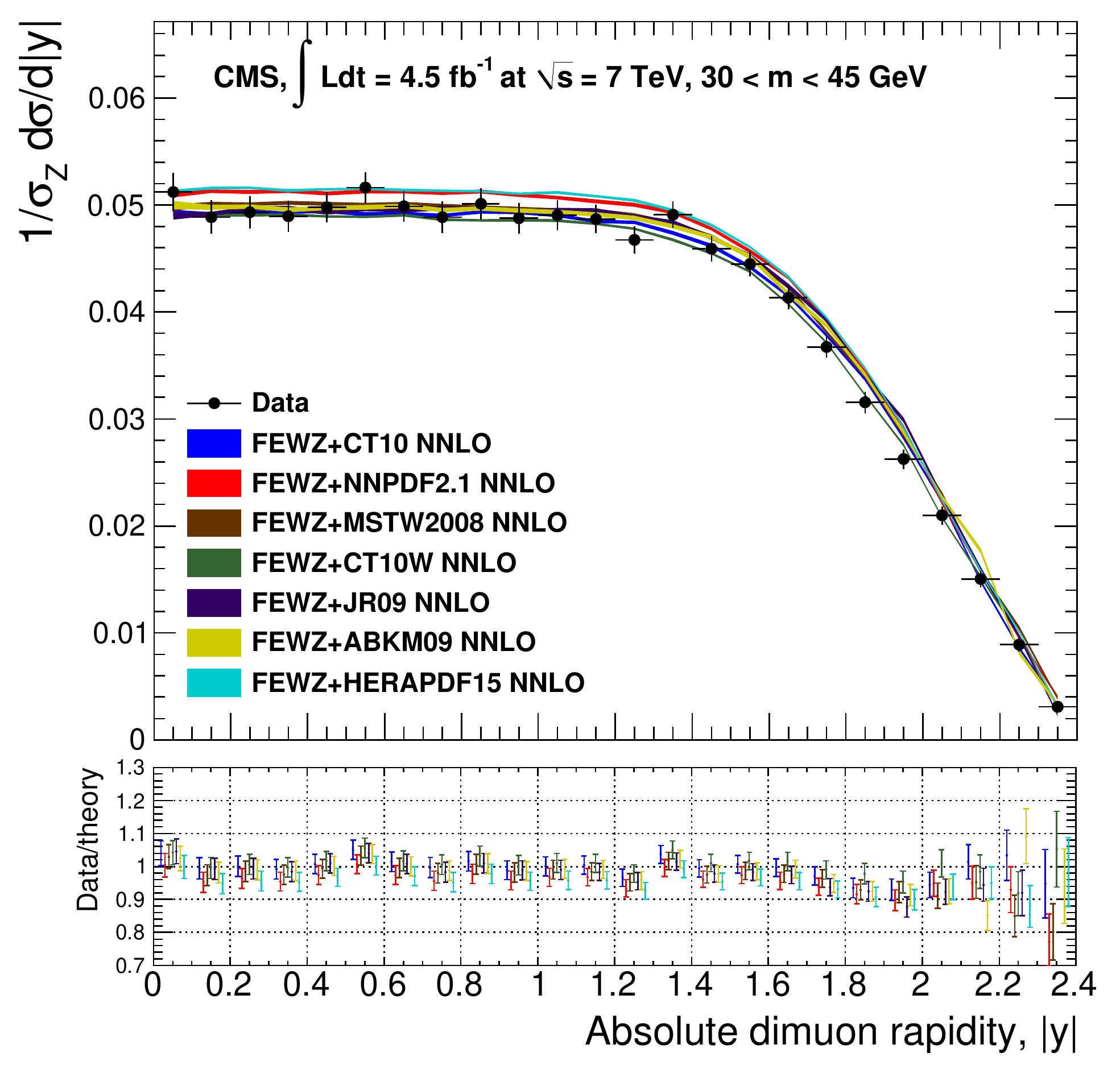}
\includegraphics[width=0.45\textwidth]{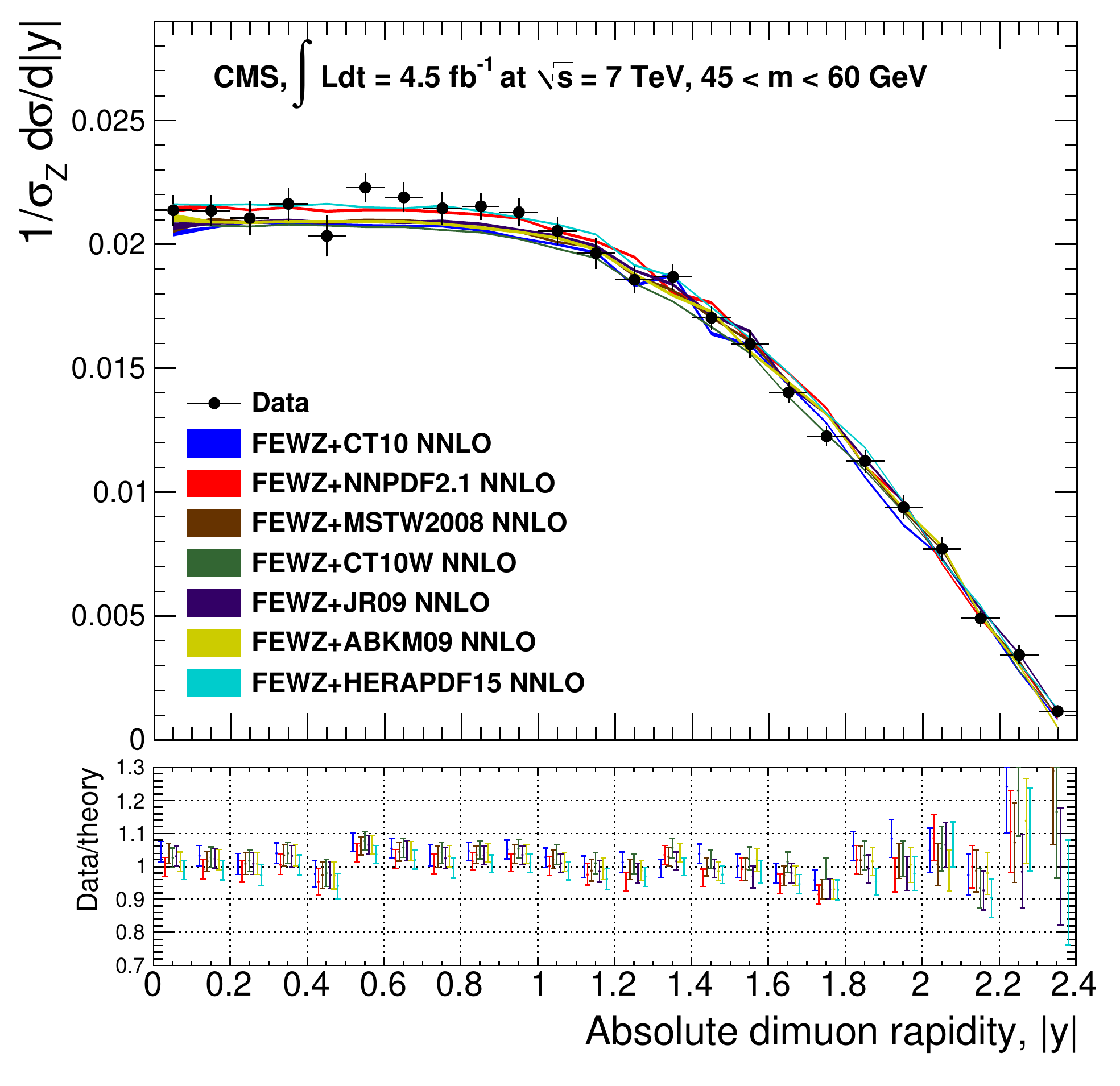}
\includegraphics[width=0.45\textwidth]{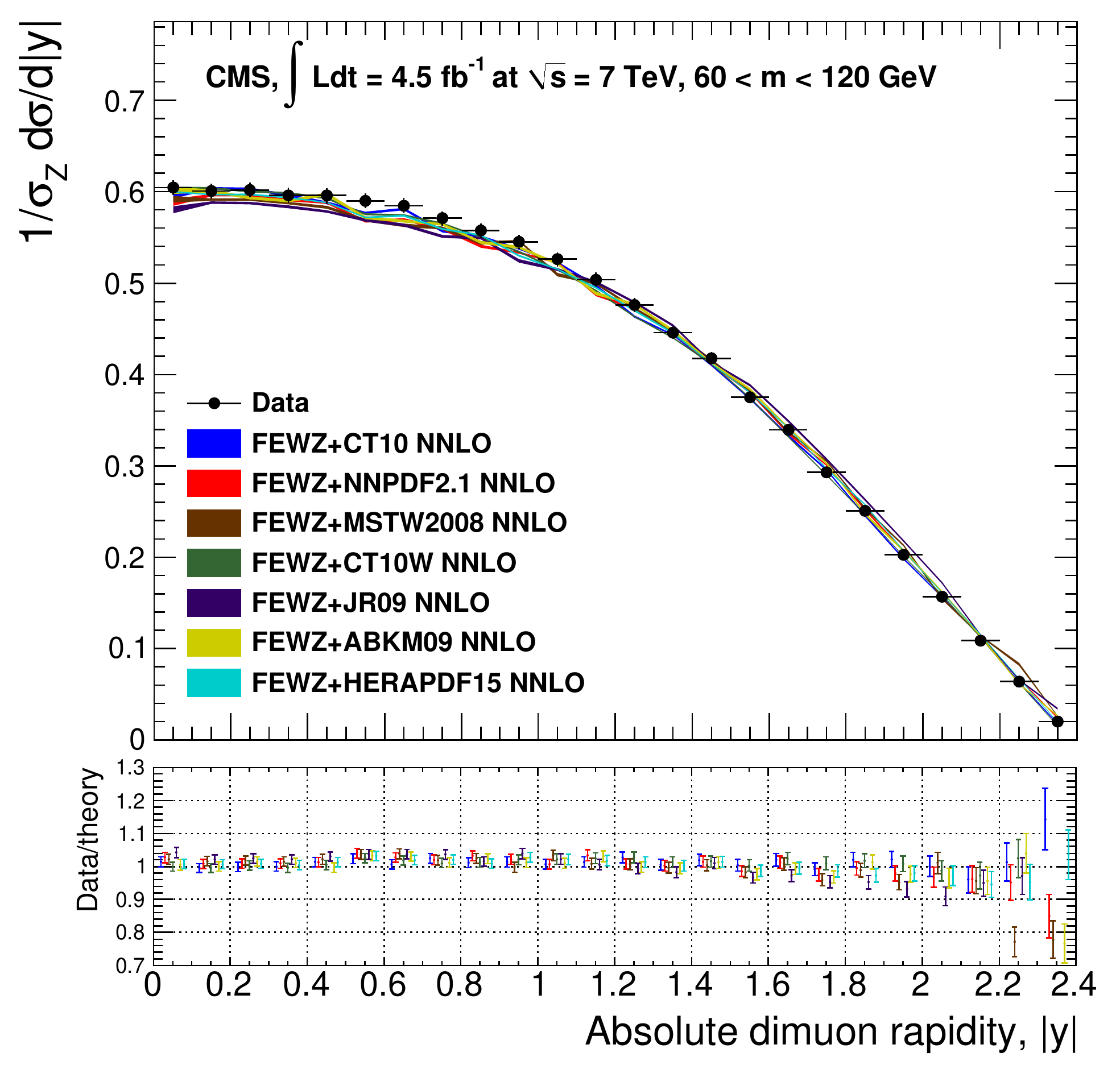}
\includegraphics[width=0.45\textwidth]{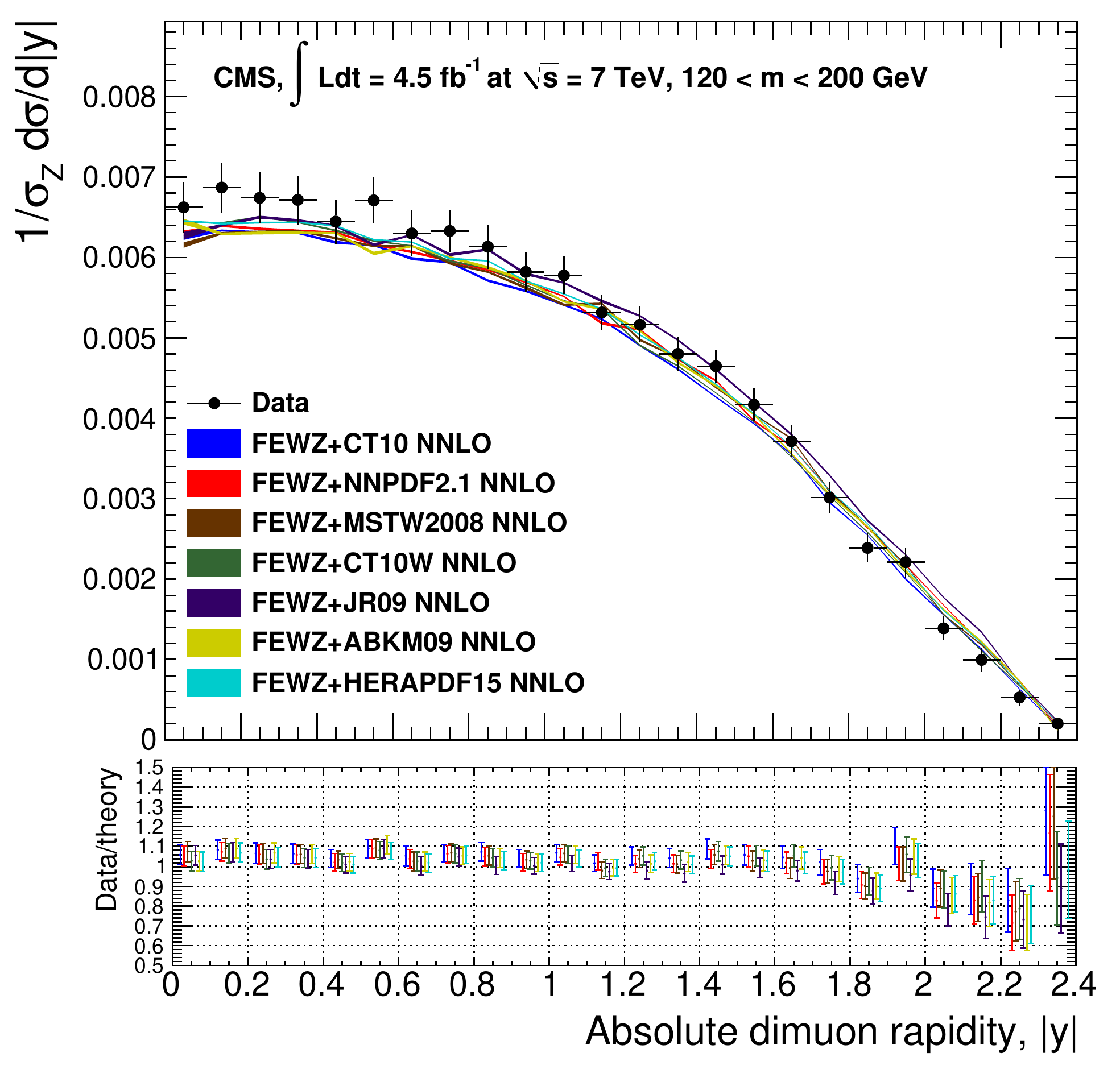}
\includegraphics[width=0.45\textwidth]{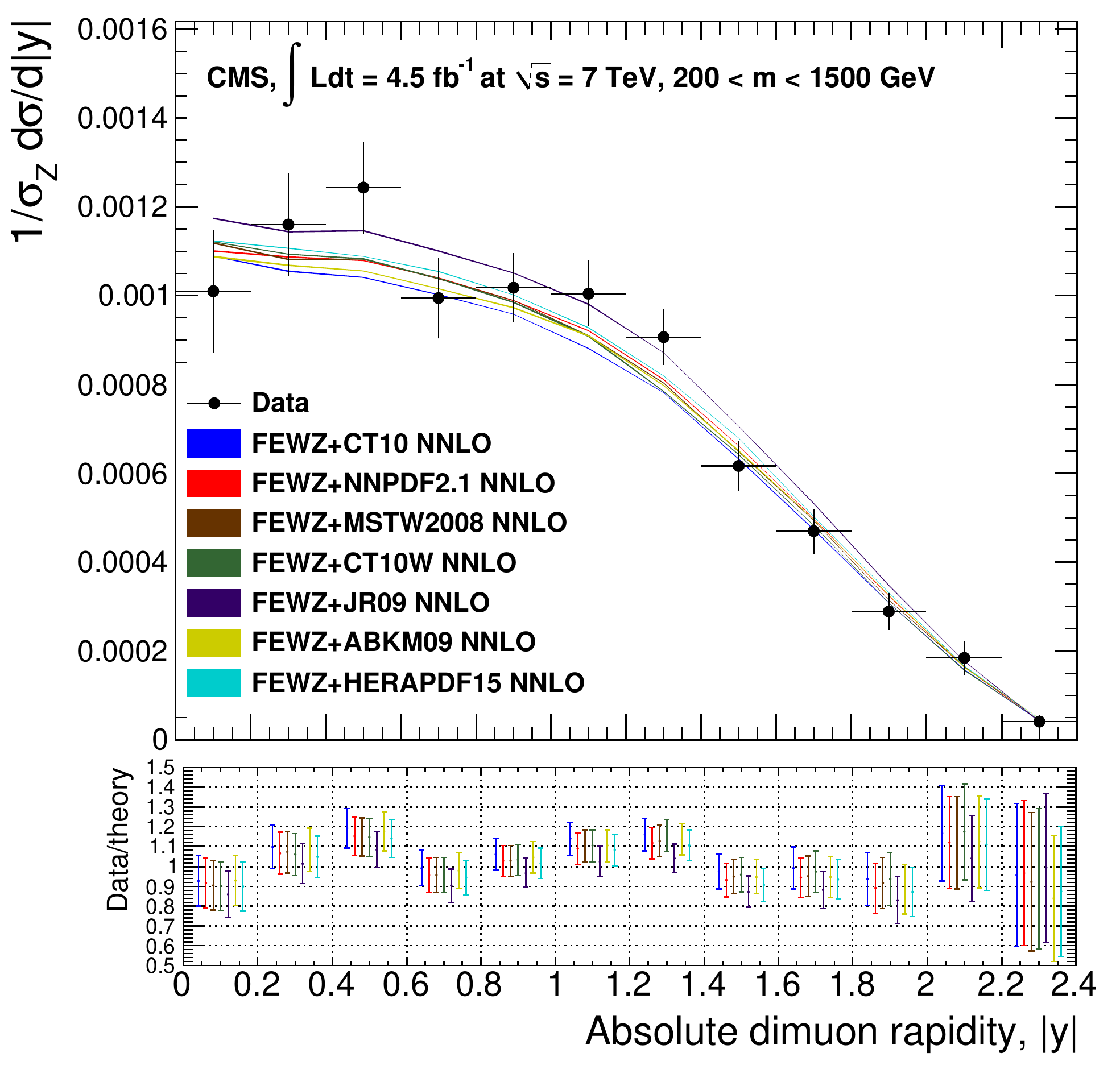}
\caption{
\label{fig:2Dfinal}
The DY rapidity spectrum normalized to the $\Z$-peak region $(1/\sigma_{\Z}\,\rd^2\sigma/\rd\abs{y})$,
compared to theoretical expectations using various PDF sets.
The uncertainty bands in the theoretical predictions
indicate the statistical uncertainty only.
The smaller plots show the ratio of data to theoretical expectation. The error bars include
the experimental uncertainty in the data and statistical
uncertainty in the theoretical expectation, combined quadratically.
}
}
\end{figure}

In order to assess the sensitivity of the double-differential cross section measurement to the PDF uncertainties,
we perform a comparison with the theoretical expectations calculated with various PDF sets.
Figure~\ref{fig:2Dfinal} shows the comparison
with currently available NNLO PDFs, most of which are from the pre-LHC era: CT10, CT10W, NNPDF2.1, HERAPDF15, MSTW2008, JR09, and ABKM09.

As seen in Fig.~\ref{fig:2Dfinal}, the predictions of various existing PDF sets are rather different,
especially in the low- and high-mass regions.
Given the uncertainties, the measurements provide sufficient sensitivity
to different PDFs and can be used to calculate a new generation of PDFs.
The uncertainty bands in the theoretical expectations in the figure indicate the statistical
uncertainty from the \FEWZ calculation.
Table~\ref{tab:normFactor2D} shows the statistical and the PDF uncertainties separately.

In the low-mass region (20--45\GeV), we observe that the values of the double-differential cross section
calculated with the NNPDF2.1 are higher than the values calculated with other PDF sets.
The NNPDF2.1 calculation shows good agreement with the measurement
result in the 20--30\GeV region, but it deviates from the measurement in the 30--45\GeV region by about 10\%.
In the peak region, all predictions are relatively close to each other and
agree well with the measurements.
At high mass the JR09 PDF calculation predicts significantly larger values
than other PDF sets. The statistical uncertainties in the measurements
for $m>200$\GeV are of the order of the spread in the theoretical predictions.

\begin{table} [htb]
\begin{center}
\topcaption{
  The DY dimuon rapidity spectrum within the detector acceptance,
  normalized to the $\Z$-peak region,
  ${r}_\text{pre-FSR, det} = (1/\sigma_{\Z}\,\rd\sigma/\rd\abs{y})$,
  tabulated for different mass regions.
  The rows are the dimuon rapidity bins and the columns are mass bins (in \GeVns).
  The uncertainties are the total experimental uncertainties.
}
\label{tab:rshape2D1}
\begin{tabular}{|c|cccccc|}
\hline
$\abs{y}$ & 20--30 & 30--45 & 45--60 & 60--120 & 120--200 & 200--1500 \\
& ($10^{-2}$) & ($10^{-2}$) & ($10^{-2}$) & ($10^{-1}$) & ($10^{-3}$) & ($10^{-4}$) \\
\hline
0.0--0.1 & 3.40 $\pm$ 0.22 & 5.12 $\pm$ 0.18 & 2.14 $\pm$ 0.06 & 6.05 $\pm$ 0.08 & 6.62 $\pm$ 0.31 & \multirow{2}{*}{10.1 $\pm$ 1.4} \\
0.1--0.2 & 3.37 $\pm$ 0.21 & 4.89 $\pm$ 0.16 & 2.13 $\pm$ 0.06 & 6.01 $\pm$ 0.08 & 6.87 $\pm$ 0.31 &  \\
\hline
0.2--0.3 & 3.28 $\pm$ 0.21 & 4.93 $\pm$ 0.15 & 2.11 $\pm$ 0.07 & 6.02 $\pm$ 0.08 & 6.74 $\pm$ 0.32 & \multirow{2}{*}{11.6 $\pm$ 1.2} \\
0.3--0.4 & 3.36 $\pm$ 0.20 & 4.89 $\pm$ 0.15 & 2.16 $\pm$ 0.07 & 5.96 $\pm$ 0.08 & 6.71 $\pm$ 0.30 &  \\
\hline
0.4--0.5 & 3.40 $\pm$ 0.19 & 4.98 $\pm$ 0.14 & 2.03 $\pm$ 0.08 & 5.96 $\pm$ 0.08 & 6.45 $\pm$ 0.27 & \multirow{2}{*}{12.4 $\pm$ 1.0} \\
0.5--0.6 & 3.45 $\pm$ 0.18 & 5.16 $\pm$ 0.14 & 2.23 $\pm$ 0.06 & 5.90 $\pm$ 0.08 & 6.71 $\pm$ 0.28 &  \\
\hline
0.6--0.7 & 3.51 $\pm$ 0.17 & 4.99 $\pm$ 0.14 & 2.19 $\pm$ 0.06 & 5.85 $\pm$ 0.08 & 6.30 $\pm$ 0.29 & \multirow{2}{*}{9.94 $\pm$ 0.91} \\
0.7--0.8 & 3.46 $\pm$ 0.15 & 4.89 $\pm$ 0.15 & 2.15 $\pm$ 0.07 & 5.71 $\pm$ 0.08 & 6.33 $\pm$ 0.26 &  \\
\hline
0.8--0.9 & 3.44 $\pm$ 0.13 & 5.01 $\pm$ 0.15 & 2.15 $\pm$ 0.06 & 5.57 $\pm$ 0.08 & 6.13 $\pm$ 0.28 & \multirow{2}{*}{10.17 $\pm$ 0.78} \\
0.9--1.0 & 3.40 $\pm$ 0.12 & 4.88 $\pm$ 0.14 & 2.13 $\pm$ 0.06 & 5.45 $\pm$ 0.08 & 5.82 $\pm$ 0.24 &  \\
\hline
1.0--1.1 & 3.34 $\pm$ 0.11 & 4.90 $\pm$ 0.14 & 2.05 $\pm$ 0.06 & 5.26 $\pm$ 0.08 & 5.78 $\pm$ 0.23 & \multirow{2}{*}{10.04 $\pm$ 0.74} \\
1.1--1.2 & 3.33 $\pm$ 0.11 & 4.87 $\pm$ 0.13 & 1.96 $\pm$ 0.06 & 5.04 $\pm$ 0.08 & 5.32 $\pm$ 0.22 &  \\
\hline
1.2--1.3 & 3.22 $\pm$ 0.11 & 4.67 $\pm$ 0.13 & 1.86 $\pm$ 0.06 & 4.76 $\pm$ 0.08 & 5.17 $\pm$ 0.22 & \multirow{2}{*}{9.07 $\pm$ 0.63} \\
1.3--1.4 & 3.23 $\pm$ 0.13 & 4.91 $\pm$ 0.13 & 1.87 $\pm$ 0.05 & 4.46 $\pm$ 0.07 & 4.80 $\pm$ 0.21 &  \\
\hline
1.4--1.5 & 3.11 $\pm$ 0.14 & 4.59 $\pm$ 0.12 & 1.70 $\pm$ 0.05 & 4.18 $\pm$ 0.07 & 4.65 $\pm$ 0.21 & \multirow{2}{*}{6.16 $\pm$ 0.56} \\
1.5--1.6 & 3.14 $\pm$ 0.14 & 4.45 $\pm$ 0.12 & 1.60 $\pm$ 0.06 & 3.75 $\pm$ 0.06 & 4.17 $\pm$ 0.21 &  \\
\hline
1.6--1.7 & 2.95 $\pm$ 0.14 & 4.13 $\pm$ 0.11 & 1.40 $\pm$ 0.04 & 3.39 $\pm$ 0.06 & 3.72 $\pm$ 0.20 & \multirow{2}{*}{4.70 $\pm$ 0.50} \\
1.7--1.8 & 2.90 $\pm$ 0.15 & 3.67 $\pm$ 0.10 & 1.22 $\pm$ 0.04 & 2.93 $\pm$ 0.06 & 3.01 $\pm$ 0.19 &  \\
\hline
1.8--1.9 & 2.68 $\pm$ 0.14 & 3.16 $\pm$ 0.10 & 1.13 $\pm$ 0.05 & 2.51 $\pm$ 0.05 & 2.39 $\pm$ 0.18 & \multirow{2}{*}{2.89 $\pm$ 0.41} \\
1.9--2.0 & 2.40 $\pm$ 0.13 & 2.63 $\pm$ 0.09 & 0.94 $\pm$ 0.05 & 2.03 $\pm$ 0.05 & 2.21 $\pm$ 0.19 &  \\
\hline
2.0--2.1 & 2.10 $\pm$ 0.11 & 2.10 $\pm$ 0.09 & 0.77 $\pm$ 0.05 & 1.57 $\pm$ 0.05 & 1.39 $\pm$ 0.15 & \multirow{2}{*}{1.84 $\pm$ 0.38} \\
2.1--2.2 & 1.60 $\pm$ 0.09 & 1.50 $\pm$ 0.08 & 0.49 $\pm$ 0.03 & 1.09 $\pm$ 0.04 & 1.00 $\pm$ 0.14 &  \\
\hline
2.2--2.3 & 1.04 $\pm$ 0.08 & 0.89 $\pm$ 0.06 & 0.34 $\pm$ 0.04 & 0.64 $\pm$ 0.04 & 0.53 $\pm$ 0.10 & \multirow{2}{*}{0.41 $\pm$ 0.16} \\
2.3--2.4 & 0.39 $\pm$ 0.04 & 0.31 $\pm$ 0.03 & 0.12 $\pm$ 0.02 & 0.20 $\pm$ 0.01 & 0.20 $\pm$ 0.05 &  \\
\hline
\end{tabular}
\end{center}
\end{table}

In addition to the ${r}^{ij}_\text{pre-FSR, det}$ measurement we report the cross section without
FSR correction, ${r}^{ij}_\text{post-FSR,~det}$.
The corresponding definition is

\begin{equation}
{r}^{ij}_\text{post-FSR,~det} = \frac{1}{\Delta y^{j}} \cdot \left(\left. \frac{N_\mathrm{u}^{'ij}}{\epsilon^{ij}\rho^{'ij}} \middle/ \frac{N_\mathrm{u}^{'\text{norm}}}{\epsilon^\text{norm}\rho^{'\text{norm}}} \right.\right),
\end{equation}

where $N_\mathrm{u}^{'ij}, N_\mathrm{u}^{'\text{norm}}, \rho^{'ij}$, and $\rho^{'\text{norm}}$ do not contain the FSR correction.
All the ${r}$ shape measurements are summarized in Tables~\ref{tab:rshape2D1}--\ref{tab:rshape2D2}.

\begin{table} [htb]
\begin{center}
\topcaption{
  The DY dimuon rapidity spectrum within the detector acceptance,
  normalized to the $\Z$-peak region,
  ${r}_\text{post-FSR, det} = (1/\sigma_{\Z}\,\rd\sigma/\rd\abs{y})$,
  tabulated for different mass regions.
  The rows are the dimuon rapidity bins and the columns are mass bins (in \GeVns).
  The uncertainties are the total experimental uncertainties.
}
\label{tab:rshape2D2}
\begin{tabular}{|c|cccccc|}
\hline
$\abs{y}$ & 20--30 & 30--45 & 45--60 & 60--120 & 120--200 & 200--1500 \\
& ($10^{-2}$) & ($10^{-2}$) & ($10^{-2}$) & ($10^{-1}$) & ($10^{-3}$) & ($10^{-4}$) \\
\hline
0.0--0.1 & 3.37 $\pm$ 0.22 & 5.16 $\pm$ 0.18 & 2.57 $\pm$ 0.07 & 6.03 $\pm$ 0.08 & 6.32 $\pm$ 0.35 & \multirow{2}{*}{9.6 $\pm$ 1.5} \\
0.1--0.2 & 3.33 $\pm$ 0.21 & 4.91 $\pm$ 0.16 & 2.57 $\pm$ 0.07 & 5.99 $\pm$ 0.08 & 6.51 $\pm$ 0.35 &  \\
\hline
0.2--0.3 & 3.26 $\pm$ 0.21 & 4.98 $\pm$ 0.15 & 2.53 $\pm$ 0.07 & 6.00 $\pm$ 0.08 & 6.42 $\pm$ 0.35 & \multirow{2}{*}{11.1 $\pm$ 1.3} \\
0.3--0.4 & 3.33 $\pm$ 0.20 & 4.94 $\pm$ 0.14 & 2.60 $\pm$ 0.07 & 5.94 $\pm$ 0.08 & 6.36 $\pm$ 0.33 &  \\
\hline
0.4--0.5 & 3.38 $\pm$ 0.19 & 5.01 $\pm$ 0.14 & 2.49 $\pm$ 0.07 & 5.94 $\pm$ 0.08 & 6.15 $\pm$ 0.31 & \multirow{2}{*}{11.9 $\pm$ 1.2} \\
0.5--0.6 & 3.42 $\pm$ 0.18 & 5.20 $\pm$ 0.15 & 2.66 $\pm$ 0.07 & 5.89 $\pm$ 0.08 & 6.34 $\pm$ 0.32 &  \\
\hline
0.6--0.7 & 3.48 $\pm$ 0.17 & 5.02 $\pm$ 0.15 & 2.62 $\pm$ 0.07 & 5.84 $\pm$ 0.08 & 5.97 $\pm$ 0.32 & \multirow{2}{*}{9.5 $\pm$ 1.0} \\
0.7--0.8 & 3.42 $\pm$ 0.15 & 4.93 $\pm$ 0.15 & 2.57 $\pm$ 0.07 & 5.70 $\pm$ 0.08 & 6.01 $\pm$ 0.30 &  \\
\hline
0.8--0.9 & 3.41 $\pm$ 0.14 & 5.04 $\pm$ 0.15 & 2.57 $\pm$ 0.07 & 5.56 $\pm$ 0.08 & 5.86 $\pm$ 0.32 & \multirow{2}{*}{9.71 $\pm$ 0.91} \\
0.9--1.0 & 3.37 $\pm$ 0.13 & 4.91 $\pm$ 0.15 & 2.53 $\pm$ 0.07 & 5.44 $\pm$ 0.08 & 5.57 $\pm$ 0.28 &  \\
\hline
1.0--1.1 & 3.30 $\pm$ 0.12 & 4.94 $\pm$ 0.15 & 2.46 $\pm$ 0.07 & 5.25 $\pm$ 0.08 & 5.50 $\pm$ 0.28 & \multirow{2}{*}{9.64 $\pm$ 0.88} \\
1.1--1.2 & 3.30 $\pm$ 0.12 & 4.89 $\pm$ 0.14 & 2.37 $\pm$ 0.07 & 5.03 $\pm$ 0.08 & 5.08 $\pm$ 0.26 &  \\
\hline
1.2--1.3 & 3.19 $\pm$ 0.12 & 4.70 $\pm$ 0.13 & 2.22 $\pm$ 0.07 & 4.75 $\pm$ 0.07 & 4.90 $\pm$ 0.26 & \multirow{2}{*}{8.62 $\pm$ 0.77} \\
1.3--1.4 & 3.19 $\pm$ 0.14 & 4.92 $\pm$ 0.13 & 2.20 $\pm$ 0.06 & 4.45 $\pm$ 0.07 & 4.53 $\pm$ 0.25 &  \\
\hline
1.4--1.5 & 3.09 $\pm$ 0.14 & 4.60 $\pm$ 0.13 & 2.02 $\pm$ 0.06 & 4.17 $\pm$ 0.07 & 4.40 $\pm$ 0.25 & \multirow{2}{*}{5.86 $\pm$ 0.65} \\
1.5--1.6 & 3.09 $\pm$ 0.15 & 4.45 $\pm$ 0.12 & 1.88 $\pm$ 0.05 & 3.75 $\pm$ 0.06 & 3.97 $\pm$ 0.24 &  \\
\hline
1.6--1.7 & 2.91 $\pm$ 0.15 & 4.12 $\pm$ 0.12 & 1.65 $\pm$ 0.05 & 3.39 $\pm$ 0.06 & 3.53 $\pm$ 0.23 & \multirow{2}{*}{4.49 $\pm$ 0.59} \\
1.7--1.8 & 2.86 $\pm$ 0.15 & 3.66 $\pm$ 0.11 & 1.45 $\pm$ 0.05 & 2.93 $\pm$ 0.06 & 2.86 $\pm$ 0.22 &  \\
\hline
1.8--1.9 & 2.64 $\pm$ 0.15 & 3.14 $\pm$ 0.10 & 1.31 $\pm$ 0.04 & 2.50 $\pm$ 0.05 & 2.28 $\pm$ 0.20 & \multirow{2}{*}{2.77 $\pm$ 0.50} \\
1.9--2.0 & 2.36 $\pm$ 0.13 & 2.61 $\pm$ 0.09 & 1.09 $\pm$ 0.04 & 2.02 $\pm$ 0.05 & 2.09 $\pm$ 0.21 &  \\
\hline
2.0--2.1 & 2.05 $\pm$ 0.11 & 2.08 $\pm$ 0.09 & 0.89 $\pm$ 0.04 & 1.56 $\pm$ 0.05 & 1.32 $\pm$ 0.17 & \multirow{2}{*}{1.77 $\pm$ 0.46} \\
2.1--2.2 & 1.56 $\pm$ 0.10 & 1.48 $\pm$ 0.07 & 0.58 $\pm$ 0.03 & 1.09 $\pm$ 0.04 & 0.94 $\pm$ 0.15 &  \\
\hline
2.2--2.3 & 1.00 $\pm$ 0.08 & 0.88 $\pm$ 0.07 & 0.39 $\pm$ 0.02 & 0.64 $\pm$ 0.04 & 0.50 $\pm$ 0.11 & \multirow{2}{*}{0.39 $\pm$ 0.20} \\
2.3--2.4 & 0.37 $\pm$ 0.04 & 0.30 $\pm$ 0.03 & 0.13 $\pm$ 0.01 & 0.20 $\pm$ 0.01 & 0.19 $\pm$ 0.06 &  \\
\hline
\end{tabular}
\end{center}
\end{table}

These double-differential DY measurements will impose constraints on the quark
and antiquark PDFs in a wide range of $x$, and in particular, should allow the replacement of fixed-target
DY data with modern collider data in PDF analyses. Such replacement would be advantageous
because fixed-target data were taken at low energies (thus being affected by larger theoretical uncertainties),
mostly on nuclear targets (requiring nuclear corrections), and the full experimental covariance matrices are not provided.
Fixed-target DY data has been instrumental to constrain quark flavor
separation in global PDF analyses in the last 20 years, thus present measurements should become a
crucial source of information on quark and antiquark PDFs in future global fits.
\section{Summary}
\label{sec:summary}
This paper presents measurements of the Drell--Yan differential cross section $\rd\sigma/\rd{}m$ in the dimuon and dielectron channels for the mass range
$15 < m < 1500$\GeV
 and the double-differential cross section $\rd^2\sigma/\rd{}m\,\rd\abs{y}$
 in the dimuon channel  for the mass range $20 < m < 1500$\GeV in proton-proton collisions at $\sqrt{s} = 7$\TeV.
The inclusive Z cross section measurements in the mass range $60 < m < 120$\GeV are also presented and
these are the most precise measurements of the Z cross section at a hadron collider.

The differential cross section measurements are normalized to the $\Z$-peak region (60--120\GeV),
canceling the uncertainty
in the integrated luminosity and reducing the PDF uncertainty in the acceptance, the pileup effect in the
reconstruction efficiency, and the uncertainty of the efficiency. The measurements are corrected
for the effects of resolution, which cause event migration between bins in mass and rapidity.
The observed dilepton mass is also corrected for final-state photon radiation.
The $\rd\sigma/\rd{}m$ differential cross section results
are given separately for both lepton flavors in the fiducial region and are extrapolated to the full phase space.
Since the electron and muon results are consistent, they are combined.
The results are in good agreement with the
standard model predictions, calculated at NNLO with the program \FEWZ using the CT10 PDF set.

The $\rd^2\sigma/\rd{}m\,\rd\abs{y}$ measurement is compared to the NLO prediction calculated with \FEWZ using the CT10 PDFs
and the NNLO theoretical predictions as computed with \FEWZ
 using the CT10, NNPDF2.1, MSTW2008, HERAPDF15, JR09, ABKM09, and CT10W PDFs.
This is the first double-differential Drell--Yan cross section measurement with a hadron collider
and will provide precise inputs to update the PDF sets.

\section*{Acknowledgments}

\hyphenation{Bundes-ministerium Forschungs-gemeinschaft Forschungs-zentren}
We would like to thank the authors of \FEWZ for the fruitful discussions,
cooperation, and cross-checks in performing the theoretical calculations for our analysis.

\hyphenation{Bundes-ministerium Forschungs-gemeinschaft Forschungs-zentren} We congratulate our colleagues in the CERN accelerator departments for the excellent performance of the LHC and thank the technical and administrative staffs at CERN and at other CMS institutes for their contributions to the success of the CMS effort. In addition, we gratefully acknowledge the computing centres and personnel of the Worldwide LHC Computing Grid for delivering so effectively the computing infrastructure essential to our analyses. Finally, we acknowledge the enduring support for the construction and operation of the LHC and the CMS detector provided by the following funding agencies: the Austrian Federal Ministry of Science and Research and the Austrian Science Fund; the Belgian Fonds de la Recherche Scientifique, and Fonds voor Wetenschappelijk Onderzoek; the Brazilian Funding Agencies (CNPq, CAPES, FAPERJ, and FAPESP); the Bulgarian Ministry of Education and Science; CERN; the Chinese Academy of Sciences, Ministry of Science and Technology, and National Natural Science Foundation of China; the Colombian Funding Agency (COLCIENCIAS); the Croatian Ministry of Science, Education and Sport; the Research Promotion Foundation, Cyprus; the Ministry of Education and Research, Recurrent financing contract SF0690030s09 and European Regional Development Fund, Estonia; the Academy of Finland, Finnish Ministry of Education and Culture, and Helsinki Institute of Physics; the Institut National de Physique Nucl\'eaire et de Physique des Particules~/~CNRS, and Commissariat \`a l'\'Energie Atomique et aux \'Energies Alternatives~/~CEA, France; the Bundesministerium f\"ur Bildung und Forschung, Deutsche Forschungsgemeinschaft, and Helmholtz-Gemeinschaft Deutscher Forschungszentren, Germany; the General Secretariat for Research and Technology, Greece; the National Scientific Research Foundation, and National Innovation Office, Hungary; the Department of Atomic Energy and the Department of Science and Technology, India; the Institute for Studies in Theoretical Physics and Mathematics, Iran; the Science Foundation, Ireland; the Istituto Nazionale di Fisica Nucleare, Italy; the Korean Ministry of Education, Science and Technology and the World Class University program of NRF, Republic of Korea; the Lithuanian Academy of Sciences; the Mexican Funding Agencies (CINVESTAV, CONACYT, SEP, and UASLP-FAI); the Ministry of Business, Innovation and Employment, New Zealand; the Pakistan Atomic Energy Commission; the Ministry of Science and Higher Education and the National Science Centre, Poland; the Funda\c{c}\~ao para a Ci\^encia e a Tecnologia, Portugal; JINR, Dubna; the Ministry of Education and Science of the Russian Federation, the Federal Agency of Atomic Energy of the Russian Federation, Russian Academy of Sciences, and the Russian Foundation for Basic Research; the Ministry of Education, Science and Technological Development of Serbia; the Secretar\'{\i}a de Estado de Investigaci\'on, Desarrollo e Innovaci\'on and Programa Consolider-Ingenio 2010, Spain; the Swiss Funding Agencies (ETH Board, ETH Zurich, PSI, SNF, UniZH, Canton Zurich, and SER); the National Science Council, Taipei; the Thailand Center of Excellence in Physics, the Institute for the Promotion of Teaching Science and Technology of Thailand, Special Task Force for Activating Research and the National Science and Technology Development Agency of Thailand; the Scientific and Technical Research Council of Turkey, and Turkish Atomic Energy Authority; the Science and Technology Facilities Council, UK; the US Department of Energy, and the US National Science Foundation.

Individuals have received support from the Marie-Curie programme and the European Research Council and EPLANET (European Union); the Leventis Foundation; the A. P. Sloan Foundation; the Alexander von Humboldt Foundation; the Belgian Federal Science Policy Office; the Fonds pour la Formation \`a la Recherche dans l'Industrie et dans l'Agriculture (FRIA-Belgium); the Agentschap voor Innovatie door Wetenschap en Technologie (IWT-Belgium); the Ministry of Education, Youth and Sports (MEYS) of Czech Republic; the Council of Science and Industrial Research, India; the Compagnia di San Paolo (Torino); the HOMING PLUS programme of Foundation for Polish Science, cofinanced by EU, Regional Development Fund; and the Thalis and Aristeia programmes cofinanced by EU-ESF and the Greek NSRF.
\newpage
\bibliography{auto_generated}   % will be created by the tdr script.

\cleardoublepage \appendix\section{The CMS Collaboration \label{app:collab}}\begin{sloppypar}\hyphenpenalty=5000\widowpenalty=500\clubpenalty=5000\textbf{Yerevan Physics Institute,  Yerevan,  Armenia}\\*[0pt]
S.~Chatrchyan, V.~Khachatryan, A.M.~Sirunyan, A.~Tumasyan
\vskip\cmsinstskip
\textbf{Institut f\"{u}r Hochenergiephysik der OeAW,  Wien,  Austria}\\*[0pt]
W.~Adam, T.~Bergauer, M.~Dragicevic, J.~Er\"{o}, C.~Fabjan\cmsAuthorMark{1}, M.~Friedl, R.~Fr\"{u}hwirth\cmsAuthorMark{1}, V.M.~Ghete, N.~H\"{o}rmann, J.~Hrubec, M.~Jeitler\cmsAuthorMark{1}, W.~Kiesenhofer, V.~Kn\"{u}nz, M.~Krammer\cmsAuthorMark{1}, I.~Kr\"{a}tschmer, D.~Liko, I.~Mikulec, D.~Rabady\cmsAuthorMark{2}, B.~Rahbaran, C.~Rohringer, H.~Rohringer, R.~Sch\"{o}fbeck, J.~Strauss, A.~Taurok, W.~Treberer-Treberspurg, W.~Waltenberger, C.-E.~Wulz\cmsAuthorMark{1}
% ...
\vskip\cmsinstskip
% ...
1:~~Also at Vienna University of Technology, Vienna, Austria\\
2:~~Also at CERN, European Organization for Nuclear Research, Geneva, Switzerland\\
% ...

\end{sloppypar}
\end{document}